\documentclass[12pt]{article}
\pdfoutput=1
\usepackage[utf8x]{inputenc}
\usepackage[colorlinks,linkcolor=blue,citecolor=blue,bookmarks,bookmarksnumbered]{hyperref}
\usepackage[scaled=0.85]{helvet}
\usepackage{amsmath,amssymb,accents,mathrsfs,XoohmE}
\usepackage{graphicx,color}
\usepackage{booktabs}
\usepackage{multirow}
\usepackage{placeins}
\usepackage{amsmath}

\usepackage{cite}

\usepackage{enumitem}

\usepackage{XoohmE}

\def\rD{{\rm D}}
\def\rP{{\rm P}}

\def\bDb{\hbox{\kern2pt\vrule height10pt depth-9.2pt width6pt\kern-9pt{$\boldsymbol D$}}\mkern-2mu}

\def\bQb{\hbox{\kern2pt\vrule height10pt depth-9.2pt width6pt\kern-9pt{$\boldsymbol Q$}}}
\def\bA{{\bs{A}}}
\def\bB{{\bs{B}}}
\def\bF{{\bs{F}}}
\def\bG{{\bs{G}}}

\def\bH{{\bs{H}}}

\def\bT{{\bs{T}}}

\def\bX{{\bs{X}}}
\def\bbX{{\Bar{\bs{X}}}}

\def\Bj{{\bs\psi}}
\def\Bl{{\bs\lambda}}
\def\Bvp{{\bs\varphi}}
\def\Bc{{\bs\chi}}

\def\BF{{\bs\Phi}}
\def\bBF{{\Bar{\bs{\Phi}}}}

\def\brd{{\bm{\rm d}}}

\def\a{\alpha}
\def\b{\beta}
\def\c{\chi}
\def\d{\delta}
\def\e{\epsilon}

\def\g{\gamma}

\def\i{\iota}
\def\j{\psi}
\def\k{\kappa}
\def\l{\lambda}
\def\m{\mu}
\def\n{\nu}
\def\o{\omega}

\def\r{\rho}
\def\s{\sigma}
\def\t{\tau}

\def\L{\Lambda}
\def\O{\Omega}
\def\P{\Pi}

\def\S{\Sigma}

\def\ca{{\cal A}}
\def\cb{{\cal B}}
\def\cc{{\cal C}}
\def\cd{{\cal D}}
\def\ce{{\cal E}}
\def\cf{{\cal F}}
\def\cg{{\cal G}}

\def\cj{{\cal J}}

\def\cl{{\cal L}}

\def\cn{{\cal N}}
\def\co{{\cal O}}
\def\cp{{\cal P}}

\def\cw{{\cal W}}

\def\cy{{\cal Y}}

\def\tca{\Tilde{\cal A}}
\def\tcb{\Tilde{\cal B}}
\def\tcc{\Tilde{\cal C}}
\def\tcd{\Tilde{\cal D}}
\def\vca{\Check{\cal A}}
\def\vcb{\Check{\cal B}}
\def\vcc{\Check{\cal C}}
\def\vcd{\Check{\cal D}}
\def\vce{\Check{\cal E}}

\def\hcb{\Hat{\cal B}}

\definecolor{Green}  {rgb}{0.10,0.70,0.10} 
\definecolor{Orange} {rgb}{1.00,0.50,0.15} 
\definecolor{Red}    {rgb}{0.90,0.00,0.12} 
\definecolor{Purple} {rgb}{0.50,0.25,0.55} 
\definecolor{Turque} {rgb}{0.00,0.65,0.85} 
\definecolor{Blue}   {rgb}{0.00,0.00,1.00} 
\definecolor{Magenta}{rgb}{1.00,0.00,1.00} 
\definecolor{Gold}   {rgb}{1.00,0.75,0.25} 
\definecolor{Seaweed}{rgb}{0.01,0.24,0.09} 
\definecolor{Brown}  {rgb}{0.43,0.26,0.32} 
\definecolor{grey1}  {rgb}{0.20,0.20,0.20} 
\definecolor{grey2}  {rgb}{0.40,0.40,0.40} 
\definecolor{grey3}  {rgb}{0.60,0.60,0.60} 
\definecolor{grey4}  {rgb}{0.80,0.80,0.80} 
\definecolor{grey5}  {rgb}{0.90,0.90,0.90} 
\def\C#1#2{{\ifcase#1\or
             \color{Green}\or \color{Orange}\or \color{Red}\or
              \color{Purple}\or \color{Turque}\or \color{Blue}\or
               \color{Magenta}\or \color{Gold}\or \color{Seaweed}\or
                \color{Brown}\or\color{grey1}\or\color{grey2}\or
                 \color{grey3}\else\color{grey4}\fi#2}}

\definecolor{Slate} {rgb}{0.00,0.45,0.55}

\usepackage[enableskew,vcentermath]{youngtab}

\definecolor{Hey}{rgb}{.9,.05,.4}
\definecolor{orange}{rgb}{1,.5,0}
\definecolor{plum}{rgb}{.4,0,.6}
\definecolor{R}{rgb}{1,0,0}
\definecolor{G}{rgb}{0.1,0.7,0}
\definecolor{B}{rgb}{0,0,1}
\usepackage{lipsum}
\usepackage{listings}
\definecolor{MyDarkGreen}{rgb}{0.0,0.4,0.0} 
\def\fracm#1#2{\hbox{\large{${\frac{{#1}}{{#2}}}$}}}

\def\be{\begin{equation}}
\def\ee{\end{equation}}
\newcommand{\bea}{\begin{eqnarray}}
\newcommand{\eea}{\end{eqnarray}}
\newcommand{\ena}{\end{eqnarray}}

\def\pp{{\mathchoice
          {
              \kern 1pt%
              \raise 1pt
              \vbox{\hrule width5pt height0.4pt depth0pt
                    \kern -2pt
                    \hbox{\kern 2.3pt
                          \vrule width0.4pt height6pt depth0pt
                          }
                    \kern -2pt
                    \hrule width5pt height0.4pt depth0pt}%
                    \kern 1pt
           }
            {
              \kern 1pt%
              \raise 1pt
              \vbox{\hrule width4.3pt height0.4pt depth0pt
                    \kern -1.8pt
                    \hbox{\kern 1.95pt
                          \vrule width0.4pt height5.4pt depth0pt
                          }
                    \kern -1.8pt
                    \hrule width4.3pt height0.4pt depth0pt}%
                    \kern 1pt
            }
            {
              \kern 0.5pt%
              \raise 1pt
              \vbox{\hrule width4.0pt height0.3pt depth0pt
                    \kern -1.9pt  
                    \hbox{\kern 1.85pt
                          \vrule width0.3pt height5.7pt depth0pt
                          }
                    \kern -1.9pt
                    \hrule width4.0pt height0.3pt depth0pt}%
                    \kern 0.5pt
            }
            {
              \kern 0.5pt%
              \raise 1pt
              \vbox{\hrule width3.6pt height0.3pt depth0pt
                    \kern -1.5pt
                    \hbox{\kern 1.65pt
                          \vrule width0.3pt height4.5pt depth0pt
                          }
                    \kern -1.5pt
                    \hrule width3.6pt height0.3pt depth0pt}%
                    \kern 0.5pt
            }
        }}

\def\mm{{\mathchoice
                       {
                             \kern 1pt
               \raise 1pt    \vbox{\hrule width5pt height0.4pt depth0pt
                                  \kern 2pt
                                  \hrule width5pt height0.4pt depth0pt}
                             \kern 1pt}
                       {
                            \kern 1pt
               \raise 1pt \vbox{\hrule width4.3pt height0.4pt depth0pt
                                  \kern 1.8pt
                                  \hrule width4.3pt height0.4pt depth0pt}
                             \kern 1pt}
                       {
                            \kern 0.5pt
               \raise 1pt
                            \vbox{\hrule width4.0pt height0.3pt depth0pt
                                  \kern 1.9pt
                                  \hrule width4.0pt height0.3pt depth0pt}
                            \kern 1pt}
                       {
                           \kern 0.5pt
             \raise 1pt  \vbox{\hrule width3.6pt height0.3pt depth0pt
                                  \kern 1.5pt
                                  \hrule width3.6pt height0.3pt depth0pt}
                           \kern 0.5pt}
                       }}

\def\ad{{\kern0.5pt
                   \alpha \kern-5.05pt \raise5.8pt\hbox{$\textstyle.$}\kern
0.5pt}}

\def\qd{{\kern0.5pt
                   q \kern-5.05pt \raise5.8pt\hbox{$\textstyle.$}\kern
0.5pt}}
\def\Dot#1{{\kern0.5pt
     {#1} \kern-5.05pt \raise5.8pt\hbox{$\textstyle.$}\kern
0.5pt}}

\catcode`@=11
\def\un#1{\relax\ifmmode\@@underline#1\else
        $\@@underline{\hbox{#1}}$\relax\fi}
\catcode`@=12

\def\dslash{\not{\hbox{\kern-2pt $\partial$}}}
\def\Dslash{\not{\hbox{\kern-4pt $D$}}}
\def\pslash{\not{\hbox{\kern-2.3pt $p$}}}
 \newtoks\slashfraction
 \slashfraction={.13}
 \def\slash#1{\setbox0\hbox{$ #1 $}
 \setbox0\hbox to \the\slashfraction\wd0{\hss \box0}/\box0 }
 
\def\kcr{{\hbox{\ro \char'170}}}                
\def\ktl{{\hbox{\ro \char'170}}}        
\def\ktr{{\hbox{\ro \char'170}}}        
\def\kbl{{\hbox{\ro \char'170}}}        
\def\kbr{{\hbox{\ro \char'170}}}        

\def\plpl{\raise-2pt\hbox{$\raise3pt\hbox{$_+$}\hskip-6.67pt\raise0.0pt
\hbox{$^+$}\hskip 0.01pt$}}
\def\mimi{\raise-2pt\hbox{$\raise3pt\hbox{$_-$}\hskip-6.67pt\raise0.0pt
\hbox{$^-$}\hskip 0.01pt$}}

\def\bo{{\raise.15ex\hbox{\large$\Box$}}}               
\def\pa{\partial}                                       
\def\TH{{\raise.2ex\hbox{$\displaystyle \bigodot$}\mskip-4.7mu \llap H \;}}
\def\face{{\raise.2ex\hbox{$\displaystyle \bigodot$}\mskip-2.2mu \llap {$\ddot
        \smile$}}}                                      

\def\dt#1{\on{\hbox{\bf .}}{#1}}                
\def\Dot#1{\dt{#1}}

\def\Tilde#1{\widetilde{#1}}                    
\def\Hat#1{\widehat{#1}}                        
\def\Bar#1{\overline{#1}}                       
\def\leftrightarrowfill{$\mathsurround=0pt \mathord\leftarrow \mkern-6mu
        \cleaders\hbox{$\mkern-2mu \mathord- \mkern-2mu$}\hfill
        \mkern-6mu \mathord\rightarrow$}
\def\dvec#1{\vbox{\ialign{##\crcr
        \leftrightarrowfill\crcr\noalign{\kern-1pt\nointerlineskip}
        $\hfil\displaystyle{#1}\hfil$\crcr}}}           
\def\dt#1{{\buildrel {\hbox{\LARGE .}} \over {#1}}}     

\def\fracm#1#2{\hbox{\large{${\frac{{#1}}{{#2}}}$}}}
\def\sfrac#1#2{{\vphantom1\smash{\lower.5ex\hbox{\small$#1$}}\over
        \vphantom1\smash{\raise.4ex\hbox{\small$#2$}}}} 
\def\bfrac#1#2{{\vphantom1\smash{\lower.5ex\hbox{$#1$}}\over
        \vphantom1\smash{\raise.3ex\hbox{$#2$}}}}       
\def\afrac#1#2{{\vphantom1\smash{\lower.5ex\hbox{$#1$}}\over#2}}    

\def\pa{\partial}      
\let\bm\relax
\newcommand{\bm}[1]{{\boldsymbol{#1}}}

\def\ad{{\Dot{\alpha}}}

 \font\rOpe=cmsy10                        
 \def\ktl{{\hbox{\rOpe\char'170}}}        
 \def\kbl{{\hbox{\rOpe\char'170}}}        
 \def\kcr{{\reflectbox{\rOpe\char'170}}}        
 \def\ktr{{\reflectbox{\rOpe\char'170}}}        
 \def\kbr{{\reflectbox{\rOpe\char'170}}}        
 \def\Border{\vbox{\hsize0pt
        \setlength{\unitlength}{1mm}
        \newcount\xco
        \newcount\yco
        \xco=-21
        \yco=12
        \begin{picture}(0,0)(-7.5,0)
        \put(\xco,\yco){$\ktl$}
        \advance\yco by-1
        {\loop
        \put(\xco,\yco){$\kcr$}
        \advance\yco by-2
        \ifnum\yco>-240
        \repeat
        \put(\xco,\yco){$\kbl$}}
        \xco=170
        \yco=12
        \put(\xco,\yco){$\ktr$}
        \advance\yco by-1
        {\loop
        \put(\xco,\yco){$\kcr$}
        \advance\yco by-2
        \ifnum\yco>-240
        \repeat
        \put(\xco,\yco){$\kbr$}}
        \put(-19.5,13){\scalebox{.6065}{%
         University of Maryland Center for String and Particle  Theory \&\ Physics Department%
        |University of Maryland Center for String and Particle  Theory \&\ Physics Department}}
        \put(-19.5,-241.5){\scalebox{.5835}{%
         ****University of Maryland * Center for String and
         Particle  Theory* Physics Department****University of Maryland *Center
        for String and Particle  Theory* Physics Department}}
        \end{picture}
        \par\vskip-8mm}}
\definecolor{UMred}{rgb}{.9,.05,.2}
\definecolor{HUblue}{rgb}{.0,.3,.7}

\definecolor{Red}    {rgb}{0.90,0.00,0.12} 
\definecolor{Blue}   {rgb}{0.00,0.00,1.00} 
\definecolor{Green}  {rgb}{0.10,0.70,0.10} 
\definecolor{Turque} {rgb}{0.00,0.65,0.85} 
\definecolor{Orange} {rgb}{1.00,0.50,0.15} 
\definecolor{Magenta}{rgb}{1.00,0.00,1.00} 
\definecolor{Gold}   {rgb}{1.00,0.75,0.25} 
\definecolor{Seaweed}{rgb}{0.01,0.24,0.09} 
\definecolor{Purple} {rgb}{0.50,0.25,0.55} 
\definecolor{Brown}  {rgb}{0.43,0.26,0.32} 
\definecolor{grey1}  {rgb}{0.20,0.20,0.20} 
\definecolor{grey2}  {rgb}{0.40,0.40,0.40} 
\definecolor{grey3}  {rgb}{0.60,0.60,0.60} 
\definecolor{grey4}  {rgb}{0.80,0.80,0.80} 
\definecolor{grey5}  {rgb}{0.90,0.90,0.90} 
\def\C#1#2{{\ifcase#1\or
             \color{Red}\or \color{Green}\or \color{Blue}\or\
              \color{Turque}\or \color{Orange}\or \color{Magenta}\or
               \color{Gold}\or \color{Seaweed}\or \color{Purple}\or
                \color{Brown}\or\color{grey1}\or\color{grey2}\or
                 \color{grey3}\else\color{grey4}\fi#2}}

\definecolor{Slate} {rgb}{0.00,0.45,0.55}

\newdimen\parshift\parshift=\parindent
\catcode`@=11
 \long\def\@footnotetext#1{\insert\footins{\reset@font\footnotesize
           \interlinepenalty\interfootnotelinepenalty\splittopskip%
            \footnotesep\splitmaxdepth\dp\strutbox\floatingpenalty\@MM%
             \hsize\columnwidth\addtolength{\hsize}{-2\parindent}
              \@parboxrestore\protected@edef\@currentlabel%
              {\csname p@footnote\endcsname\@thefnmark}%
                \color@begingroup%
                 \@makefntext{\rule\z@\footnotesep\ignorespaces#1%
                  \@finalstrut\strutbox}%
                \color@endgroup}}
 \long\def\@makefntext#1{\hglue\parshift%
           \vbox{\noindent\baselineskip=11pt plus.5pt minus.5pt\hb@xt@0em{\hss\@makefnmark\kern1pt}#1}}
\catcode`@=12

\newskip\humongous \humongous=0pt plus 1000pt minus 1000pt
\def\caja{\mathsurround=0pt}
\def\eqalign#1{\,\vcenter{\openup2\jot \caja
        \ialign{\strut \hfil$\displaystyle{##}$&$
        \displaystyle{{}##}$\hfil\crcr#1\crcr}}\,}
\newif\ifdtup

\makeatletter
\def\section{\@startsection{section}{1}{\z@}
        {3ex plus-1ex minus-.2ex}{1pt plus1pt}{\large\sf\bfseries\boldmath}}
\def\subsection{\@startsection{subsection}{2}{\z@}
         {1.5ex plus-1ex minus-.2ex}{0.01pt plus1pt}{\sf\slshape}}
\def\subsubsection{\@startsection{subsubsection}{3}{\z@}
          {1.5ex plus-1ex minus-.2ex}{0.01pt plus0.2pt}{\sf\boldmath}}
\def\paragraph{\@startsection{paragraph}{4}{\z@}
           {.75ex \@plus.5ex \@minus.2ex}{-2mm}{\sf\bfseries\boldmath}}
\makeatother

\allowdisplaybreaks
\seceq

\begin{document}

\thispagestyle{empty}
\noindent{\small
\hfill{$~~$}  \\ 
{}
}
\begin{center}
{\large \bf
On 1D, $\bm {\cal N}$ = 4 Supersymmetric SYK-Type Models (I)  
}   \\   [8mm]
{\large {
S.\ James Gates, Jr.\footnote{sylvester$_-$gates@brown.edu}${}^{,a, b}$,
Yangrui Hu\footnote{yangrui$_-$hu@brown.edu}${}^{,a,b}$, and
S.-N. Hazel Mak\footnote{sze$_-$ning$_-$mak@brown.edu}${}^{,a,b}$
}}
\\*[6mm]
\emph{
\centering
$^{a}$Brown Theoretical Physics Center,
\\[1pt]
Box S, 340 Brook Street, Barus Hall,
Providence, RI 02912, USA
\\[10pt]
and
\\[10pt]
$^{b}$Department of Physics, Brown University,
\\[1pt]
Box 1843, 182 Hope Street, Barus \& Holley,
Providence, RI 02912, USA
}
 \\*[65mm]
{ ABSTRACT}\\[05mm]
\parbox{142mm}{\parindent=2pc\indent\baselineskip=14pt plus1pt
Proposals are made to describe 1D, $\cal N$ = 4 supersymmetrical systems that extend SYK models by compactifying from 4D, $\cal N$ = 1 supersymmetric Lagrangians involving chiral, vector, and tensor supermultiplets.  
Quartic fermionic vertices are generated via integrals over the whole superspace, while $2(q-1)$-point fermionic vertices are generated via superpotentials. 
The coupling constants in the superfield Lagrangians are arbitrary, and can be chosen to be Gaussian random. In that case, these 1D, $\cal N$ = 4 supersymmetric SYK models would exhibit Wishart-Laguerre randomness, which share the same feature among other 1D supersymmetric SYK models in literature.
One difference with 1D, ${\cal N} = 1$ and ${\cal N} = 2$ models though, is our models contain dynamical bosons, but this is consistent with other 1D, ${\cal N} = 4$ and 2D, ${\cal N} = 2$ models in literature.  Added conjectures on duality and possible mirror symmetry realizations in these models is noted. 
} \end{center}
\vfill
\noindent PACS: 11.30.Pb, 12.60.Jv\\
Keywords: supersymmetry, superfields, off-shell, SYK models
\vfill
\clearpage

\newpage
{\hypersetup{linkcolor=black}
\tableofcontents
}

\newpage
\section{Introduction}
\label{sec:NTRO}


The Sachdev-Ye-Kitaev (SYK) model was first proposed by Sachdev and Ye to describe a random quantum Heisenberg magnet \cite{SY}, and later modified by Kitaev to the present commonly used form \cite{K1,K2}.
The model consists of random all-to-all quartic interactions among $N$ Majorana fermions in 1D, where $N$ is large. 
In graph theory language, if we imagine fermions as vertices and interactions as edges, it is a complete hypergraph.


SYK models exhibit many interesting properties at large $N$, and in the IR limit where the model is strongly coupled. 
First of all, it is solvable \cite{SY,K1,K2,SYK-solv1,SYK-solv2,SYK-solv3,SYK-solv4,SYK-solv5,SYK-solv6,SYK-solv7,SYK-solv8}. At leading order in $1/N$, melonic Feynman diagrams dominate and one can write the Schwinger-Dyson equations in bilocal fields. At low energies one can solve these exactly.
Secondly, it is a quenched disorder system which is maximally chaotic \cite{K1,K2,SYK-solv3,SYK-solv4,SYK-solv7}. More specifically, it saturates the bound of the Lyapunov exponent. Notably, black holes have the same property \cite{SYK-bh1,SYK-bh2,SYK-bh3,SYK-bh4,SYK-bh5,SYK-bh6,SYK-bh7,SYK-bh8,SYK-bh9,SYK-bh10,SYK-sbh1,SYK-sbh2}.
Thirdly, it has emergent conformal symmetry in IR \cite{SYK-solv1,SYK-solv3,SYK-solv8,SYK-cft1,SYK-cft2,SYK-cft3,SYK-cft4,SYK-cft5,SYK-cft6}. From this 1D CFT one can explore quantum gravity in AdS$_{2}$ (AdS$_{2}$/CFT$_{1}$ or nAdS$_{2}$/nCFT$_{1}$ correspondence) \cite{SYK-bh1,SYK-bh2,SYK-bh3,SYK-bh4,SYK-cft1,SYK-cft2,SYK-ads1,SYK-ads2,SYK-ads3,SYK-ads4,SYK-ads5,SYK-ads6,SYK-ads7,SYK-ads8,SYK-ads9,SYK-ads10,SYK-ads11,SYK-ads12}. In particular, the appearance of a Schwarzian action hints at the possibility of it being holographically dual to Jackiw-Teitelboim (JT) dilaton gravity in AdS$_{2}$ or its variants \cite{SYK-solv3,SYK-ads4,SYK-ads6,SYK-ads7,SYK-jt1,SYK-jt2,SYK-jt3,SYK-jt4,SYK-jt5,SYK-jt6}.

This unique combination of properties of the SYK model has attracted great interests and allows it to describe many physical systems, including strange metals \cite{SYK-bh4,SYK-cft2,SYK-cmt5,SYK-cmt7}, traversable wormholes \cite{SYK-wh1,SYK-wh2,SYK-wh3,SYK-wh4,SYK-wh5,SYK-wh6}, and other condensed matter phenomena \cite{SYK-bh2,SYK-bh5,SYK-bh7,SYK-cft3,SYK-ads9,SYK-cmt1,SYK-cmt2,SYK-cmt3,SYK-cmt4,SYK-cmt6}.


There are many generalizations of the SYK model.
One introduces global symmetries like U(1) (complex fermions) \cite{SYK-bh4,SYK-ads9,SYK-U(1)1,SYK-U(1)2} or SO($M$) \cite{SYK-SU1,SYK-SU2}. 
Another introduces more flavors \cite{SYK-U(1)2,SYK-flavor1}.
Such theories have additional zero modes in the IR limit which correspond not to Schwarzians, but to actions for a particle moving on group manifolds corresponding to those global symmetries.
There also exist a considerable amount of studies on the Gurau-Witten tensor model, a SYK-like model without quenched disorder \cite{SYK-tensor1,SYK-tensor2,SYK-tensor3,SYK-tensor4,SYK-tensor5}. The removal of quenched disorder permits the model to go from an average of an ensemble of theories to a true quantum theory, and this allows easier probes of the bulk. Instead of vectors, the model consists of quartic interactions of rank-3 tensors. It is also dominated by melonic diagrams, and the 2-pt correlation functions are exactly the same as those in SYK at the leading order of $1/N$. A supersymmetric version of SYK-like tensor model can be found in \cite{SUSYSYK4}. 

SYK-type models in higher dimensions have also attracted studies, as one might wonder if it is possible to obtain higher dimensional examples of AdS/CFT \cite{SYK-highd1,SYK-highd2,SYK-highd3}. For example, in 2D, one can construct quartic (or 2$(q-1)$-pt) fermionic vertices \cite{SYK-highd1,SYK-highd2}. However, the interaction would be marginal (for 4-pt) or irrelevant (for 2$(q-1)$-pt). 
One could replace fermions by bosons \cite{SYK-tensor3,SUSYSYK6,SYK-highd3}, so that the interaction becomes relevant (and thus strongly coupled in IR). However, the resulting potential can possess negative directions and thus the model has no well-defined vacuum.
Finally, one can consider 2D, $\mathcal{N}=2$ supersymmetric SYK analogs \cite{SUSYSYK6}, which avoid problems arising from pure fermionic or pure bosonic models. In IR, the emergent reparametrization invariance does not cause divergences, and conformal symmetry is preserved. The model is thus easier to analyze, but the chaos bound is not saturated and the bulk theory does not correspond to a dilaton gravity theory, as in 1D SYK theories.


Another class of generalizations correspond to 1D supersymmetric SYK models.
They were first introduced in \cite{SUSYSYK2} with $\cn = 4$ supersymmetries. Then 1D, $\cn = 1,2$ supersymmetric SYK models were investigated in \cite{SUSYSYK1,SUSYSYK3,SUSYSYK4,SUSYSYK5,SUSYSYK6,SUSYSYK7,SUSYSYK8,SUSYSYK9,SUSYSYK10,SUSYSYK11,SUSYSYK12}. 
The studies of 1D, $\cn = 1,2$ supersymmetric SYK models attract a lot of interests because of the following.
Instead of random Hamiltonian, one chooses the supercharges to be Gaussian random, and the Hamiltonian is the anticommutator of the supercharges. Roughly speaking, ``squaring'' (in the $\mathcal{N}=1$ context) the random distribution is like ``folding'' up the eigenvalue distributions and forcing them to all be non-negative.
In more technical terms, the coupling thus goes from Gaussian random to Wishart-Laguerre random \cite{rmt1,rmt2,SUSYSYK9,SUSYSYK10}, and the eigenvalue distribution goes from Wigner's semi-circle
\cite{rmt-w} to the Marchenko-Pastur distribution \cite{rmt-mp}. This feature is drastically different from ordinary SYK.
Unlike higher dimensional supersymmetric models, it is maximally chaotic \cite{SUSYSYK1,SUSYSYK5} in 1D and it flows to super-Schwarzians in IR. 
This provides evidence for holographic duality between supersymmetric JT gravity and supersymmetric SYK \cite{SYK-jt2,SYK-jt3}.


In 1D, $\mathcal{N}=1$ and $\mathcal{N}=2$ supersymmetric SYK models, superconformal symmetry occurs in IR. In the $\mathcal{N}=1$ case, the ground state energy is non-zero but approaches zero in the large $N$ limit, thus supersymmetry is broken non-perturbatively. In the $\mathcal{N}=2$ case, there are many zero energy states and supersymmetry is unbroken \cite{SUSYSYK1}. 
A further difference is the emergence of new reparametrization symmetry in addition to the super-Schwarzian in the 1D, $\mathcal{N}=2$ case \cite{SUSYSYK1}. Both symmetries together resemble conformal symmetries in AdS$_{2}$ space.
The near-horizon limit of 4D, $\mathcal{N}=2$ extremal black hole can be described by AdS$_{2}$ space, and it is probably holographically dual to 1D, $\mathcal{N}=2$ SYK. However, asymptotically flat supersymmetric black holes in 4D have $\cn = 4$ supersymmetries \cite{SYK-sbh1,SYK-sbh2,SYK-sbh3}, which makes the study of $\cn = 4$ SYK more interesting. 
With supersymmetry one is able to count the microscopic states and find the zero temperature entropy \cite{SYK-sbh2}.


Discussions of supersymmetric SYK models in the literature are largely focused on one dimensional models with $\mathcal{N}=1$ and $\mathcal{N}=2$ supersymmetries \cite{SUSYSYK1,SUSYSYK3,SUSYSYK4,SUSYSYK5,SUSYSYK6,SUSYSYK7,SUSYSYK8,SUSYSYK9,SUSYSYK10,SUSYSYK11,SUSYSYK12}, though there is a study on $\cn = 4$ models \cite{SUSYSYK2}.
This naturally raises a question, ``Why is it interesting to construct such models with higher degrees of extended supersymmetry?''
One could find some hints
in the history of the relationship between quantum conformal symmetry and the number 
$\cal N$ of extended supersymmetries that can be realised in models.

It may be recalled the 4D, $\cal N$ = 4 supersymmetrical Yang-Mills theory \cite{N4D4a,N4D4b}  
was the first QFT discovered that is a finite field theory at low orders 
\cite{N4D4c1,N4D4c2,N4D4d1,N4D4d2,N4D4e1,N4D4e2} in perturbation theory and it remains an impelling 
driver of research in QFT to this very day \cite{AmTHdR,amp}.  This was an early demonstration of
the power of combining SUSY with conformal symmetry.

In a similar manner, the 3D, $\cal N$ = 6 supersymmetrical Chern-Simons theory plays
an important role currently.  It is an interesting historical note that the field 
spectrum and two different descriptions in terms of Lagrangians\footnote{In fact, two 
different Lagrangians were given in this first description of the action.  One was compatible
with a 3D, $\cal N$ = 2 superfield formulation.} for this theory were described 
as a special case of an action that appeared in a 1991 paper \cite{CST1}, though the 
full $\cal N $ = 6 SUSY variations was not presented.  Next the indications that an $\cal N$ = 6 
theory would necessarily be related in a special way to conformal symmetry appeared in 
the work of \cite{CST2}.  Eleven years later, the full implications of these 
observations appeared in the work of \cite{CST3} which has been used to define 
M-Theory scattering amplitudes in papers \cite{PF1,PF2} where explorations beyond the
supergravity limit are probed.

With this history as background motivation, in this work we will explore SYK models
that possess $ \cal N$ $>$ 2 SUSY which is the current state of the art. 
In \cite{SUSYSYK2}, one 1D, $\mathcal{N}=4$ SYK model is built entirely from 1D chiral supermultiplet, and another one is built from both chiral and vector supermultiplets. 
In our work, we build more $\cn = 4$ SYK models with 4D, $\cn = 1$ chiral, vector and tensor supermultiplets.
We start from 4D superfield Lagrangians\footnote{The superfield approach enables the construction of effective actions via bilocal superfields as formulated in \cite{SUSYSYK1,SUSYSYK8}, although not calculated in this paper. For our models, this should be a much neater method.}.
In particular,
we are interested in showing the existence of 4D, $\cal N$ = 1 theories with the
property that under simple compactification\footnote{This is in essence the same idea as ``SUSY holography'', ``SUSY QFT/QM correspondence'', or ``0-brane reduction''. See
\cite{GRana2,ENUF,adnk1,adnk2,adnk3,adnk4,adnk5,adnk6,adnk7,adnk8,adnk9,AdnkKoR,adnk10,adnk11}.}
lead to 1D, $\cal N$ = 4 theories of
the form of SYK models.



The chiral supermultiplet appeared in foundational works \cite{GL1,GL2,WZ} that 
established SUSY.  It grew as an active research topic initiated by early efforts \cite{Wss,Fy8},
and the vector supermultiplet \cite{VSM1,VSM2} appeared at essentially
the same time in the western literature.

One other ingredient, that is important for our exploration, is given by the real linear/2-form/tensor supermultipet
as discovered by Siegel \cite{C-GuLL}.  This supermultiplet, like the chiral supermultiplet
only possess physical degrees of freedom with spins of one-half or zero.  However, 
one of the spin-0 degrees of freedom is a two-form gauge field.

We organize our paper in the following manner.

In the second chapter, we review the previous works \cite{SUSYSYK1,SUSYSYK2,SUSYSYK3,SUSYSYK4,SUSYSYK5,SUSYSYK6}
from a perspective that provides a foundation for our supersymmetrical exploration
of possible larger SUSY extensions.

The third chapter is devoted to setting in place our conventions for discussing
the chiral supermultiplet, the vector supermultiplet, and the tensor
supermultiplet at
the level of 2-point vertices in Lagrangians.  
This discussion
is in the language of superfields using superspace framework and hence supersymmetry is manifest.
Furthermore, as the discussion
is situated in four dimensions,
it is also relevant to the
supersymmetrical extensions of
older similar models \cite{ThRR,NmBJL1,NmBJL2,GN}
with four fermion couplings.

The fourth and fifth chapters introduce 3-point and $q$-point superfield interactions respectively. 4-point SYK-type vertices emerge in both cases when we go on-shell.

Our sixth chapter is devoted to the introduction of higher $q$-point superfield interactions which gives $2(q-1)$-point fermionic interactions on-shell.

The seventh chapter includes a discussion of the results for one dimensional
Lagrangians that follow from the compactification of the Lagrangians constructed
in four dimensions.  The emergence of $\cal N$ = 4 extended supersymmetry
is made manifest.

Finally, there is one chapter that describe the whole story, and give comments and conclusions. We follow the presentation of our work with four appendices and the bibliography.

\newpage
\section{A Brief Review of 1D SUSY SYK Theory Lagrangians}

\subsection{1D, $\mathcal{N}=1$ SUSY SYK}

In 1D, $\mathcal{N}=1$ supersymmetric SYK model, consider Majorana fermions $\psi^{i}$, $i = 1, \dots, N$, which satisfy
\begin{equation}
    \{ \psi^{i}, \psi^{j} \} ~=~ \d^{ij} ~~~.
\end{equation}
The supercharge can be written as \cite{SUSYSYK1}
\begin{equation}
    Q ~=~ i \sum_{1\leq i<j<k \leq N} C_{ijk} \psi^{i} \psi^{j} \psi^{k} ~~~,
\end{equation}
where $C_{ijk}$ is a real antisymmetric tensor. We take $C_{ijk}$ to be independent Gaussian random variables with zero mean and variance specified by a constant $J > 0$ with units of energy,
\begin{equation}
    \langle C_{ijk} \rangle ~=~ 0 ~~~,~~~
    \langle C_{ijk}^{2} \rangle ~=~ \frac{2J}{N^{2}} ~~~.
\end{equation}
The Hamiltonian is given by
\begin{equation}
    {\cal H} ~=~ Q^{2} ~=~ E_{0} ~+~ \sum_{1\leq i<j<k<l \leq N} J_{ijkl}  \psi^{i} \psi^{j} \psi^{k} \psi^{l} ~~~,
\end{equation}
where
\begin{equation}
     E_{0} ~=~  \fracm18\,\sum_{1\leq i<j<k \leq N} C_{ijk}^{2} ~~~,~~~
     J_{ijkl} ~=~ -\fracm18\,  \sum_{a} C_{a[ij} C_{kl]a} ~~~.
\label{eqn:Jform}
\end{equation}
Here our (\ref{eqn:Jform}) agrees with 
Equation (2.14) in \cite{SUSYSYK9}, while does not agree with Equation (1.5) in \cite{SUSYSYK1}. Therefore we report our explicit calculations in Appendix \ref{appen:Hcalculation}.
Note that now the independent variables $C_{ijk}$ follows the Gaussian distribution, instead of the variables $J_{ijkl}$, and this is formally the only difference between a non-supersymmetric SYK model and a supersymmetric one \cite{SUSYSYK1}. The couplings $J_{ijkl}$ now follow a Wishart-Laguerre random distribution \cite{rmt1,rmt2,SUSYSYK9,SUSYSYK10}.

Let $b^{i}$ be non-dynamical auxiliary bosons. The off-shell Lagrangian is
\begin{equation}
    \cl ~=~ \sum_{i} \Big[~ \fracm12 \psi^{i} \pa_{\t} \psi^{i} ~-~ \fracm12 b^{i} b^{i} ~+~ i \sum_{1 \leq j<k \leq N} C_{ijk} b^{i} \psi^{j} \psi^{k} ~\Big] ~~~.
\end{equation}
It is possible to embed the components in this Lagrangian into superfields, 
\begin{equation}
    \Psi^{i}(\t,\theta) ~=~ \psi^{i}(\t) ~+~ \theta \, b^{i}(\t) ~~~,
\end{equation}
and we can introduce the supercovariant derivative
\begin{equation}
    \rD_{\theta} ~=~ \pa_{\theta} ~+~ \theta \, \pa_{\t} ~~~.
\end{equation}
Then we could rewrite the component Lagrangian as a superfield Lagrangian,
\begin{equation}
    \cl ~=~ \int d\theta ~ \Big[~ -~ \fracm12 \sum_{i} \Psi^{i} \rD_{\theta} \Psi^{i} ~+~ i \sum_{1 \leq i<j<k \leq N} C_{ijk} \Psi^{i} \Psi^{j} \Psi^{k}  ~\Big] ~~~,
\end{equation}
which is manifestly supersymmetric.

It is also possible to generalize this model to interactions among $2(q-1)$ Majorana fermions. The supercharge can be written as \cite{SUSYSYK9,SUSYSYK10}
\begin{equation}
    Q ~=~ i^{\frac{q-1}{2}} \sum_{1 \leq i_{1} < i_{2} < \cdots < i_{q} \leq N }  C_{i_{1}i_{2} \cdots i_{q}} \psi^{i_{1}} \psi^{i_{2}} \cdots \psi^{i_{q}} ~~~,
\end{equation}
and we will recover the quartic fermionic interaction by taking $q=3$. The mean and variance of the variables $C_{i_{1}i_{2} \cdots i_{q}}$ are 
\begin{equation}
    \langle C_{i_{1}i_{2} \cdots i_{q}} \rangle ~=~ 0 ~~~,~~~
    \langle C_{i_{1}i_{2} \cdots i_{q}}^{2} \rangle ~=~ \frac{(q-1)!J}{N^{q-1}} ~~~.
\end{equation}
The off-shell Lagrangian would be
\begin{equation}
    \cl ~=~ \sum_{i} \Big[~ \fracm12 \psi^{i} \pa_{\t} \psi^{i} ~-~ \fracm12 b^{i} b^{i} ~+~ i^{\frac{q-1}{2}} \sum_{1 \leq j_{1} < \cdots < j_{q-1} \leq N} C_{i j_{1} \cdots j_{q-1}} b^{i} \psi^{j_{1}} \cdots \psi^{j_{q-1}} ~\Big] ~~~,
\end{equation}
and the superfield Lagrangian would be
\begin{equation}
    \cl ~=~ \int d\theta ~ \Big[~ -~ \fracm12 \sum_{i} \Psi^{i} \rD_{\theta} \Psi^{i} ~+~ i^{\frac{q-1}{2}} \sum_{1 \leq i_{1} < i_{2} < \cdots < i_{q} \leq N}  C_{i_{1} i_{2} \cdots i_{q}} \Psi^{i_{1}} \Psi^{i_{2}} \cdots \Psi^{i_{q}}  ~\Big] ~~~.
\end{equation}

\subsection{1D, $\mathcal{N}=2$ SUSY SYK}

In 1D, $\mathcal{N}=2$ supersymmetric SYK model, consider complex fermions $\psi^{i}$ and $\Bar{\psi}_{i}$, $i = 1, \dots, N$, which satisfy 
\begin{equation}
    \{ \psi^{i}, \Bar{\psi}^{j} \} ~=~ \d^{ij} ~~~,~~~
    \{ \psi^{i}, \psi^{j} \} = 0 ~~~,~~~
    \{ \Bar{\psi}^{i}, \Bar{\psi}^{j} \} = 0 ~~~.
\end{equation}
The supercharges can be written as \cite{SUSYSYK1}
\begin{equation}
    Q ~=~ i \sum_{1\leq i<j<k \leq N} C_{ijk} \psi^{i} \psi^{j} \psi^{k} ~~~,~~~
    \Bar{Q} ~=~ i \sum_{1\leq i<j<k \leq N} \Bar{C}_{ijk} \Bar{\psi}^{i} \Bar{\psi}^{j} \Bar{\psi}^{k} ~~~,
\end{equation}
where $C_{ijk}$ and $\Bar{C}_{ijk}$ are taken as independent Gaussian random complex numbers with zero mean and variance specified by a constant $J > 0$ with units of energy,
\begin{equation}
    \langle C_{ijk} \rangle ~=~ \langle \Bar{C}_{ijk} \rangle ~=~ 0 ~~~,~~~
    \langle C_{ijk} \Bar{C}_{ijk} \rangle ~=~ \frac{2J}{N^{2}} ~~~.
\end{equation}
The Hamiltonian is given by
\begin{equation}
    {\cal H} ~=~ \{ Q , \Bar{Q} \} ~\sim~ |C|^{2} ~+~ \sum_{i,j,k,l} J_{ijkl} \psi^{i} \psi^{j} \Bar{\psi}^{k} \Bar{\psi}^{l} ~~~,
\end{equation}
where
\begin{equation}
     J_{ijkl} ~\sim~ \sum_{a} C_{aij} \Bar{C}_{kla} ~~~.
\label{eqn:Jform2}
\end{equation}
Note that $J_{ijkl}$ are not Gaussian random. 

Let $b^{i}$ and $\Bar{b}^{i}$ be non-dynamical auxiliary bosons. The off-shell Lagrangian is
\begin{equation}
    \cl ~=~ \sum_{i} \Big[~ \fracm12 \Bar{\psi}^{i} \pa_{\t} \psi^{i} ~-~ \fracm12 \Bar{b}^{i} b^{i} ~+~ i \sum_{1 \leq j<k \leq N} C_{ijk} \Bar{b}^{i} \psi^{j} \psi^{k} ~+~ i \sum_{1 \leq j<k \leq N} \Bar{C}_{ijk} b^{i} \Bar{\psi}^{j} \Bar{\psi}^{k} ~\Big] ~~~.
    \label{equ:N2Laganrangian-comp}
\end{equation}
One could introduce supercovariant derivative operators\footnote{Here we adopt different conventions compared with the one in \cite{SUSYSYK11}, so that $\{\rD,\Bar{\rD}\} = \pa_{\t}$ and get the exact (\ref{equ:N2Laganrangian-comp}) from (\ref{equ:N2Laganrangian}).}
\begin{equation}
    \rD_{\theta} ~=~ \pa_{\theta} ~+~ \fracm12\,\Bar{\theta} \, \pa_{\t} ~~~,~~~
    \Bar{\rD}_{\Bar{\theta}} ~=~ \pa_{\Bar{\theta}} ~+~ \fracm12\,\theta \, \pa_{\t} ~~~,
\end{equation}
and embed the above components into chiral superfields $\Psi^{i}$ and anti-chiral superfields $\Bar{\Psi}^{i}$, i.e.
\begin{equation}
    \Bar{\rD}_{\Bar{\theta}} \Psi^{i} ~=~ 0 ~~~,~~~
    \rD_{\theta} \Bar{\Psi}_{i} ~=~ 0 ~~~,
\label{eq:N2syk}    
\end{equation}
which are solved by \cite{SUSYSYK1}\footnote{In\cite{SUSYSYK1}, their original superfield is 
$\Psi^{i} (\t,\theta,\Bar{\theta}) ~=~ \psi^{i} (\t + \theta\Bar{\theta}) ~+~ \theta \, b^{i} (\t)$,
which is written in the chiral basis that appears in a lot of old supersymmetry literature. Here we adopt a different convention for $\rD$ and $\Bar{\rD}$, thus $\psi^{i} (\t + \theta\Bar{\theta})$ is replaced by $\psi^{i} (\t + \frac12\theta\Bar{\theta})$, and we expand it so that the argument does not contain Grassmann variables. In addition, we define the bosonic component field as $\Bar{b}^{i}$, such that the interaction terms in the superfield Lagrangian would recover those in the component Lagrangian.}
\begin{equation}
    \Psi^{i} (\t,\theta,\Bar{\theta}) ~=~ \psi^{i} (\t) ~+~ \theta \, \Bar{b}^{i} (\t) ~+~ \fracm12 \, \theta\Bar{\theta} \, \pa_{\t} \psi^{i} (\t)  ~~~.
\end{equation}
The superfield Lagrangian is thus\cite{SUSYSYK11}
\begin{equation}
\begin{split}
    \cl ~=&~ \int d\Bar{\theta} ~ \sum_{i} \Big[~ -~ \fracm12 \Bar{\Psi}^{i} \rD_{\theta} \Psi^{i} ~\Big] 
    ~+~ \int d\theta \sum_{1 \leq i<j<k \leq N} \Big[~ i C_{ijk} \Psi^{i} \Psi^{j} \Psi^{k}  ~\Big] \\
    &~~~~~~~~~~~~~~~ ~+~ \int d\Bar{\theta} \sum_{1 \leq i<j<k \leq N} \Big[~ i \Bar{C}_{ijk} \Bar{\Psi}^{i} \Bar{\Psi}^{j} \Bar{\Psi}^{k}  ~\Big]  ~~~.
\end{split}
\label{equ:N2Laganrangian}
\end{equation}

One could also generalize this model to $2(q-1)$-point fermionic interactions. The supercharges can be written as
\begin{equation}
\begin{split}
    Q ~=&~ i^{\frac{q-1}{2}} \sum_{1 \leq i_{1} < i_{2} < \cdots < i_{q} \leq N }  C_{i_{1}i_{2} \cdots i_{q}} \psi^{i_{1}} \psi^{i_{2}} \cdots \psi^{i_{q}} ~~~, \\
    \Bar{Q} ~=&~ i^{\frac{q-1}{2}} \sum_{1 \leq i_{1} < i_{2} < \cdots < i_{q} \leq N }  \Bar{C}_{i_{1}i_{2} \cdots i_{q}} \Bar{\psi}^{i_{1}} \Bar{\psi}^{i_{2}} \cdots \Bar{\psi}^{i_{q}} ~~~,
\end{split}
\end{equation}
and we will recover the quartic fermionic interaction by taking $q=3$. The mean and variance of the variables $C_{i_{1}i_{2} \cdots i_{q}}$ and $\Bar{C}_{i_{1}i_{2} \cdots i_{q}}$ are 
\begin{equation}
    \langle C_{i_{1}i_{2} \cdots i_{q}} \rangle ~=~ \langle \Bar{C}_{i_{1}i_{2} \cdots i_{q}} \rangle ~=~ 0 ~~~,~~~
    \langle C_{i_{1}i_{2} \cdots i_{q}} \Bar{C}_{i_{1}i_{2} \cdots i_{q}} \rangle ~=~ \frac{(q-1)!J}{N^{q-1}} ~~~.
\end{equation}
The off-shell Lagrangian in components would be \cite{SUSYSYK5}
\begin{equation}
\begin{split}
    \cl ~=&~ \sum_{i} \Big[~ \fracm12 \Bar{\psi}^{i} \pa_{\t} \psi^{i} ~-~ \fracm12 \Bar{b}^{i} b^{i} ~+~ i^{\frac{q-1}{2}} \sum_{1 \leq j_{1} < \cdots < j_{q-1} \leq N} C_{i j_{1} \cdots j_{q-1}} \Bar{b}^{i} \psi^{j_{1}} \cdots \psi^{j_{q-1}} \\
    &~~~~~~~~~~~~~~~ ~+~ i^{\frac{q-1}{2}} \sum_{1 \leq j_{1} < \cdots < j_{q-1} \leq N} \Bar{C}_{i j_{1} \cdots j_{q-1}} b^{i} \Bar{\psi}^{j_{1}} \cdots \Bar{\psi}^{j_{q-1}} ~\Big] ~~~,
\end{split}
\end{equation}
and the superfield Lagrangian would be \cite{SUSYSYK11}
\begin{equation}
\begin{split}
    \cl ~=&~ \int d\Bar{\theta} ~ \sum_{i} \Big[~ -~ \fracm12 \Bar{\Psi}^{i} \rD_{\theta} \Psi^{i} ~\Big] 
    ~+~ \int d\theta \sum_{1 \leq i_{1} < i_{2} < \cdots < i_{q} \leq N} \Big[~ i^{\frac{q-1}{2}} C_{i_{1} i_{2} \cdots i_{q}}  \Psi^{i_{1}} \Psi^{i_{2}} \cdots \Psi^{i_{q}}  ~\Big] \\
    &~~~~~~~~~~~~~~~ ~+~ \int d\Bar{\theta} \sum_{1 \leq i_{1} < i_{2} < \cdots < i_{q} \leq N} \Big[~ i^{\frac{q-1}{2}} \Bar{C}_{i_{1} i_{2} \cdots i_{q}} \Bar{\Psi}^{i_{1}} \Bar{\Psi}^{i_{2}} \cdots \Bar{\Psi}^{i_{q}}  ~\Big] ~~~.
\end{split}
\end{equation}

It should be noted that the constraint in 
(\ref{eq:N2syk}) also implies the leading term in the superfield Lagrangian be rewritten according to
\begin{equation}
\int d\Bar{\theta} ~ \sum_{i} \Big[~ -~ \fracm12 \Bar{\Psi}^{i} \rD_{\theta} \Psi^{i} ~\Big]
~=~
\int d\Bar{\theta} ~ \sum_{i} \Big[~ \fracm12 \rD_{\theta} ( \Bar{\Psi}^{i} \Psi^{i} ) ~\Big]
~=~
\int d {\theta}\,  d\Bar{\theta} ~ \sum_{i} \Big[~ -~ \fracm12 \Bar{\Psi}^{i} \Psi^{i} ~\Big] ~~~,
\end{equation}
and the total Lagrangian can be written in
terms as
\begin{equation}
    \cl ~=~  \int d {\theta}\,  d\Bar{\theta} ~K( \Bar{\Psi}^{i} , \Psi^{i} ) 
    ~+~ \int d\theta ~ \Big[~ {\cal W} (\Psi^{i})  ~\Big]
    ~+~ \int d\Bar{\theta} ~ \Big[~ {\Bar {\cal W}}( \Bar{\Psi}^{i})  ~\Big] ~~~,
\label{eqn:N2sfLform}
\end{equation}
where a K\"ahler-like potential $K$ and
superpotential-like quantity $\cal W$ are introduced.  The function $K$ must be constructed from only even powers of the 
spinorial superfields $\Psi^{i}$, and
$ {\Bar{\Psi}}^{i}$, while the function $\cal W $ must
be constructed from only odd powers of the 
spinorial superfields $\Psi^{i}$, and
$ {\Bar{\Psi}}^{i}$.  For example, a linear
term in $\cal W$ breaks the U(1) symmetry of
the model while leading to a Fayet-Iliopoulos
term.

\subsection{1D, $\mathcal{N}=4$ SUSY SYK}
\label{subsec:AAD}

The earliest work on $\cn = 4$ supersymmetric SYK (and also supersymmetric SYK) is \cite{SUSYSYK2}. It starts with the chiral multiplet $\Phi^{i}_{\a}=(\phi^{i}_{\a},\psi^{i}_{\a},F^{i}_{\a})$, where $\phi$ is a complex scalar, $\psi$ is a complex Weyl fermion with spinor index suppressed, and $F$ is a complex auxiliary scalar.
The index $\a=1,\dots,N$ indicates intersection mode connecting two branes, and the index $i = 1, \dots , q$ denotes the pair of branes connected.

The superfield interaction Lagrangian for $q=3$ is written as the chiral superspace integration of a superpotential,
\begin{equation}
    \cl_{\text{int}} ~=~ \int d^{2}\theta ~ \O_{\a\b\g} \, \phi^{1}_{\a} \phi^{2}_{\b} \phi^{3}_{\g} ~+~ {\rm h.\,c.} ~~~,
\end{equation}
where $\O_{\a\b\g}$ are Gaussian random with zero mean and variance specified by a constant $\O$,
\begin{equation}
    \langle \O_{\a\b\g} \rangle ~=~ 0 ~~~,~~~
    \langle | \O_{\a\b\g} |^{2} \rangle ~=~ \O^{2} ~~~.
\end{equation}
The component Lagrangian is
\begin{equation}
\begin{split}
    \cl ~=&~ \sum_{i,\a} \Big[~ \pa_{\t} \Bar{\phi}^{i}_{\a} \pa_{\t} \phi^{i}_{\a} ~+~ \Bar{\psi}^{i}_{\a} \pa_{\t} \psi^{i}_{\a} ~-~ \Bar{F}^{i}_{\a} F^{i}_{\a} ~\Big] \\
    &~+~ \Big[~ \sum_{\vec{\a}} \sum_{\vec{i}\in S_{3}} \O_{\a\b\g} \big(~ \phi^{i_{1}}_{\a} \phi^{i_{2}}_{\b} F^{i_{3}}_{\g} ~+~ \psi^{i_{1}}_{\a} \e \psi^{i_{2}}_{\b} \phi^{i_{3}}_{\g} ~\big) ~+~ {\rm h.\,c.} ~\Big] ~~~,
\end{split}
\label{eqn:AADcomp}
\end{equation}
where $\vec{\a} = (\a,\b,\g)$, $\vec{i} = (i_{1},i_{2},i_{3})$, and $S_{3}$ is the 3-element permutation group. All of these components live in 1D, i.e. they are functions of time only.

This is one of the models one can build with $\cn = 4$ supersymmetries in 1D. Below we will build more models from the chiral, vector and tensor supermultiplets.

\newpage
\section{Review of 4D, $\cal N$ = 1 Theories}

In 4D, $\cal N$ = 1 theories, one class of well known supermultiplets consists
of the chiral supermultiplets, usually denoted by $\Phi$. There exist
a second such class, the complex linear supermultiplet, whose field strength 
superfield can be denoted by $\Sigma$. But we will not discuss any model constructed from the complex linear supermultiplet in this paper.

In the following, we will utilize chiral projection operators that are defined by
\begin{equation}
    (\rP^{(\pm)}){}^{ab} ~=~ \fracm12 \, \big[\, C^{ab} \pm (\g^5)^{ab} \,\big] ~~~,
\end{equation}
where it satisfies 
\begin{align}
    (\rP^{(\pm)})_{a}{}^{b}  (\rP^{(\pm)})_{b}{}^{c}  ~=~  (\rP^{(\pm)})_{a}{}^{c}  ~~~,~~~
    (\rP^{(\pm)})_{a}{}^{b}  (\rP^{(\mp)})_{b}{}^{c}  ~=~  0 ~~~,
\end{align}
and
\begin{equation}
\begin{split}
    &(\rP^{(\pm)})_{a}{}^{b} (\g^{5})_{b}{}^{c}  ~=~(\g^{5})_{a}{}^{b}(\rP^{(\pm)})_{b}{}^{c}   ~=~ \pm (\rP^{(\pm)})_{a}{}^{c} ~~~,~~~ (\rP^{(\pm)})_{ab}~=~-\,(\rP^{(\pm)})_{ba} ~~~,\\
    &(\g^{\m})_{a}{}^{b} (\rP^{(\pm)})_{b}{}^{c} ~=~ (\rP^{(\mp)})_{a}{}^{b} (\g^{\m})_{b}{}^{c}  ~~~,~~~ (\rP^{(\pm)}\g^{\m})_{ab}~=~ (\rP^{(\mp)}\g^{\m})_{ba}
    ~~~, \\
    & (\g^{\m\n})_{a}{}^{b} (\rP^{(\pm)})_{b}{}^{c} ~=~ (\rP^{(\pm)})_{a}{}^{b} (\g^{\m\n})_{b}{}^{c} ~~~,~~~ (\rP^{(\pm)}\g^{\m\n})_{ab}~=~ (\rP^{(\pm)}\g^{\m\n})_{ba}~~~.
    \end{split}
\end{equation}
Another property to note is
\begin{equation}
    (\rP^{(+)})^{*} ~=~ \rP^{(-)} ~~~,~~~ (\rP^{(+)}\g^{\mu})^{*} ~=~ \rP^{(-)}\g^{\mu} ~~~,~~~ (\rP^{(+)}\g^{\mu\n})^{*} ~=~ \rP^{(-)}\g^{\mu\n} ~~~.
\end{equation}
Then we can define
\begin{equation}
    \rD^{(\pm)}_{a} ~=~ (\rP^{(\pm)}){}_{a}{}^{b} \rD_{b} ~~~,
\end{equation}
where the superfields $\Phi$ and ${\Bar{\Phi}}$ are covariantly chiral and antichiral if
\begin{equation}
    \rD^{(-)}_{a} \Phi ~=~ 0 ~~~,~~~ \rD^{(+)}_{a} \Bar{\Phi} ~=~ 0 ~~~.
\end{equation}

\subsection{Majorana Four-Component Notation CS}{\label{sec:M4CS}}

The chiral supermultiplet (CS) contains propagating fields: scalar $\bA$, pseudoscalar $\bB$, spin-$\frac12$ fermion $\Bj_{a}$; and auxiliary fields: scalar $\bF$ and pseudoscalar $\bG$.
The transformation laws are
\be
\eqalign{
{~~~~} 
\rD_a \bA ~=&~ \Bj_a  ~~~~~~~~~~~\,~~~,~~~
\rD_a \bB ~=~ - \, i \, (\g^5){}_a{}^b \, \Bj_b  ~~~, \cr
\rD_a \Bj_b ~=&~ i\, (\g^\m){}_{a \,b}\,  \pa_\m \bA 
~+~  (\g^5\g^\m){}_{a \,b} \, \pa_\m \bB ~-~ i \, C_{a\, b} 
\,\bF  ~+~  (\g^5){}_{ a \, b} \bG  ~~~, \cr
\rD_a \bF ~=&~  (\g^\m){}_a{}^b \, \pa_\m \, \Bj_b   
~~~,~~~ 
\rD_a \bG ~=~ i \,(\g^5\g^\m){}_a{}^b \, \pa_\m \,  
\Bj_b  ~~~.
} \label{QT1}
\ee
The Lagrangian is
\begin{align}
    \cl_{\rm CS} ~=~ -~ \fracm{1}{2} \partial_{\mu} \bA \partial^\mu \bA 
    ~-~ \fracm{1}{2} \partial_{\mu}\bB \partial^\mu \bB
    ~+~\fracm{1}{2} {\bF}^2  + \fracm{1}{2} {\bG}^2  
    ~+~ i \, \fracm{1}{2}(\gamma^\mu)^{ab} {\bm {\psi}}_a \partial_\mu {\bm {\psi}}_b  ~~~.
\end{align}

Let us focus on the propagating bosons and define
\begin{align}
    \BF ~=&~ \bA ~+~ i \, \bB ~~~.
    \label{eq:ComPLX1}  
\end{align}
Note that it satisfies the chiral condition
\begin{align}
    \rD_{a}^{(-)} \, \BF ~=&~  0 ~~~.
    \label{eq:xxx}
\end{align}
It can be shown that the Lagrangian above can be derived from the following expression
\begin{align}
    \cl_{\rm CS} ~=~ - \fracm{1}{32} \, \rD^{a} \, \rD_{a}^{(+)} \, \rD^{b} \rD_{b}^{(-)} \, \bBF \, \BF ~~~.
\end{align}
Although this is written in components, one could think about it as a ``superfield Lagrangian'' with superfield $\Phi$, and $\Phi| \equiv \BF$. The $\rD$-operators acting on the superfield $\Phi$ should be thought as supercovariant derivatives, while those acting on the component $\BF$ should be thought as abstract operators following the transformation laws. Similar comments apply to other Lagrangians we write down. 

If we also define
\begin{equation}
    \bX ~=~ \bF ~+~ i \, \bG ~~~,
\end{equation}
we now rewrite the bosonic transformation laws as
\begin{equation}
\begin{aligned}
    \rD_{a}^{(+)} \BF ~=&~ 2 (\rP^{(+)})_{a}{}^{b} \Bj_b ~~~,~~~ &
    \rD_{a}^{(-)} \BF ~=&~  0 ~~~, \\
    \rD_{a}^{(+)} \bX ~=&~ 0 ~~~,~~~ &
    \rD_{a}^{(-)} \bX ~=&~ 2 (\rP^{(-)} \g^{\m})_{a}{}^{b} \pa_{\m} \Bj_b ~~~,
\end{aligned}
\end{equation}
and the fermionic transformation laws as
\begin{equation}
\begin{split}
    \rD_{a}^{(+)} \Bj_{b} ~=&~ i (\rP^{(+)} \g^{\m})_{ab} \pa_{\m} \bBF ~-~ i (\rP^{(+)})_{ab} \bX ~~~, \\
    \rD_{a}^{(-)} \Bj_{b} ~=&~ i (\rP^{(-)} \g^{\m})_{ab} \pa_{\m} \BF ~-~ i (\rP^{(-)})_{ab} \bbX ~~~.
\end{split}
\label{eq:CSpm}
\end{equation}
The Lagrangian for the component fields can be rewritten as
\begin{equation}
    \cl_{\rm CS} ~=~ -~ \fracm{1}{2} \pa_{\m} \BF \pa^{\m} \bBF 
    ~+~ \fracm{1}{2} \bX \bbX 
    ~+~ i \, \fracm{1}{2} (\g^{\m})^{ab} \Bj_{a} \pa_{\m} \Bj_{b}  ~~~.
    \label{equ:chiral-L}
\end{equation}

\subsection{Majorana Four-Component Notation VS}{\label{sec:M4VS}}

The vector supermultiplet (VS) contains propagating fields: vector $\bA_{\m}$ and spin-$\frac12$ fermion $\Bl_{a}$; and auxiliary fields: scalar ${\bm {\rm d}}$.
The transformation laws are
\be
\eqalign{
{~~~~} {\rm D}_a \, \bA_{\m} ~&=~  (\g_\m){}_a {}^b \,  \Bl_b  ~~~, \cr
{\rm D}_a \Bl_b ~&=~   - \,i \, \fracm 14 ( [\, \g^{\m}\, , \,  \g^{\n} 
\,]){}_a{}_b \, (\,  \partial_\m  \, \bA_{\n}    ~-~  \pa_\n \, \bA_{\m}  \, )
~+~  (\g^5){}_{a \,b} \,   {\bm {\rm d}} ~~,  {~~~~~~~}  \cr
{\rm D}_a \, {\bm {\rm d}} ~&=~  i \, (\g^5\g^\m){}_a {}^b \, 
\,  \pa_\m \Bl_b  ~~~, \cr
}  \label{QT2}
\ee
and once more the chiral projection operators can be used to rewrite the form of
the results in (\ref{QT2}) so that these appear as,
\be
\eqalign{ {~~~~~~~~~}
{~~~~} \rD^{(\pm)}_a \, \bA_{\m} ~&=~ (\rP^{(\pm)} \g_\m){}_a {}^b \,  \Bl_b  ~~~, \cr
 \rD^{(\pm)}_a \, {\bm {\rm d}} ~&=~ \pm \,  i \, (\rP^{(\pm)} \g^\m){}_a {}^b \, 
\,  \pa_\m \Bl_b  ~~~, \cr
{\rm D}^{(\pm)}_a \Bl_b ~&=~   - \,i \, \fracm 12 ( \rP^{(\pm)} \, \g^{\m\n}\,){}_a{}_b \, \bF_{\m\n}
~\pm~ (\rP^{(\pm)}){}_{a \,b} \,   {\bm {\rm d}} ~~,  {~~~~~~~}  \cr
}  \label{eq:QT2z}
\ee
where we define $[\, \g^{\m}\, , \,  \g^{\n} 
\,] = 2\g^{\m\n}$ and the field strength $\bF_{\m\n} = \partial_\m  \, \bA_{\n} - \pa_\n \, \bA_{\m}$. 
The Lagrangian for the component fields is given by
\be
    \mathcal{L}_{\rm VS} ~=~ -~ \fracm{1}{4} \bF{}_{\mu\nu} \bF{}^{\mu\nu}  ~+~ i\,  \fracm{1}{2} \, (\gamma^\mu)^{a b} {\bm \l}_a \partial_\mu {\bm \l}_b ~+~ \fracm{1}{2} {\bm {\rm d}}^2   ~~~.
    \label{equ:VS-L}
\ee

Note that we can define a chiral current (for more details, see Section \ref{sec:npt2npt}) that satisfy the chiral condition,
\begin{equation}
    \rD^{(-)}_{a} \,\big(\, \Bl^{b} ( \rP^{(+)} \Bl )_{b} \,\big) ~=~ 0 ~~~,
\end{equation}
and construct the Lagrangian from considering the chiral half of the superspace,
\begin{equation}
    \cl_{\rm VS} ~=~ - \fracm{1}{16} \, \rD^{a} \, \rD_{a}^{(+)} \, \Bl^{b} ( \rP^{(+)} \Bl )_{b} ~+~ {\rm h.\,c.} ~~~.
\end{equation}
This expression would give precisely the component Lagrangian stated above.

\subsection{Majorana Four-Component Notation TS}{\label{sec:M4TS}}

The tensor supermultiplet (TS) contains propagating fields: scalar ${\bm \varphi}$, antisymmetric rank-2 tensor $\bB_{\m\n}$, and spin-$\frac12$ fermion $\Bc_{a}$.
The transformation laws are
\begin{align}
\begin{split}
    \rD_{a} \Bvp ~=&~ \Bc_{a} ~~~, \\
    \rD_{a} \bB_{\m\nu} ~=&~ - \fracm{1}{4} ([\g_{\m},\g_{\nu}])_{a}{}^{b} \Bc_{b} ~~~,  \\
    \rD_{a} \Bc_{b} ~=&~ i (\g^{\m})_{ab} \pa_{\m} \Bvp - \e_{\m}{}^{\r\a\b} (\g^{5}\g^{\m})_{ab} \pa_{\r} \bB_{\a\b} ~~~.
\end{split}
\end{align}
We can define the field strength of the 2-form $\bB_{\a\b}$ to be
\begin{equation}
    \bH_{\r\a\b} ~=~ \pa_{[\r} \bB_{\a\b]} ~~~,
\end{equation}
and the Hodge-dual of the field strength to be
\begin{equation}
    \bH_{\m} ~=~ \fracm{1}{3!} \e_{\m}{}^{\r\a\b} \bH_{\r\a\b} ~~~.
\end{equation}
We can rewrite the transformation laws with chiral projection operators, 
\begin{equation}
\begin{split}
    \rD^{(\pm)}_{a} \Bvp ~=&~ (\rP^{(\pm)})_{a}{}^{b} \Bc_{b} ~~~, \\
    \rD^{(\pm)}_{a} \bH_{\m} ~=&~ \mp i (\rP^{(\pm)} \g_{\m}{}^{\r})_{a}{}^{b} \pa_{\r} \Bc_{b} ~~~,  \\
    \rD^{(\pm)}_{a} \Bc_{b} ~=&~ i (\rP^{(\pm)} \g^{\m})_{ab} \,\big[~ \pa_{\m} \Bvp ~\pm~ i \bH_{\m} ~\big] ~~~.
\end{split}
\end{equation}

The component Lagrangian is
\begin{equation}
\begin{split}
    \cl_{\rm TS} ~=&~ -~ \fracm12 \pa_\m \bm{\varphi} \pa^\m \bm{\varphi} ~-~ \fracm{1}{12} \bm{H}_{\m\n\r} \bm{H}^{\m\n\r} ~+~ i \fracm12 (\g^{\m})^{bc} \bm{\chi}_b \pa_\m \bm{\chi}_c \\
    ~=&~ -~ \fracm12 \pa_\m \bm{\varphi} \pa^\m \bm{\varphi} ~+~ \fracm12 \bm{H}_{\m} \bm{H}^{\m} ~+~ i \fracm12 (\g^{\m})^{bc} \bm{\chi}_b \pa_\m \bm{\chi}_c ~~~.
\label{equ:TS-L}
\end{split}
\end{equation}
It can be rewritten as 
\begin{equation}
    \cl_{\rm TS} ~=~ \fracm{1}{32} \, \rD^{a} \, \rD_{a}^{(+)} \, \rD^{b} \rD_{b}^{(-)} \, \Bvp^{2} ~+~ {\rm h.\,c.} ~~~.
\end{equation}

\newpage

\section{From Three-Point Off-Shell Vertices to Four-Point On-Shell SYK}

In these interactions, we can introduce more than one copies of a certain supermultiplet. We use $\ca$, $\vca$, and $\tca$ to label copies of chiral supermultiplets, vector supermultiplets, and tensor supermultiplets respectively.
We denote the total copies of these three supermultiplets to be $N_{CS}$, $N_{VS}$, and $N_{TS}$.

In the following, for the integration over the whole superspace, we take the normalization
\begin{equation}
    \int d^2 \theta \, d^2 \Bar{\theta}  ~=~ \fracm{1}{16} \, \rD^{a} \rD^{(+)}_{a} \, \rD^{b} \rD^{(-)}_{b}  ~~~,
\end{equation}
while for the integration over the chiral half of the superspace, we take the normalization 
\begin{equation}
    \int d^2 \theta  ~=~ \fracm{1}{4} \, \rD^{a} \rD^{(+)}_{a}  ~~~.
\end{equation}

There is one superfield interaction by which 3-point terms may be introduced. 
By going on-shell, we see the emergence of 4-point SYK-type terms.

\subsection{CS + 3PT}


To start off, we introduce a 3-point interaction between an anti-chiral superfield and two chiral superfields.
The full superfield Lagrangian is
\begin{equation}
\begin{split}
    \cl_{\rm CS+3PT} ~=&~ \int d^2 \theta \, d^2 \Bar{\theta} ~ \Big[~
    -~ \fracm12 \, \bBF^{\ca} \, \BF_{\ca} ~\Big]
    ~+~ \Big\{~ \int d^2 \theta \, d^2 \Bar{\theta} ~ \Big[~ \fracm14 \, \k_{\ca\cb\cc} ~ \bBF^{\ca} \, \BF^{\cb} \, \BF^{\cc}
    ~\Big] ~+~ {\rm h.\, c.} ~\Big\}  \\
    ~=&~ \int d^2 \theta \, d^2 \Bar{\theta} ~ \Big[~
    -~ \fracm14 \, \bBF^{\ca} \, \BF_{\ca}
    ~+~ \fracm14 \, \k_{\ca\cb\cc} ~ \bBF^{\ca} \, \BF^{\cb} \, \BF^{\cc}
    ~\Big] ~+~ {\rm h.\, c.} ~~~,
\end{split}
\end{equation}
and the interaction term reads
\begin{equation}
    \cl_{\rm 3PT} ~=~ \fracm{1}{64} \, \k_{\ca\cb\cc} \, \rD^{a} \rD^{(+)}_{a} \, \rD^{b} \rD^{(-)}_{b} ~ 
    \Big[~  \bBF^{\ca} \, \BF^{\cb} \, \BF^{\cc}  ~\Big] ~+~ {\rm h.\, c.} ~~~.
\end{equation}
Component-wise, the interaction term is equivalent to
\begin{equation}
\begin{split}
    \cl^{\rm (off-shell)}_{\rm 3PT} ~=~ \fracm12 \,\k{}_{{\ca}{\cb}{\cc}} \Big[\,
    & ( \pa^\m \Bar{\bm\Phi}^{\ca} ) ( \pa_\m {\bm\Phi}^{\cb} ) {\bm\Phi}^{\cc}  
    ~+~ i \, 2 \, (\rP^{(-)}\g^{\m})^{ab} \, (\pa_{\mu}{\bm {\psi}}{}^{\ca}_a) {\bm {\psi}}{}^{\cb}_b {\bm\Phi}^{\cc}  \\
    &~-~ \Bar{\bX}^{\ca} \bX^{\cb} {\bm\Phi}^{\cc} 
    ~-~ i \, (\rP^{(+)})^{ab} \, \Bar{\bX}^{\ca} {\bm {\psi}}{}^{\cb}_a {\bm {\psi}}{}^{\cc}_b  \,\Big] 
    ~+~ {\rm {h.\, c.}}  ~~~.
\end{split}
\label{equ:offshell-3PT-A}
\end{equation}


To obtain the on-shell Lagrangian, we include the relevant kinetic terms and find the equations of motion of auxiliary fields, and substitute back to the Lagrangian. By doing this we can see explicitly that this off-shell 3-point interaction contains a SYK-type term on-shell. 
We put the step-by-step derivations towards on-shell Lagrangian in Appendix \ref{appen:CS+3PT-A} for interested readers to follow. 

From the off-shell Lagrangian $\cl_{\rm CS} + \cl_{\rm 3PT-A}$, one obtain the on-shell Lagrangian
\begin{equation}
\begin{split}
    \cl^{\rm (on-shell)}_{\rm CS+3PT} ~=&~ 
    -~ \fracm{1}{2} \pa_{\m} \BF^{\ca} \pa^{\m} \bBF_{\ca} 
    ~+~ i \, \fracm{1}{2} (\g^{\m})^{ab} \Bj^{\ca}_{a} \pa_{\m} \Bj_{b\ca} \\
    &~+~\fracm12 \,\k{}_{{\ca}{\cb}{\cc}} \Big[\,
    ( \pa^\m \Bar{\bm\Phi}^{\ca} ) ( \pa_\m {\bm\Phi}^{\cb} ) {\bm\Phi}^{\cc}  
    ~+~ i \, 2 \, (\rP^{(-)}\g^{\m})^{ab} \, (\pa_{\mu}{\bm {\psi}}{}^{\ca}_a) {\bm {\psi}}{}^{\cb}_b {\bm\Phi}^{\cc}   \,\Big] \\
    &~+~\fracm12 \,\k^{*}_{{\ca}{\cb}{\cc}} \Big[\,
    ( \pa^\m {\bm\Phi}^{\ca} ) ( \pa_\m \Bar{\bm\Phi}^{\cb} ) \Bar{\bm\Phi}^{\cc}  
    ~+~ i \, 2 \, (\rP^{(+)}\g^{\m})^{ab} \, (\pa_{\mu}{\bm {\psi}}{}^{\ca}_a) {\bm {\psi}}{}^{\cb}_b \Bar{\bm\Phi}^{\cc}   \,\Big] \\
    &~+~\frac12 \, \frac{\k_{\ca\cb\cc} \, \k^{*}_{\cd\ce\cf} \, (\rP^{(+)})^{ab} \, (\rP^{(-)})^{cd} }{\d_{\cd\ca} ~-~ \cy_{\cd\ca} } ~ \Bj_{a}^{\cb} \Bj_{b}^{\cc} \Bj_{c}^{\ce} \Bj_{d}^{\cf} ~~~,
    \end{split}
    \label{equ:onshell-3PT-A}
\end{equation}
where 
\begin{equation}
    \cy_{\ca\cb} ~=~ \k^{*}_{\ca\cb\cc} {\Bar{\bm\Phi}}^{\cc} ~+~ \k_{\cb\ca\cc} {\bm \Phi}^{\cc} ~~~,
\end{equation}
is linear in $\BF$.
Note that we can do the following expansion,
\begin{equation}
\begin{split}
    \frac{1}{\d_{\ca\cb} ~-~ \cy_{\ca\cb}} ~=&~ \sum_{j=0}^{\infty} \big(\, \cy^{j} \,\big)^{\cb\ca} \\
    ~=&~ \d^{\cb\ca} ~+~ \cy^{\cb\ca} ~+~ \cy^{\cb\cc} \cy_{\cc}{}^{\ca} ~+~ \cy^{\cb\cc} \cy_{\cc\cd} \cy^{\cd\ca} ~+~ \cdots ~~~,
\end{split}
\end{equation}
where we treat $\ca$, $\cb$ indices as the row and column indices of the $\cy$ ``matrix''. 
Then the last term in the on-shell Lagrangian above can be rewritten as 
\begin{equation}
    \frac12 \, \k_{\ca\cb\cc} \, \k^{*}_{\cd\ce\cf} \, (\rP^{(+)})^{ab} \, (\rP^{(-)})^{cd} ~ \sum_{j=0}^{\infty} \big(\, \cy^{j} \,\big)^{\ca\cd} ~ \Bj_{a}^{\cb} \Bj_{b}^{\cc} \Bj_{c}^{\ce} \Bj_{d}^{\cf} ~~~,
\end{equation}
and the first term in this expansion is clearly the SYK term containing four Majorana fermions. 

In the following table, we list the collection of bosons and fermions in this model, where we have $N_{CS}$ copies of chiral multiplets. Also since we have imposed on-shell condition, $\bm{\psi}^{\ca}$ satisfies the Dirac equation. 
\begin{table}[htp!]
    \centering
    \begin{tabular}{|c|c|c|c|}
        \hline
        Bosons & total \# & Fermions & total \# \\\hline
        $\bA^{\ca}$, $\bB^{\ca}$ & $2\times N_{CS}$ & $\bm{\psi}^{\ca}_a$ & $2\times N_{CS}$ \\\hline
    \end{tabular}
    \label{tab:bnf_dof}
\end{table}
\newline \noindent

\newpage

\section{From $q$-Point Off-Shell Vertices to Four-Point On-Shell SYK\label{sec:n-point}}

In this chapter, we explore off-shell $q$-point interactions that are obtained by integrating over the whole superspace (nCS-A, nTS-A). These interactions also give 4-point SYK terms on-shell.

\subsection{CS + nCS-A}


An $q$-point superfield interaction among one chiral and a polynomial of anti-chiral superfields together with kinetic terms can be written in the form
\begin{equation}
\begin{split}
    \cl_{\rm CS+nCS-A} ~=~ \int d^2 \theta \, d^2 \Bar{\theta} ~ \Big[~
    -~ \fracm14 \, \bBF^{\ca} \, \BF_{\ca} 
    ~+~ \fracm12 \, \BF^{\ca} \cp_{\ca} (\bBF)
    ~\Big] ~+~ {\rm h.\, c.} ~~~.
\end{split}
\end{equation}
The interaction term is
\begin{equation}
    \cl_{\rm nCS-A} ~=~ \fracm{1}{32} ~ \rD^{a} \rD_{a}^{(+)} \, \rD^{b} \rD_{b}^{(-)} ~ \Big[~ 
    \BF^{\ca} \cp_{\ca} (\bBF) ~\Big] ~+~ {\rm h.\, c.}  ~~~,
\end{equation}
and the polynomial is defined by
\begin{equation}
    \cp_{\ca}(\Bar{\bm \Phi}) ~=~ \sum_{i=2}^{P} \k^{(i)}_{\ca\cb_1\cdots\cb_i} \prod_{k=1}^{i} {\Bar{\bm \Phi}}^{\cb_k}  ~~~,
\label{eqn:PdefAAAA}
\end{equation}
where $\k^{(i)}_{\ca\cb_1\cdots\cb_{i}}$'s are arbitrary coefficients, and the degree of the polynomial is $P$. 
Note that we start the polynomial from degree 2, i.e. the coupling terms start at cubic order, as the quadratic order term has the form of the kinetic term.
Obviously, $\cb_1$ to $\cb_i$ indices for any $1 \leq i \leq P$ on the coefficient $\k^{(i)}_{\ca\cb_{1}\cdots\cb_{i}}$ are symmetric.
We then have
\begin{align}
    \cp'_{\ca\cb_1}({\Bar{\bm \Phi}}) ~=&~  \sum_{j=2}^{P} j \, \k^{(j)}_{\ca\cb_1\cdots\cb_{j}} \prod_{k=2}^{j} {\Bar{\bm \Phi}}^{\cb_k}  ~~~,  \\
    \cp''_{\ca\cb_1\cb_2}({\Bar{\bm \Phi}}) ~=&~ 2\, \k^{(2)}_{\ca\cb_1\cb_2}  ~+~ \sum_{j=3}^{P} j(j-1) \, \k^{(j)}_{\ca\cb_1\cdots\cb_{j}} \prod_{k=3}^{j} {\Bar{\bm \Phi}}^{\cb_k}  ~~~.  
\end{align}
In terms of components, the interaction can be written as
\begin{equation}
\begin{split}
    \cl^{\rm (off-shell)}_{n{\rm {CS-A}}} 
    ~=&~  
    -~ \fracm12 \cp'_{\ca\cb}({\Bar{\bm \Phi}})\,\BF^{\ca}\,\Box\,\Bar{\BF}^{\cb}
    ~-~ \fracm12 \cp''_{\ca\cb_1\cb_2}({\Bar{\bm \Phi}})\,\BF^{\ca}\,(\pa_{\m}\Bar{\BF}^{\cb_1})\,(\pa^{\m}\Bar{\BF}^{\cb_2})\\
    &~-~ i\,\cp'_{\ca\cb}({\Bar{\bm \Phi}})\,(\rP^{(+)}\g^{\m})^{ab}\,{\bm \psi}_a^{\ca}\,\pa_{\mu}{\bm \psi}_b^{\cb}\\
    &~-~ i\,\cp''_{\ca\cb_1\cb_2}({\Bar{\bm \Phi}})\,(\rP^{(+)}\g^{\m})^{ab}\,{\bm \psi}_a^{\ca}\,(\pa_{\mu}{\bm \bBF}^{\cb_1})\,{\bm \psi}_b^{\cb_2}\\
    &~-~ i\,\fracm12\,\cp''_{\ca\cb_1\cb_2}({\Bar{\bm \Phi}})\,(\rP^{(-)})^{ab}\,\bX^{\ca}\,{\bm \psi}_a^{\cb_1}\,{\bm \psi}_b^{\cb_2} \\
    &~-~ \fracm12\,\cp'_{\ca\cb}({\Bar{\bm \Phi}})\,\bX^{\ca}\,\Bar{\bX}^{\cb}
    ~+~ {\rm {h.\, c.}} ~~~.
\end{split}
\label{equ:offshell-nCS-A}
\end{equation}

There are only two auxiliary fields $\bX$ and $\bbX$ in this model. The step-by-step derivations towards the on-shell Lagrangian is in Appendix \ref{appen:CS+nCS-A}. 
From the off-shell Lagrangian $\cl_{{\rm {CS}}}  + \cl_{n{\rm {CS-A}}}$,
the on-shell one is 
\begin{equation}
\begin{split}
    \cl^{\rm (on-shell)}_{{\rm CS} + n{\rm {CS-A}}} 
    ~=&~  
    -~  \fracm{1}{2} \, \pa^{\m} \BF^{\ca} \pa_{\m} \bBF_{\ca}
    ~+~ i  \, \fracm{1}{2} \,(\g^\m)^{ab} \Bj^{\ca}_a \pa_\m  \Bj_{b \, \ca} \\
    &~+~\Big[ ~-~ \fracm12 \cp'_{\ca\cb}({\Bar{\bm \Phi}})\,\BF^{\ca}\,\Box\,\Bar{\BF}^{\cb}
    ~-~ \fracm12 \cp''_{\ca\cb_1\cb_2}({\Bar{\bm \Phi}})\,\BF^{\ca}\,(\pa_{\m}\Bar{\BF}^{\cb_1})\,(\pa^{\m}\Bar{\BF}^{\cb_2}) \\
    &~~~~~~~ ~-~ i\,\cp'_{\ca\cb}({\Bar{\bm \Phi}})\,(\rP^{(+)}\g^{\m})^{ab}\,{\bm \psi}_a^{\ca}\,\pa_{\mu}{\bm \psi}_b^{\cb} \\
    &~~~~~~~ ~-~ i\,\cp''_{\ca\cb_1\cb_2}({\Bar{\bm \Phi}})\,(\rP^{(+)}\g^{\m})^{ab}\,{\bm \psi}_a^{\ca}\,(\pa_{\mu}{\bm \bBF}^{\cb_1})\,{\bm \psi}_b^{\cb_2} 
    ~+~ {\rm h.\, c.} \, \Big] \\
    &~+~ \frac12 ~ \frac{ \Bar{\cp}''_{\ca\cb_{1}\cb_{2}} (\BF) \, \cp''_{\cd\cc_{1}\cc_{2}} (\bBF) }{ \d_{\ca\cd} ~-~ \Bar{\cp}'_{\ca\cd} (\BF) ~-~ \cp'_{\cd\ca} (\bBF) } ~ (\rP^{(+)})^{ab} \, (\rP^{(-)})^{cd} \, \Bj_{a}^{\cb_{1}} \Bj_{b}^{\cb_{2}} \Bj_{c}^{\cc_{1}} \Bj_{d}^{\cc_{2}}
    ~~~.
\end{split}
\label{equ:onshell-nCS-A}
\end{equation}
If we expand the last term in the Lagrangian, the first term would be an arbitrary constant $\sim \k^{(2)* \, \ca}{}_{\cb_{1}\cb_{2}} \k^{(2)}_{\ca\cc_{1}\cc_{2}}$ multipled with some projection operators and 4 Majorana fermions. So it is the 4-point SYK-type term. 


In the following table, we list the collection of bosons and fermions in this model, where we have $N_{CS}$ copies of chiral multiplets. Also since we have imposed on-shell condition, $\bm{\psi}^{\ca}$ satisfies the Dirac equation. 
\begin{table}[htp!]
    \centering
    \begin{tabular}{|c|c|c|c|}
        \hline
        Bosons & total \# & Fermions & total \# \\\hline
        $\bA^{\ca}$, $\bB^{\ca}$ & $2\times N_{CS}$ & $\bm{\psi}^{\ca}_a$ & $2\times N_{CS}$ \\\hline
    \end{tabular}
    \label{tab:bnf_dof}
\end{table}

\newpage
\subsection{CS + TS + nTS-A}

An $q$-point superfield interaction among one chiral and a polynomial of $\Bvp$ with kinetic terms can be written in the form
\begin{equation}
\begin{split}
    \cl_{\rm CS+TS+nTS-A} ~=~ \int d^2 \theta \, d^2 \Bar{\theta} ~ \Big[~
    -~ \fracm14 \, \bBF^{\ca} \, \BF_{\ca} 
    ~+~ \fracm12 \, \Bvp^{\tca} \, \Bvp_{\tca} 
    ~+~ 2 \, \BF^{\ca} \cp_{\ca} (\Bvp)
    ~\Big] ~+~ {\rm h.\, c.} ~~~.
\end{split}
\end{equation}
The interaction term can be expressed as
\begin{equation}
    \cl_{\rm nTS-A} ~=~ \fracm{1}{8} ~ \rD^{a} \rD_{a}^{(+)} \, \rD^{b} \rD_{b}^{(-)} ~ \Big[~ 
    \BF^{\ca} \cp_{\ca} (\Bvp) ~\Big] ~+~ {\rm h.\, c.}  ~~~.
\end{equation}
We set
\begin{equation}
    \cp_{\ca}(\Bvp) ~=~ \sum_{i=2}^{P} \k^{(i)}_{\ca\tcb_1\cdots\tcb_i} \prod_{k=1}^{i} \Bvp^{\tcb_k}  ~~~,
\label{eqn:PdefAAAA}
\end{equation}
where $\k^{(i)}_{\ca\tcb_1\cdots\tcb_{i}}$'s are arbitrary coefficients, and the degree of the polynomial is $P$. 
Obviously, $\tcb_1$ to $\tcb_i$ indices for any $1\leq i \leq P$ on the coefficient $\k^{(i)}_{\ca\tcb_{1}\cdots\tcb_{i}}$ are symmetric.
We then have
\begin{align}
    \cp''_{\ca\tcb_1\tcb_2}(\Bvp) ~=&~ 2\, \k^{(2)}_{\ca\tcb_1\tcb_2}  ~+~ \sum_{j=3}^{P} j(j-1) \, \k^{(j)}_{\ca\tcb_1\cdots\tcb_{j}} \prod_{k=3}^{j} \Bvp^{\tcb_k}  ~~~,  \\
    \cp'''_{\ca\tcb_1\tcb_2\tcb_3}(\Bvp) ~=&~ 6\, \k^{(3)}_{\ca\tcb_1\tcb_2\tcb_3}  ~+~ \sum_{j=4}^{P} j (j-1) (j-2) \k^{(j)}_{\ca\tcb_1\cdots\tcb_{j}} \prod_{k=4}^{j} \Bvp^{\tcb_k}  ~~~, \\
    \cp''''_{\ca\tcb_1\tcb_2\tcb_3\tcb_4}(\Bvp) ~=&~ 24\,\k^{(4)}_{\ca\tcb_1\tcb_2\tcb_3\tcb_4}  ~+~ \sum_{j=5}^{P} j (j-1) (j-2) (j-3) \k^{(j)}_{\ca\tcb_1\cdots\tcb_{j}} \prod_{k=5}^{j} \Bvp^{\tcb_k}  ~~~.
\end{align}
In terms of components, the interaction can be written as
\begin{equation}
\begin{split}
    \cl^{\rm (off-shell)}_{n{\rm {TS-A}}} 
    ~=&~ 
    -~ i \fracm12 \, ({\rP}^{(-)})^{ab}  \bX^{\ca} \, \cp''_{\ca\tcb_1\tcb_2}(\Bvp) \, \Bc^{\tcb_1}_{a} \Bc^{\tcb_2}_{b}  \\
    &~+~ \fracm12 \, ({\rP}^{(+)})^{ab}  \,  ({\rP}^{(-)})^{cd} \, \cp'''_{\ca\tcb_1\tcb_2\tcb_3}(\Bvp) \, \Bj^{\ca}_{a} \Bc^{\tcb_1}_{b} \Bc^{\tcb_2}_{c} \Bc^{\tcb_3}_{d}  \\
    &~-~ i \,  ({\rP}^{(+)}\g^{\m})^{ab} \, \cp''_{\ca\tcb_1\tcb_2}(\Bvp) \, \bT_{\m}^{\tcb_1} \Bj^{\ca}_{a} \Bc^{\tcb_2}_{b}  \\
    &~+~ \fracm18 ({\rP}^{(+)})^{ab}  \,  ({\rP}^{(-)})^{cd}  \, \BF^{\ca} \, \cp''''_{\ca\tcb_1\tcb_2\tcb_3\tcb_4}(\Bvp) \, \Bc^{\tcb_1}_{a} \Bc^{\tcb_2}_{b} \Bc^{\tcb_3}_{c} \Bc^{\tcb_4}_{d}  \\
    &~-~ i \fracm14 \, (\g^{5}\g^{\m})^{ab} \, \BF^{\ca} \, \cp'''_{\ca\tcb_1\tcb_2\tcb_3}(\Bvp) \, \bT_{\m}^{\tcb_1} \, \Bc^{\tcb_2}_{a} \Bc^{\tcb_3}_{b}  \\
    &~+~ i \, ({\rP}^{(-)}\g^{\m})^{ab} \, \BF^{\ca} \, \cp''_{\ca\tcb_1\tcb_2}(\Bvp) \, \Bc^{\tcb_1}_{a} \pa_{\m} \Bc^{\tcb_2}_{b}  \\
    &~-~ \fracm12 \, \BF^{\ca} \, \cp''_{\ca\tcb_1\tcb_2}(\Bvp) \, \bT_{\m}^{\tcb_1} \bT^{\m \, \tcb_2} 
    ~+~ {\rm {h.\, c.}} ~~~,
\end{split}
\label{equ:offshell-nTS-A}
\end{equation}
where
\begin{equation}
    \bT_{\m}^{\tcb} ~=~ \pa_{\m} \Bvp^{\tcb} ~+~ i \bH_{\m}^{\tcb} ~~~.
\end{equation}


From off-shell $\cl_{{\rm {CS}}} + \cl_{{\rm {TS}}} + \cl_{{\rm {nTS-A}}}$, the final on-shell Lagrangian is
\begin{equation}
\begin{split}
    \cl^{(\rm on-shell)}_{{\rm {CS+ TS + nTS-A}}}  ~=&
    ~-~  \fracm{1}{2} \, \pa^{\m} \BF^{\ca} \pa_{\m} \bBF_{\ca}
    ~+~ i  \, \fracm{1}{2} \,(\g^\m)^{ab} \Bj^{\ca}_a \pa_\m  \Bj_{b \, \ca} \\
    &~-~ \fracm12 \pa_\m \bm{\varphi}^{\tca} \pa^\m \bm{\varphi}_{\tca} ~+~ \fracm12 \bm{H}_{\m}^{\tca} \bm{H}^{\m}_{\tca}
    ~+~ i \fracm12 (\g^{\m})^{ab} \Bc_a^{\tca} \pa_\m \Bc_{b \tca} \\
    &~+~\Big[\, \fracm12 \, ({\rP}^{(+)})^{ab}  \,  ({\rP}^{(-)})^{cd} \, \cp'''_{\ca\tcb_1\tcb_2\tcb_3}(\Bvp) \, \Bj^{\ca}_{a} \Bc^{\tcb_1}_{b} \Bc^{\tcb_2}_{c} \Bc^{\tcb_3}_{d}  \\
    &~~~~~~ ~-~ i \,  ({\rP}^{(+)}\g^{\m})^{ab} \, \cp''_{\ca\tcb_1\tcb_2}(\Bvp) \, \bT_{\m}^{\tcb_1} \Bj^{\ca}_{a} \Bc^{\tcb_2}_{b}  \\
    &~~~~~~ ~+~ \fracm18 \, ({\rP}^{(+)})^{ab}  \,  ({\rP}^{(-)})^{cd}  \, \BF^{\ca} \, \cp''''_{\ca\tcb_1\tcb_2\tcb_3\tcb_4}(\Bvp) \, \Bc^{\tcb_1}_{a} \Bc^{\tcb_2}_{b} \Bc^{\tcb_3}_{c} \Bc^{\tcb_4}_{d}  \\
    &~~~~~~ ~-~ i \fracm14 \, (\g^{5}\g^{\m})^{ab} \, \BF^{\ca} \, \cp'''_{\ca\tcb_1\tcb_2\tcb_3}(\Bvp) \, \bT_{\m}^{\tcb_1} \, \Bc^{\tcb_2}_{a} \Bc^{\tcb_3}_{b}  \\
    &~~~~~~ ~+~ i \, ({\rP}^{(-)}\g^{\m})^{ab} \, \BF^{\ca} \, \cp''_{\ca\tcb_1\tcb_2}(\Bvp) \, \Bc^{\tcb_1}_{a} \pa_{\m} \Bc^{\tcb_2}_{b}  \\
    &~~~~~~ ~-~ \fracm12 \, \BF^{\ca} \, \cp''_{\ca\tcb_1\tcb_2}(\Bvp) \, \bT_{\m}^{\tcb_1} \bT^{\m \, \tcb_2} ~+~ {\rm h.\, c.}\, \Big] \\
    &~+~ \fracm12 \, {\Bar\cp}''^{\ca}{}_{\tcb_1\tcb_2}({\bm \varphi}) \, \cp''_{\ca\tcb_3\tcb_4}({\bm \varphi}) ~ (\rP^{(+)}){}^{ab} \,  (\rP^{(-)}){}^{cd} ~ {\bm\chi}_{a}^{\tcb_1} {\bm\chi}_{b}^{\tcb_2} {\bm\chi}_{c}^{\tcb_3} {\bm\chi}_{d}^{\tcb_4}  ~~~.
\end{split}
\label{equ:onshell-nTS-A}
\end{equation}

There are two interaction terms that are purely fermionic. The first term comes from the auxiliary Lagrangian $\cl_{\bX}$,
\begin{equation}
    \fracm12 \, {\Bar\cp}''^{\ca}{}_{\tcb_1\tcb_2}({\bm \varphi}) \, \cp''_{\ca\tcb_3\tcb_4}({\bm \varphi}) ~ (\rP^{(+)}){}^{ab} \,  (\rP^{(-)}){}^{cd} ~ {\bm\chi}_{a}^{\tcb_1} {\bm\chi}_{b}^{\tcb_2} {\bm\chi}_{c}^{\tcb_3} {\bm\chi}_{d}^{\tcb_4} ~~~.
\end{equation}
Since the first terms in the polynomials $\cp''$ and $\Bar{\cp}''$ are constants, 
the leading term in its expansion is the 4-point SYK term.
The second term comes from the nTS-A propagating terms,
\begin{equation}
\begin{split}
    & ~ \fracm12 \, \cp'''_{\ca\tcb_1\tcb_2\tcb_3}(\Bvp) \,  ({\rP}^{(+)})^{ab}  \,  ({\rP}^{(-)})^{cd} \, \Bj^{\ca}_{a} \Bc^{\tcb_1}_{b} \Bc^{\tcb_2}_{c} \Bc^{\tcb_3}_{d} ~+~ {\rm h.\,c.}  \\
    ~=&~ 3 \, \k^{(3)}_{\ca\tcb_1\tcb_2\tcb_3} \, ({\rP}^{(+)})^{ab}  \,  ({\rP}^{(-)})^{cd} \, \Bj^{\ca}_{a} \Bc^{\tcb_1}_{b} \Bc^{\tcb_2}_{c} \Bc^{\tcb_3}_{d} ~+~ {\rm h.\,c.} ~+~ \cdots ~~~,
\end{split}
\end{equation}
which involves interactions between CS fermions and TS fermions.

The table below includes the collection of bosons and fermions involved in this interaction. 
\begin{table}[htp!]
    \centering
    \begin{tabular}{|c|c|c|c|}
        \hline
        Bosons & total \# & Fermions & total \# \\\hline
        $\bA^{\ca}$, $\bB^{\ca}$ & $2\times N_{CS}$ & $\bm{\psi}^{\ca}_a$ & $2\times N_{CS}$ \\\hline
        $\bm{\varphi}^{\tcb}$, $\bm{H}_{\mu}^{\tcb}$ & $4\times N_{TS}$ & $\bm{\chi}_a^{\tcb}$ & $4\times N_{TS}$ \\
        \hline
    \end{tabular}
    \label{tab:bnf_dof}
\end{table}
\newline \noindent
In this model we have $N_{CS}$ copies of chiral multiplets and $N_{TS}$ copies of tensor multiplets. Also since we have imposed on-shell condition, $\bm{\psi}^{\ca}$ satisfy the Dirac equation.

\newpage

\section{From $q$-Point Off-Shell Vertices to 2$(q-1)$-Point On-Shell SYK}
\label{sec:npt2npt}

In the following, we explore off-shell $q$-point interactions that are obtained by integrating superpotentials over the chiral half of the superspace (nVS-B, nTS-B).

As the superpotentials that involves half the superspace integrations require chiral superfields, let us concentrate on the propagating fermions ($\Bl_a$, $\Bc_a$, ${\bm \psi}_a$) in VS, TS, and CS respectively. 

Define the currents 
\begin{equation}
\label{equ:current}
    \begin{split}
        \cj_{fg} ~=~ f^{a} \, ( \rP^{(-)} \, g )_{a} ~~~,&~~~ \Bar{\cj}_{fg} ~=~ f^{a} \, ( \rP^{(+)} \, g )_{a} ~~~,\\
        \cj^{\mu}_{fg} ~=~ f^{a} \, ( \rP^{(-)}\g^{\mu} \, g )_{a} ~~~,&~~~ \Bar{\cj}^{\mu}_{fg} ~=~ f^{a} \, ( \rP^{(+)}\g^{\mu} \, g )_{a} ~~~,\\
        \cj^{\mu\n}_{fg} ~=~ f^{a} \, ( \rP^{(-)}\g^{\mu\n} \, g )_{a} ~~~,&~~~ \Bar{\cj}^{\mu\n}_{fg} ~=~ f^{a} \, ( \rP^{(+)}\g^{\mu\n} \, g )_{a} ~~~,\\
    \end{split}
\end{equation}
where $f,g =$ ($\Bl$, $\Bc$, ${\bm \psi}$). Also, $\cj_{fg} = \cj_{gf}$, $\Bar{\cj}_{fg}=\Bar{\cj}_{gf}$, $\cj^{\mu}_{fg} = -\Bar{\cj}^{\mu}_{gf}$, ${\cj}^{\mu\n}_{fg} = -{\cj}^{\mu\n}_{gf}$, and $\Bar{\cj}^{\mu\n}_{fg} = -\Bar{\cj}^{\mu\n}_{gf}$. 
Then in order to construct the proper superpotentials in terms of polynomials of these currents, we examine the chiral condition 
\begin{equation}
    \rD^{(-)}_{a} \cj ~=~ 0 ~~~,
\end{equation}
and antichiral condition 
\begin{equation}
    \rD^{(+)}_{a} \cj ~=~ 0 ~~~,
\end{equation}
on all of them. 

Note that we have
\begin{equation}
\begin{split}
    \rD_{a}^{(-)} ( \rP^{(\pm)} ){}_b{}^c
    \Bj_{c} ~\ne&~ 0 ~~~,
\end{split}
\label{eq:CSpm1}
\end{equation}
which implies that any current involved $\Bj_a$ is neither chiral or antichiral. 
Table \ref{tab:sum_current} is the summary of chiral/antichiral properties of all currents defined in Eq. (\ref{equ:current}). 
\begin{table}[htp!]
    \centering
    \begin{tabular}{|c|c|c|c|c|c|c|}
    \hline
       $(f,g)$  & $\cj_{fg}$ & $\Bar{\cj}_{fg}$ & $\cj^{\mu}_{fg}$ & $\Bar{\cj}^{\mu}_{fg}$ & $\cj^{\mu\n}_{fg}$ & $\Bar{\cj}^{\mu\n}_{fg}$ \\ \hline
       $(\Bl,\Bl)$  & AC & C & N/A & N/A & 0 & 0\\ \hline
       $(\Bl,\Bc)$  & N/A & N/A & AC & C & N/A & N/A\\ \hline
       $(\Bc,\Bc)$  & C & AC & N/A & N/A & 0 & 0\\ \hline
    \end{tabular}
    \caption{Summary of chiral/antichiral properties of all currents defined in Eq. (\ref{equ:current}). C = Chiral, AC = Anti-Chiral, N/A = neither chiral or anti-chiral, 0 = this current is identically zero.}
    \label{tab:sum_current}
\end{table}


Moreover, if we consider the following Fierz Identities,
\begin{align}
    (\rP^{(\pm)}\g^{\m})^{ab}\, (\rP^{(\pm)}\g_{\m})^{cd} ~=&~ -2\, (\rP^{(\pm)})^{ac}\, (\rP^{(\mp)})^{bd} ~~~, \\
    (\rP^{(\pm)}\g^{\m})^{ab}\, (\rP^{(\mp)}\g_{\m})^{cd} ~=&~ -\, (\rP^{(\pm)}\g^{\r})^{ac}\, (\rP^{(\mp)}\g_{\r})^{bd} ~~~, \\
    (\rP^{(\pm)}\g^{\m\n})^{ab}\, (\rP^{(\pm)}\g_{\m\n})^{cd} ~=&~ 6\, (\rP^{(\pm)})^{ac}\, (\rP^{(\pm)})^{bd} -\fracm12\,(\rP^{(\pm)}\g^{\m\n})^{ac}\,(\rP^{(\pm)}\g_{\m\n})^{bd} ~~~, \\
    (\rP^{(\pm)}\g^{\m\n})^{ab}\, (\rP^{(\mp)}\g_{\m\n})^{cd} ~=&~ 0 ~~~,
\end{align}
We will obtain equivalences between chiral currents as indicated below,
\begin{align}
    \cj^{\mu}_{fg}\,\cj_{\mu}{}_{lm} ~=&~ 2\,\cj_{fl}\,\Bar{\cj}_{gm} ~~~, \\
    \Bar{\cj}^{\mu}_{fg}\,\Bar{\cj}_{\mu}{}_{lm} ~=&~ 2\,\Bar{\cj}_{fl}\,{\cj}_{gm} ~~~, \\
    \cj^{\mu}_{fg}\,\Bar{\cj}_{\mu}{}_{lm} ~=&~ \cj^{\r}_{fl}\,\Bar{\cj}_{\r}{}_{gm} ~~~, \\
    \cj^{\mu\n}_{fg}\,\cj_{\mu\n}{}_{lm} ~=&~ -6\,\cj_{ff}\cj_{gg} ~~~, \\
    \Bar{\cj}^{\mu\n}_{fg}\,\Bar{\cj}_{\mu\n}{}_{fg} ~=&~ -6\,\Bar{\cj}_{ff}\Bar{\cj}_{gg}  ~~~, \\
    \cj^{\mu\n}_{fg}\,\Bar{\cj}_{\mu\n}{}_{lm} ~=&~ 0 ~~~,
\end{align}
where $f,g,l,m=$ ($\Bl$, $\Bc$, ${\bm \psi}$).

In the following sections, we will explicitly discuss three models constructed by polynomials of 
\begin{equation}
    \begin{split}
        \cj_{11} ~\equiv&~ \Bar{\cj}_{\Bl\Bl}~~~,\\
        \cj_{22} ~\equiv&~ {\cj}_{\Bc\Bc}~~~,\\
    \end{split}
\end{equation}
which satisfy the chiral condition. The other chiral currents also lead to possible superpotentials, which will not be explicitly constructed below. 
We will consider interactions of the form
\begin{equation}
    \cl_{\rm int} ~=~ 
    \int d^{2} \theta ~ \Big[~ \cw (\BF,\cj) ~\Big] ~+~ {\rm {h.\, c.}}   ~~~,
\end{equation}
where the superpotential is a function of chiral superfield $\BF$ from chiral supermultiplet and the chiral current $\cj$ in focus, and it takes the form
\begin{equation}
    \cw (\BF,\cj) ~=~ \fracm12 \, \BF^{\ca} \cf_{\ca} (\cj) ~~~,
\end{equation}
where $\cf(\cj)$ is a polynomial in $\cj$.

\newpage
\subsection{CS + VS + nVS-B}


Recall the current defined by
\begin{equation}
    \cj_{11}^{\vcb_{1}\vcb_{2}} ~=~ \Bl^{a \, \vcb_{1}} (\rP^{(+)} \Bl^{\vcb_{2}})_{a} ~~~,
\end{equation}
and that it satisfies the chiral condition
\begin{equation}
    {\rm D}^{(-)}_{a} \,\cj_{11}^{\vcb_{1}\vcb_{2}} ~=~ 0 ~~~.
\end{equation}
Define also 
\begin{equation}
    \Bar{\cj}_{11}^{\vcb_{1}\vcb_{2}} ~=~ \Bl^{a \, \vcb_{1}} (\rP^{(-)} \Bl^{\vcb_{2}})_{a} ~~~,
\end{equation}
and note that $(\cj_{11})^{*} = - \Bar{\cj}_{11}$.
The full interaction Lagrangian with kinetic terms can be constructed as
\begin{equation}
\begin{split}
    \cl_{\rm CS+VS+nVS-B} ~=&~ \int d^{2} \theta \, d^{2} \Bar{\theta} ~ \Big[~ -~ \fracm12 \, \bBF^{\ca} \, \BF_{\ca} ~\Big] \\
    &~+~ \Big\{~ \int d^{2} \theta ~ \Big[~ -~ \fracm14 \, \Bl^{a \, \vcb} (\rP^{(+)} \Bl_{\vcb})_{a} ~+~ \fracm12 \, \BF^{\ca} \cf_{\ca} (\cj_{11}) ~\Big] ~+~ {\rm h.\,c.} ~\Big\} ~~~,
\end{split}
\end{equation}
and the nVS-B interaction piece can be expressed as
\be
    \cl_{\rm nVS-B} ~=~ 
    \fracm18 \, {\rm D}^{a}{\rm D}^{(+)}_{a} ~ 
    \Big[~ \BF^{\ca} \cf_{\ca} (\cj_{11}) ~\Big]  ~+~ {\rm {h.\, c.}}  ~~~,
\ee
where the polynomial function is defined as
\begin{equation}
\begin{split}
    \cf_{\ca} (\cj_{11}) ~=&~  \sum_{i=1}^{P} \k^{(i)}_{\ca\vcb_1\vcb_2\cdots\vcb_{2i-1}\vcb_{2i}} \prod_{k=1}^{i} \cj_{11}^{\vcb_{2k-1}\vcb_{2k}}  ~~~, \\
    \Bar{\cf}_{\ca} (\Bar{\cj}_{11}) ~=&~  \sum_{i=1}^{P} (-1)^{i} \k^{(i) *}_{\ca\vcb_1\vcb_2\cdots\vcb_{2i-1}\vcb_{2i}} \prod_{k=1}^{i} \Bar{\cj}_{11}^{\vcb_{2k-1}\vcb_{2k}} ~~~,
\label{eqn:PdefnVM}
\end{split}
\end{equation}
in which the index pairs $\vcb_1\vcb_2, \dots, \vcb_{2P-1}\vcb_{2P}$ are symmetric. 
From the above definitions we obtain
\begin{align}
    \cf'_{\ca\vcb_1\vcb_2}(\cj_{11}) ~=&~  \k^{(1)}_{\ca\vcb_1\vcb_2}  ~+~ \sum_{j=2}^{P} j \, \k^{(j)}_{\ca\vcb_1\vcb_2\cdots\vcb_{2j-1}\vcb_{2j}} \prod_{k=2}^{j} \cj_{11}^{\vcb_{2k-1}\vcb_{2k}}  ~~~,  \\
    \cf''_{\ca\vcb_1\vcb_2\vcb_3\vcb_4}(\cj_{11}) ~=&~ 2\,\k^{(2)}_{\ca\vcb_1\vcb_2\vcb_3\vcb_4}  ~+~ \sum_{j=3}^{P} j (j-1) \, \k^{(j)}_{\ca\vcb_1\vcb_2\cdots\vcb_{2j-1}\vcb_{2j}} \prod_{k=3}^{j} \cj_{11}^{\vcb_{2k-1}\vcb_{2k}}  ~~~.
\end{align}
Then the component description of the action is 
\begin{equation}
\begin{split}
    \cl^{\rm (off-shell)}_{\rm nVS-B} ~=&~
    ~i \fracm12 \, \bX^{\ca} \, \cf_{\ca} (\cj_{11}) 
    ~+~ i \fracm12 \,\cf'_{\ca\vcb_1\vcb_2}(\cj_{11})\,(\rP^{(+)}\g^{\m\n})^{ab}\,\bm{\psi}_a^{\ca}\, \bF_{\m\n}^{\vcb_1}\,\Bl_b^{\vcb_2} \\
    &~-~ \cf'_{\ca\vcb_1\vcb_2}(\cj_{11})\,(\rP^{(+)})^{ab}\,\bm{\psi}_a^{\ca}\, {\bm {\rm d}}^{\vcb_1}\,\Bl_b^{\vcb_2} \\
    &~+~ \bm{\Phi}^{\ca}\,\cf''_{\ca\vcb_1\vcb_2\vcb_3\vcb_4}(\cj_{11})\,\Big[\, -~ \fracm18 (\rP^{(+)}\g^{\m\n}\g^{\a\b})^{ab}\,\bF_{\m\n}^{\vcb_1}\,\bF_{\a\b}^{\vcb_3} \\
    &~+~ i \fracm12 \,(\rP^{(+)}\g^{\a\b})^{ab}\,{\bm {\rm d}}^{\vcb_1}\,\bF_{\a\b}^{\vcb_3}
    ~-~ \fracm12 \, (\rP^{(+)})^{ab}\,{\bm {\rm d}}^{\vcb_1}\,{\bm {\rm d}}^{\vcb_3}\,
    \Big]\,\Bl_a^{\vcb_2}\,\Bl_b^{\vcb_4} \\
    &~+~ \bm{\Phi}^{\ca}\,\cf'_{\ca\vcb_1\vcb_2}(\cj_{11})\, \Big[~ i \,(\rP^{(-)}\g^{\m})^{ab}\,(\pa_{\mu}\Bl_a^{\vcb_1})\,\Bl_b^{\vcb_2} ~-~ \fracm12 \,{\bm {\rm d}}^{\vcb_1}\,{\bm {\rm d}}^{\vcb_2}\\
    &~+~ \fracm14 \,\bF_{\m\n}^{\vcb_1}\,\bF^{\m\n}{}^{\vcb_2}
    ~-~ i \fracm18 \,\e^{\m\n\a\b}\,\bF_{\m\n}^{\vcb_1}\,\bF_{\a\b}^{\vcb_2} ~\Big] ~+~ {\rm {h.\, c.}} ~~~.
\end{split}
\label{equ:offshell-nVS-B}
\end{equation}


From $\cl_{{\rm {CS}}} + \cl_{{\rm {VS}}}+ \cl_{n{\rm {VS-B}}}$, follow the standard approach, presented explicitly in Appendix \ref{appen:CS+VS+nVS-B}, and the on-shell Lagrangian can be obtained as
\begin{equation}
\begin{split}
    \cl^{(\rm on-shell)}_{{\rm {CS+ VS + nVS-B}}}  ~=&~
    -~  \fracm{1}{2} \, \pa^{\m} \BF^{\ca} \pa_{\m} \bBF_{\ca}
    ~+~ i  \, \fracm{1}{2} \,(\g^\m)^{ab} \Bj^{\ca}_a \pa_\m  \Bj_{b \, \ca} \\
    &~-~ \fracm{1}{4} \bF^{\vcb}_{\mu\nu} \bF_{\vcb}^{\mu\nu}  ~+~ i\,  \fracm{1}{2} \, (\gamma^\mu)^{a b} {\bm \l}^{\vcb}_a \partial_\mu {\bm \l}_b{}_{\vcb}  \\
    &~+~ \Big[\,  i \fracm12 \,\cf'_{\ca\vcb_1\vcb_2}(\cj_{11})\,(\rP^{(+)}\g^{\m\n})^{ab}\,\bm{\psi}_a^{\ca}\, \bF_{\m\n}^{\vcb_1}\,\Bl_b^{\vcb_2} \\
    &~~~~~~ ~-~ \fracm18 \bm{\Phi}^{\ca}\,\cf''_{\ca\vcb_1\vcb_2\vcb_3\vcb_4}(\cj_{11})\, (\rP^{(+)}\g^{\m\n}\g^{\a\b})^{ab}\,\bF_{\m\n}^{\vcb_1}\,\bF_{\a\b}^{\vcb_3} \Bl_a^{\vcb_2}\,\Bl_b^{\vcb_4}\\
    &~~~~~~ ~+~ i \,\bm{\Phi}^{\ca}\,\cf'_{\ca\vcb_1\vcb_2}(\cj_{11})\,(\rP^{(-)}\g^{\m})^{ab}\,(\pa_{\mu}\Bl_a^{\vcb_1})\,\Bl_b^{\vcb_2} \\
    &~~~~~~ ~+~ \fracm14 \,\bm{\Phi}^{\ca}\,\cf'_{\ca\vcb_1\vcb_2}(\cj_{11})\, \Big(\, \bF_{\m\n}^{\vcb_1}\,\bF^{\m\n}{}^{\vcb_2} 
    ~-~ i \fracm12 \, \e^{\m\n\a\b}\,\bF_{\m\n}^{\vcb_1}\,\bF_{\a\b}^{\vcb_2}   \,\Big) \\
    &~~~~~~ 
    ~+~ {\rm h.\,c.}     \,\Big] \\
    &~-~  \fracm12 \, \sum_{i,j=1}^{P}\,  (-1)^{i} \,  \k^{(i)\ca}{}_{\vcb_1\cdots\vcb_{2i}}\, \k^{(j)*}_{\ca\vcc_1\cdots\vcc_{2j}}\, \\
    &~~~~~~ ~\times~\prod_{k=1}^{i} \prod_{l=1}^{j} (\rP^{(+)})^{ a_{k}b_{k}} (\rP^{(-)})^{c_{l}d_{l}} \Bl_{a_{k}}^{\vcb_{2k-1}} \Bl_{b_{k}}^{\vcb_{2k}} \Bl_{c_{l}}^{\vcc_{2l-1}}\Bl_{d_{l}}^{\vcc_{2l}}\\
    &~-~ \fracm12\,\frac{\Omega_{\vca}\Omega_{\vcb}}{\d_{\vca\vcb}-\cy_{\vca\vcb}}  ~~~,
\end{split}
\label{equ:onshell-nVS-B}
\end{equation}
where 
\begin{equation}
    \begin{split}
        \cy_{\vcb\vcd} ~=&~ 
        -~ \bm{\Phi}^{\ca}\,\cf''_{\ca\vcb\vcb_2\vcd\vcb_4}(\cj_{11})\,\cj_{11}^{\vcb_2\vcb_4} 
        ~+~ \bm{\Phi}^{\ca}\,\cf'_{\ca\vcb\vcd}(\cj_{11}) \\
        &~+~ \Bar{\bm{\Phi}}^{\ca}\,\Bar{\cf}''_{\ca\vcb\vcb_2\vcd\vcb_4}(\Bar{\cj}_{11})\,\Bar{\cj}_{11}^{\vcb_2\vcb_4} 
        ~+~ \Bar{\bm{\Phi}}\,\Bar{\cf}'_{\ca\vcb\vcd}(\Bar{\cj}_{11}) ~~~,
    \end{split}
\end{equation}
satisfying 
\begin{equation}
    \cy_{\vcb\vcd} ~=~ \cy_{\vcd\vcb} ~~~,~~~ (\cy_{\vcb\vcd})^* ~=~ \cy_{\vcb\vcd} ~~~,
\end{equation}
and
\begin{equation}
    \begin{split}
        \Omega_{\vcb}~=&~ \Big[\,\cf'_{\ca\vcb\vcc}(\cj_{11})\,(\rP^{(+)})^{ab}-\Bar{\cf}'_{\ca\vcb\vcc}(\Bar{\cj}_{11})\,(\rP^{(-)})^{ab}\,\Big]\, \Bj_a^{\ca}\,\Bl_b^{\vcc} \\
        &~-~ i \fracm12 \,\Big[\,\bm{\Phi}^{\ca}\,\cf''_{\ca\vcb\vcc_2\vcc_3\vcc_4}(\cj_{11})\,(\rP^{(+)}\g^{\m\n})^{ab}+\Bar{\bm{\Phi}}^{\ca}\,\Bar{\cf}''_{\ca\vcb\vcc_2\vcc_3\vcc_4}(\Bar{\cj}_{11})\,(\rP^{(-)}\g^{\m\n})^{ab}\,\Big]\,\bF_{\m\n}^{\vcc_3}\Bl_a^{\vcc_2}\Bl_b^{\vcc_4} ~~~.
    \end{split}
\end{equation}

There are two interaction terms that are purely fermionic. The first one comes from $\cl_\bX$,
\begin{equation}
    ~-~ \fracm12 \, \sum_{i,j=1}^{P}\,  (-1)^{i} \,  \k^{(i)\ca}{}_{\vcb_1\cdots\vcb_{2i}}\, \k^{(j)*}_{\ca\vcc_1\cdots\vcc_{2j}} ~ 
    \prod_{k=1}^{i} \prod_{l=1}^{j} (\rP^{(+)})^{ a_{k}b_{k}} (\rP^{(-)})^{c_{l}d_{l}} \Bl_{a_{k}}^{\vcb_{2k-1}} \Bl_{b_{k}}^{\vcb_{2k}} \Bl_{c_{l}}^{\vcc_{2l-1}}\Bl_{d_{l}}^{\vcc_{2l}} ~~~,
\end{equation}
which are the 2$(q-1)$-point SYK interactions we are looking for. 
The second term comes from the auxiliary Lagrangian $\cl_{\bm{\rm d}}$ and involves Majorana fermions from both CS and VS,
\begin{equation}
\begin{split}
    & ~-~ \fracm12\,\frac{\Omega_{\vca}\Omega_{\vcb}}{\d_{\vca\vcb}-\cy_{\vca\vcb}} \\
    ~=&~ -~ \fracm12\,\Big[\,\k^{(1)}_{\ca\vcb\vcc}(\rP^{(+)})^{ab}-\k^{(1)*}_{\ca\vcb\vcc}(\rP^{(-)})^{ab}\,\Big]\Big[\,\k^{(1)}{}_{\cd}{}^{\vcb}{}_{\vce}(\rP^{(+)})^{cd}-\k^{(1)*}{}_{\cd}{}^{\vcb}{}_{\vce}(\rP^{(-)})^{cd}\,\Big]\, \Bj_a^{\ca}\,\Bl_b^{\vcc}\, \Bj_c^{\cd}\,\Bl_d^{\vce} ~+~ \cdots 
\end{split}
\end{equation}

The collection of bosons and fermions for this theory is listed below.
\begin{table}[h!]
    \centering
    \begin{tabular}{|c|c|c|c|}
        \hline
        Bosons & total \# & Fermions & total \# \\\hline
        $\bA^{\ca}$, $\bB^{\ca}$ & $2\times N_{CS}$ & $\bm{\psi}^{\ca}_a$ & $2\times N_{CS}$ \\\hline
         $\bm{A}_{\m}^{\vcb}$ & $2\times N_{VS}$ & $\Bl_a^{\tcb}$ & $2\times N_{VS}$ \\
        \hline
    \end{tabular}
    \label{tab:bnf_dof}
\end{table}
In this model we have $N_{CS}$ copies of chiral multiplets and $N_{VS}$ copies of vector multiplets.

\newpage
\subsection{CS + TS + nTS-B}


Recall the current defined by
\begin{equation}
    \cj_{22}^{\tcb_{1}\tcb_{2}} ~=~ \Bc^{a \, \tcb_{1}} (\rP^{(-)} \Bc^{\tcb_{2}})_{a} ~~~,
\end{equation}
and that it satisfies the chiral condition
\begin{equation}
    {\rm D}^{(-)}_{a} \,\cj_{22}^{\tcb_{1}\tcb_{2}} ~=~ 0 ~~~.
\end{equation}
Define also 
\begin{equation}
    \Bar{\cj}_{22}^{\tcb_{1}\tcb_{2}} ~=~ \Bc^{a \, \tcb_{1}} (\rP^{(+)} \Bc^{\tcb_{2}})_{a} ~~~,
\end{equation}
and note that $(\cj_{22})^{*} = - \Bar{\cj}_{22}$.
The full superfield Lagrangian is
\begin{equation}
\begin{split}
    \cl_{\rm CS+TS+nTS-B} ~=&~ \int d^{2} \theta \, d^{2} \Bar{\theta} ~ \Big[~ -~ \fracm14 \, \bBF^{\ca} \, \BF_{\ca} 
    ~+~ \fracm12 \, \Bvp^{\tca} \Bvp_{\tca} ~\Big] \\
    &~+~ \int d^{2} \theta ~ \Big[~ \fracm12 \, \BF^{\ca} \cf_{\ca} (\cj_{22}) ~\Big] ~+~ {\rm h.\,c.} ~~~,
\end{split}
\end{equation}
and the nTS-B interaction is
\be
    \cl_{\rm nTS-B} ~=~ 
    \fracm18 \, {\rm D}^{a}{\rm D}^{(+)}_{a} ~ 
    \Big[~ \BF^{\ca} \cf_{\ca} (\cj_{22}) ~\Big]  ~+~ {\rm {h.\, c.}}  ~~~,
\ee
where the polynomial function is defined as
\begin{equation}
\begin{split}
    \cf_{\ca} (\cj_{22}) ~=&~  \sum_{i=1}^{P} \k^{(i)}_{\ca\tcb_1\tcb_2\cdots\tcb_{2i-1}\tcb_{2i}} \prod_{k=1}^{i} \cj_{22}^{\tcb_{2k-1}\tcb_{2k}}  ~~~, \\
    \Bar{\cf}_{\ca} (\Bar{\cj}_{22}) ~=&~  \sum_{i=1}^{P} (-1)^{i} \k^{(i) *}_{\ca\tcb_1\tcb_2\cdots\tcb_{2i-1}\tcb_{2i}} \prod_{k=1}^{i} \Bar{\cj}_{22}^{\tcb_{2k-1}\tcb_{2k}} ~~~,
\label{eqn:PdefnTM}
\end{split}
\end{equation}
where the index pairs $\tcb_1\tcb_2, \dots, \tcb_{2P-1}\tcb_{2P}$ are symmetric. 
From the above definitions we obtain
\begin{align}
    \cf'_{\ca\tcb_1\tcb_2}(\cj_{22}) ~=&~  \k^{(1)}_{\ca\tcb_1\tcb_2}  ~+~ \sum_{j=2}^{P} j \, \k^{(j)}_{\ca\tcb_1\tcb_2\cdots\tcb_{2j-1}\tcb_{2j}} \prod_{k=2}^{j} \cj_{22}^{\tcb_{2k-1}\tcb_{2k}}  ~~~,  \\
    \cf''_{\ca\tcb_1\tcb_2\tcb_3\tcb_4}(\cj_{22}) ~=&~ 2\,\k^{(2)}_{\ca\tcb_1\tcb_2\tcb_3\tcb_4}  ~+~ \sum_{j=3}^{P} j (j-1) \, \k^{(j)}_{\ca\tcb_1\tcb_2\cdots\tcb_{2j-1}\tcb_{2j}} \prod_{k=3}^{j} \cj_{22}^{\tcb_{2k-1}\tcb_{2k}}  ~~~.
\end{align}
The component description of the Lagrangian is 
\begin{equation}
\begin{split}
    \cl^{\rm (off-shell)}_{n{\rm {TS-B}}} ~=&~
    i \fracm12 \, \bX^{\ca} \, \cf_{\ca} (\cj_{22})
    ~-~ i \,({\rP}^{(+)}\g^{\m})^{ab} \, \cf'_{\ca\tcb_1\tcb_2} (\cj_{22}) \,\bm{\psi}^{\ca}_a \, \bT_{\m}^{\tcb_1} \,  \Bc^{\tcb_2}_b \\
    &~-~ \fracm12 \, (\rP^{(-)}){}^{ab} \, \cf''_{\ca\tcb_1\tcb_2\tcb_3\tcb_4} (\cj_{22}) \, \BF^{\ca} \, \bT_{\m}^{\tcb_1} \, \bT^{\m \, \tcb_3} \, \Bc^{\tcb_2}_a\, \Bc^{\tcb_4}_b \\ 
    &~+~ i \, ({\rP}^{(-)}\g^{\m})^{ab} \, \cf'_{\ca\tcb_1\tcb_2} (\cj_{22}) \, \BF^{\ca}\,  \Bc^{\tcb_1}_a\, (\pa_{\m} \Bc^{\tcb_2}_b) \\
    &~-~ \fracm12 \, \cf'_{\ca\tcb_1\tcb_2} (\cj_{22}) \, \BF^{\ca}\, \bT_{\m}^{\tcb_1} \, \bT^{\m \, \tcb_2}
    ~+~ {\rm {h.\, c.}} ~~~,
\end{split}
\label{equ:offshell-nTS-B}
\end{equation}
where again
\begin{equation}
    \bT_{\m}^{\tcb} ~=~ \pa_{\m} \Bvp^{\tcb} ~+~ i \bH_{\m}^{\tcb} ~~~.
\end{equation}


From $\cl_{{\rm {CS}}} + \cl_{{\rm {TS}}}+ \cl_{n{\rm {TS-B}}}$, the final on-shell Lagrangian can be written as
\begin{equation}
\begin{split}
    \cl^{(\rm on-shell)}_{{\rm {CS+ TS + nTS-B}}}  ~=&~
    -~ \fracm12 \,  \sum_{i,j=1}^{P}\, (-1)^{i} \, \k^{(i)\ca}{}_{\tcb_1\cdots\tcb_{2i}}\, \k^{(j)*}_{\ca\tcc_1\cdots\tcc_{2j}}\,  \\
    &~~~ ~\times~ \prod_{k=1}^{i} \prod_{l=1}^{j} (\rP^{(-)})^{a_{k}b_{k}} (\rP^{(+)})^{ c_{l}d_{l}} \Bc_{a_{k}}^{\tcb_{2k-1}} \Bc_{b_{k}}^{\tcb_{2k}} \Bc_{c_{l}}^{\tcc_{2l-1}} \Bc_{d_{l}}^{\tcc_{2l}}    \\
    &~-~  \fracm{1}{2} \, \pa^{\m} \BF^{\ca} \pa_{\m} \bBF_{\ca}
    ~+~ i  \, \fracm{1}{2} \,(\g^\m)^{cd} \Bj^{\ca}_c \pa_\m  \Bj_{d \, \ca} \\
    &~-~ \fracm12 \pa_\m \bm{\varphi}^{\tca} \pa^\m \bm{\varphi}_{\tca} ~+~ \fracm12 \bm{H}_{\m}^{\tca} \bm{H}^{\m}_{\tca}
    ~+~ i \fracm12 (\g^{\m})^{bc} \Bc_b^{\tca} \pa_\m \Bc_{c \tca} \\
    &~+~ \Big[ ~-~ i \, ({\rP}^{(+)}\g^{\m})^{ab}  \, \cf'_{\ca\tcb_1\tcb_2}(\cj_{22}) \,\bm{\psi}^{\ca}_a \, \bT_{\m}^{\tcb_1} \,\Bc^{\tcb_2}_b \\
    &~~~~~~~ ~-~ \fracm12 \, (\rP^{(-)}){}^{ab} \, \cf''_{\ca\tcb_1\tcb_2\tcb_3\tcb_4}(\cj_{22}) \, \BF^{\ca} \, \bT^{\m}{}^{\tcb_1} \, \bT_{\m}^{\tcb_3} \,\Bc^{\tcb_2}_a \,\Bc^{\tcb_4}_b \\ 
    &~~~~~~~ ~+~i \, ({\rP}^{(-)}\g^{\m})^{ab}  \, \cf'_{\ca\tcb_1\tcb_2}(\cj_{22}) \, \BF^{\ca} \, \Bc^{\tcb_1}_a \, (\pa_{\m}\Bc^{\tcb_2}_b) \\
    &~~~~~~~ ~-~ \fracm12  \, \cf'_{\ca\tcb_1\tcb_2}(\cj_{22}) \, \BF^{\ca} \, \bT^{\m}{}^{\tcb_1} \, \bT_{\m}^{\tcb_2}
    ~+~ {\rm {h.\, c.}}\,\Big]  ~~~.
\end{split}
\label{equ:onshell-nTS-B}
\end{equation}

The first term in the action comes from $\cl_\bX$ and is the term which we are interested in.
In general, the number of fermions in the term labelled by $(i,j)$ is $2(i+j)$ which is an even number. These are the 2$(q-1)$-point Majorana fermion interactions that are SYK-like.

The table below lists the collection of bosons and fermions in this interaction. 
\begin{table}[htp!]
    \centering
    \begin{tabular}{|c|c|c|c|}
        \hline
        Bosons & total \# & Fermions & total \# \\\hline
        $\bA^{\ca}$, $\bB^{\ca}$ & $2\times N_{CS}$ & $\bm{\psi}^{\ca}_a$ & $2\times N_{CS}$ \\\hline
        $\bm{\varphi}^{\tcb}$, $\bm{H}_{\m}^{\tcb}$ & $4\times N_{TS}$ & $\bm{\chi}_a^{\tcb}$ & $4\times N_{TS}$ \\
        \hline
    \end{tabular}
    \label{tab:bnf_dof}
\end{table}
\newline \noindent
In this model we have $N_{CS}$ copies of chiral multiplets and $N_{TS}$ copies of tensor multiplets. Also since we have imposed on-shell condition, $\bm{\psi}^{\ca}$ satisfies the Dirac equation.

\newpage
\section{1D, $\mathcal{N}=4$ SYK Models}

In this chapter, we will present one dimensional $\mathcal{N} = 4$ off-shell and on-shell Lagrangians that include SYK-type terms, obtained by dimensional reduction. This dimensional reduction procedure is a simple one and has been applied as a foundation of building adinkras for 4D multiplets \cite{GRana2,ENUF,adnk1,adnk2,adnk3,adnk4,adnk5,adnk6,adnk7,adnk8,adnk9,AdnkKoR,adnk10,adnk11}. 

We start from the 4D, $\mathcal{N} = 1$ Lagrangians and set all spatial coordinates as zero. 
All component fields will only depend on the time coordinate, while their name and appearances will not change. Since only temporal derivative is non-vanished, some components of field strengths $\bF^{\m\n}$ and $\bH^\m$ also vanish. 

The explicit 1D projection relations of partial derivatives and field strengths are listed below. 
\begin{equation}
    \begin{split}
   & \partial_{\mu} ~\to~
        \begin{cases}
        \partial_0 = \partial_{\tau}  \\
        \partial_i = 0
        \end{cases}~~,~~
        \partial^{\mu} ~\to~
        \begin{cases}
        \partial^0 = \partial^{\tau} = -\pa_{\tau}  \\
        \partial^i = 0
        \end{cases}\\
       & \bF^{\m\n} = \pa^{\mu}\bA^{\nu} - \pa^{\nu}\bA^{\mu} ~\to~
        \begin{cases}
        \bF^{0i} = -\bF^{i0} = \pa^{0}\bA^{i} \\
        \bF^{ij} = 0\\
        \end{cases}\\
        &\bm{H}^{\mu} = \e^{\mu\rho\a\b}\pa_{\rho}\bB_{\a\b} ~\to~ 
        \begin{cases}
        \bm{H}^{i} =-\e^{0ijk}\pa_{0}\bB_{jk} \\
         \bm{H}^{0} = 0\\
         \end{cases}\\
    \end{split}
\end{equation}
Consequently, all components of $\bT_{\m} = \pa_{\m} \Bvp + i \bH_{\m}$ are non-vanishing.

Moreover, we want to mention our conventions for gamma matrices. We follow the same conventions as in \cite{adnk1}. For example,
\begin{equation}
    \begin{split}
        &(\g^0)^{ab} ~=~ (\mathbb{I}_4)^{ab}~~,~~  (\g^1)^{ab} ~=~ -( \s^3\otimes \s^3)^{ab}~~,~~  (\g^2)^{ab} ~=~ -( \s^1\otimes \mathbb{I}_2)^{ab}~~,\\
        &(\g^3)^{ab} ~=~ ( \s^3\otimes \s^1)^{ab}~~,~~ (\g^5)^{ab} ~=~ -(\s^2\otimes \mathbb{I}_2)^{ab}~~,~~  C^{ab} ~=~ -i( \s^3\otimes \s^2)^{ab}~~. 
    \end{split}
    \label{equ:gamma-Mat}
\end{equation}
More numerical contents of matrices see Appendix \ref{appen:convention}. 

In the following sections, we present the 1D projections of all off-shell and on-shell 4D, $\mathcal{N} = 1$ Lagrangians that we constructed in the previous chapters. The number of supercharges in 1D is 4, and we still use the same indices with the same ranges in this chapter: $\mu,\nu=0,1,2,3$, $a,b=1,2,3,4$, and $i=1,2,3$, although they have different meanings from 4D. 
Namely, (1.) $\mu$ in 4D is a vector index and its range $\{0,\dots,3\}$ has the meaning of the dimension of the defining representation of the Lorentz group. In 1D, $\mu$ is just a bosonic label. Also, $i$ is just a bosonic label as well in 1D. 
(2.) $a$ in 4D is a spinor index and its range $\{1,\dots,4\}$ has the meaning of the dimension of the Majorana spinors. In 1D, $a$ is a fermionic label, and its range $\{1,\dots,4\}$ means the number of supercharges is four.

\subsection{CS + 3PT}

From Equation (\ref{equ:offshell-3PT-A}) as well as the chiral supermultiplet Lagrangian (\ref{equ:chiral-L}), we follow the above dimension reduction technique and obtain the 1D, $\mathcal{N}=4$ off-shell Lagrangian as below. 
\begin{equation}
\begin{split}
    \cl^{\rm (off-shell)}_{\rm CS+3PT} ~=&~
    ~ \fracm{1}{2} \pa_{\t} \BF^{\ca} \pa_{\t} \bBF_{\ca} 
    ~+~ i \, \fracm{1}{2}\, \d^{ab} \Bj^{\ca}_a \pa_{\t} \Bj_{b\ca}
    ~+~ \fracm{1}{2} \bX^{\ca} \bbX_{\ca} \\
    &~+~\Big\{\,\fracm12 \,\k{}_{{\ca}{\cb}{\cc}} \Big[\,
    -\,( \pa_\t \Bar{\bm\Phi}^{\ca} ) ( \pa_\t {\bm\Phi}^{\cb} ) {\bm\Phi}^{\cc}  
    ~+~ i \, 2 \, (\rP^{(-)}\g^{0})^{ab} \, (\pa_{\t}{\bm {\psi}}{}^{\ca}_a) {\bm {\psi}}{}^{\cb}_b {\bm\Phi}^{\cc}  \\
    &~~~~~~~~~~~~~~~~~~~-~ \Bar{\bX}^{\ca} \bX^{\cb} {\bm\Phi}^{\cc} 
    ~-~ i \, (\rP^{(+)})^{ab} \, \Bar{\bX}^{\ca} {\bm {\psi}}{}^{\cb}_a {\bm {\psi}}{}^{\cc}_b  \,\Big] 
    ~+~ {\rm {h.\, c.}}  \,\Big\} ~~~.
\end{split}
\label{eqn:1Doff3PTA}
\end{equation}
where we have used the convention that $(\g^{0})^{ab} $ = $\d{}^{a b}$. Note that $\d^{ab}$ is different from $C^{ab}$. Numerically $\d^{ab}$ has the same values as the identity matrix, while the numerical content of $C^{ab}$ is shown in Eq. (\ref{equ:gamma-Mat}). 

From Equation (\ref{equ:onshell-3PT-A}), the projected 1D, $\mathcal{N}=4$ on-shell Lagrangian is 
\begin{equation}
\begin{split}
    \cl^{\rm (on-shell)}_{\rm CS+3PT} ~=&~ 
    ~ \fracm{1}{2} \pa_{\tau} \BF^{\ca} \pa_{\tau} \bBF_{\ca} 
    ~+~ i \, \fracm{1}{2}\, \d^{ab} \Bj^{\ca}_a \pa_{\tau} \Bj_{b\ca} \\
    &~+~\fracm12 \,\k{}_{{\ca}{\cb}{\cc}} \Big[\,
    -\,( \pa_\tau \Bar{\bm\Phi}^{\ca} ) ( \pa_\tau {\bm\Phi}^{\cb} ) {\bm\Phi}^{\cc}  
    ~+~ i \, 2 \, (\rP^{(-)}\g^{0})^{ab} \, (\pa_{\tau}{\bm {\psi}}{}^{\ca}_a) {\bm {\psi}}{}^{\cb}_b {\bm\Phi}^{\cc}   \,\Big] \\
    &~+~\fracm12 \,\k^{*}_{{\ca}{\cb}{\cc}} \Big[\,
    -\,( \pa_\tau {\bm\Phi}^{\ca} ) ( \pa_\tau \Bar{\bm\Phi}^{\cb} ) \Bar{\bm\Phi}^{\cc}  
    ~+~ i \, 2 \, (\rP^{(+)}\g^{0})^{ab} \, (\pa_{\tau}{\bm {\psi}}{}^{\ca}_a) {\bm {\psi}}{}^{\cb}_b \Bar{\bm\Phi}^{\cc}   \,\Big] \\
    &~+~\fracm12 \, \Big[ \frac{\k_{\ca\cb\cc} \, \k^{*}_{\cd\ce\cf} \, (\rP^{(+)})^{ab} \, (\rP^{(-)})^{cd} }{\d_{\cd\ca} ~-~ \cy_{\cd\ca} }\, \Big]  ~ \Bj_{a}^{\cb} \Bj_{b}^{\cc} \Bj_{c}^{\ce} \Bj_{d}^{\cf}
    ~~~,
\end{split}
\end{equation}
where 
\begin{equation}
    \cy_{\ca\cb} ~=~ \k^{*}_{\ca\cb\cc} {\Bar{\bm\Phi}}^{\cc} ~+~ \k_{\cb\ca\cc} {\bm \Phi}^{\cc} ~~~.
\end{equation}
Thus, at the component level, a non-linear $\s$-model emerges.  we will return to this point in a later chapter.

\subsection{CS + nCS-A}

From Equation (\ref{equ:offshell-nCS-A}) as well as the chiral supermultiplet Lagrangian (\ref{equ:chiral-L}), dimension reduction gives the 1D, $\mathcal{N}=4$ off-shell Lagrangian,
\begin{equation}
\begin{split}
    \cl^{\rm (off-shell)}_{{\rm CS} + n{\rm {CS-A}}} 
    ~=&~
    ~ \fracm{1}{2} \pa_{\t} \BF^{\ca} \pa_{\t} \bBF_{\ca} 
    ~+~ \fracm{1}{2} \bX^{\ca} \bbX_{\ca}
    ~+~ i \, \fracm{1}{2} \, \d^{ab} \Bj^{\ca}_{a} \pa_{\t} \Bj_{b\ca}\\
    &~+~ \Big[  ~ \fracm12 \cp'_{\ca\cb}({\Bar{\bm \Phi}})\,\BF^{\ca}\,(\pa_\t^2\,\Bar{\BF}^{\cb})
    ~+~ \fracm12 \cp''_{\ca\cb_1\cb_2}({\Bar{\bm \Phi}})\,\BF^{\ca}\,(\pa_{\t}\Bar{\BF}^{\cb_1})\,(\pa_{\t}\Bar{\BF}^{\cb_2})\\
    &~~~~~~~ ~-~ i\,\cp'_{\ca\cb}({\Bar{\bm \Phi}})\,(\rP^{(+)}\g^{0})^{ab}\,{\bm \psi}_a^{\ca}\,\pa_{\t}{\bm \psi}_b^{\cb}\\
    &~~~~~~~ ~-~ i\,\cp''_{\ca\cb_1\cb_2}({\Bar{\bm \Phi}})\,(\rP^{(+)}\g^{0})^{ab}\,{\bm \psi}_a^{\ca}\,(\pa_{\t}{\bm \bBF}^{\cb_1})\,{\bm \psi}_b^{\cb_2}\\
    &~~~~~~~ ~-~ i\,\fracm12\,\cp''_{\ca\cb_1\cb_2}({\Bar{\bm \Phi}})\,(\rP^{(-)})^{ab}\,\bX^{\ca}\,{\bm \psi}_a^{\cb_1}\,{\bm \psi}_b^{\cb_2} \\
    &~~~~~~~ ~-~ \fracm12\,\cp'_{\ca\cb}({\Bar{\bm \Phi}})\,\bX^{\ca}\,\Bar{\bX}^{\cb}
    ~+~ {\rm {h.\, c.}}  \,\Big]  ~~~,
\end{split}
\label{eq:nCSA}
\end{equation}
where $(\pa_\t)^2 = \pa_\t\,\pa_\t$. 

From Equation (\ref{equ:onshell-nCS-A}), the projected 1D, $\mathcal{N}=4$ on-shell Lagrangian is 
\begin{equation}
\begin{split}
    \cl^{\rm (on-shell)}_{{\rm CS} + n{\rm {CS-A}}} 
    ~=&~  
    ~  \fracm{1}{2} \, \pa_{\t} \BF^{\ca} \pa_{\t} \bBF_{\ca}
    ~+~ i  \, \fracm{1}{2} \,\, \d^{ab} \Bj^{\ca}_a \pa_\t  \Bj_{b \, \ca} \\
    &~+~\Big[ ~ \fracm12 \cp'_{\ca\cb}({\Bar{\bm \Phi}})\,\BF^{\ca}\,(\pa_\t)^2\,\Bar{\BF}^{\cb}
    ~+~ \fracm12 \cp''_{\ca\cb_1\cb_2}({\Bar{\bm \Phi}})\,\BF^{\ca}\,(\pa_{\t}\Bar{\BF}^{\cb_1})\,(\pa_{\t}\Bar{\BF}^{\cb_2}) \\
    &~~~~~~~ ~-~ i\,\cp'_{\ca\cb}({\Bar{\bm \Phi}})\,(\rP^{(+)}\g^{0})^{ab}\,{\bm \psi}_a^{\ca}\,\pa_{\t}{\bm \psi}_b^{\cb} \\
    &~~~~~~~ ~-~ i\,\cp''_{\ca\cb_1\cb_2}({\Bar{\bm \Phi}})\,(\rP^{(+)}\g^{0})^{ab}\,{\bm \psi}_a^{\ca}\,(\pa_{\t}{\bm \bBF}^{\cb_1})\,{\bm \psi}_b^{\cb_2} 
    ~+~ {\rm h.\, c.} \, \Big] \\
    &~+~ \frac12 ~ \frac{ \Bar{\cp}''_{\ca\cb_{1}\cb_{2}} (\BF) \, \cp''_{\cd\cc_{1}\cc_{2}} (\bBF) }{ \d_{\ca\cd} ~-~ \Bar{\cp}'_{\ca\cd} (\BF) ~-~ \cp'_{\cd\ca} (\bBF) } ~ (\rP^{(+)})^{ab} \, (\rP^{(-)})^{cd} \, \Bj_{a}^{\cb_{1}} \Bj_{b}^{\cb_{2}} \Bj_{c}^{\cc_{1}} \Bj_{d}^{\cc_{2}}
    ~~~.
\end{split}
\end{equation}

\subsection{CS + TS + nTS-A}

From Equation (\ref{equ:offshell-nTS-A}) as well as the chiral supermultiplet Lagrangian (\ref{equ:chiral-L}) and the tensor supermultiplet Lagrangian (\ref{equ:TS-L}), simple compactification leads to the 1D, $\mathcal{N}=4$ off-shell Lagrangian below.
\begin{equation}
\begin{split}
    \cl^{\rm (off-shell)}_{{\rm {CS+ TS + nTS-A}}} 
    ~=&~
    ~ \fracm{1}{2} \pa_{\t} \BF^{\ca} \pa_{\t} \bBF_{\ca} 
    ~+~ \fracm{1}{2} \bX^{\ca} \bbX_{\ca}
    ~+~ i \, \fracm{1}{2} \, \d^{ab}\Bj^{\ca}_{a} \pa_{\t} \Bj_{b\ca}\\
    &~+~ \fracm12 \pa_\t \bm{\varphi}^{\tca} \pa_\t \bm{\varphi}_{\tca} ~+~ \fracm12 \bm{H}_{i}^{\tca} \bm{H}^{i}_{\tca}
    ~+~ i \fracm12 \, \d^{ab} \bm{\chi}^{\tca}_a \pa_\t \bm{\chi}_{b\tca} \\
    &~+~\Big[ ~-~ i \fracm12 \, ({\rP}^{(-)})^{ab}  \bX^{\ca} \, \cp''_{\ca\tcb_1\tcb_2}(\Bvp) \, \Bc^{\tcb_1}_{a} \Bc^{\tcb_2}_{b}  \\
    &~+~ \fracm12 \, ({\rP}^{(+)})^{ab}  \,  ({\rP}^{(-)})^{cd} \, \cp'''_{\ca\tcb_1\tcb_2\tcb_3}(\Bvp) \, \Bj^{\ca}_{a} \Bc^{\tcb_1}_{b} \Bc^{\tcb_2}_{c} \Bc^{\tcb_3}_{d}  \\
    &~-~ i \,  ({\rP}^{(+)}\g^{\m})^{ab} \, \cp''_{\ca\tcb_1\tcb_2}(\Bvp) \, \bT_{\m}^{\tcb_1} \Bj^{\ca}_{a} \Bc^{\tcb_2}_{b}  \\
    &~+~ \fracm18 ({\rP}^{(+)})^{ab}  \,  ({\rP}^{(-)})^{cd}  \, \BF^{\ca} \, \cp''''_{\ca\tcb_1\tcb_2\tcb_3\tcb_4}(\Bvp) \, \Bc^{\tcb_1}_{a} \Bc^{\tcb_2}_{b} \Bc^{\tcb_3}_{c} \Bc^{\tcb_4}_{d}  \\
    &~-~ i \fracm14 \, (\g^{5}\g^{\m})^{ab} \, \BF^{\ca} \, \cp'''_{\ca\tcb_1\tcb_2\tcb_3}(\Bvp) \, \bT_{\m}^{\tcb_1} \, \Bc^{\tcb_2}_{a} \Bc^{\tcb_3}_{b}  \\
    &~+~ i \, ({\rP}^{(-)}\g^{\m})^{ab} \, \BF^{\ca} \, \cp''_{\ca\tcb_1\tcb_2}(\Bvp) \, \Bc^{\tcb_1}_{a} \pa_{\m} \Bc^{\tcb_2}_{b}  \\
    &~-~ \fracm12 \, \BF^{\ca} \, \cp''_{\ca\tcb_1\tcb_2}(\Bvp) \, \bT_{\m}^{\tcb_1} \bT^{\m \, \tcb_2} 
    ~+~ {\rm {h.\, c.}}\,\Big]  ~~~.
\end{split}
\label{eq:nTSA}
\end{equation}

From Equation (\ref{equ:onshell-nTS-A}), the projected 1D, $\mathcal{N}=4$ on-shell Lagrangian is 
\begin{equation}
\begin{split}
    \cl^{(\rm on-shell)}_{{\rm {CS+ TS + nTS-A}}}  ~=&~ 
    ~  \fracm{1}{2} \, \pa_{\t} \BF^{\ca} \pa_{\t} \bBF_{\ca}
    ~+~ i  \, \fracm{1}{2} \, \d^{ab} \Bj^{\ca}_a \pa_\t  \Bj_{b \, \ca} \\
    &~+~ \fracm12 \pa_\t \bm{\varphi}^{\tca} \pa_\t \bm{\varphi}_{\tca} ~+~ \fracm12 \bm{H}_{i}^{\tca} \bm{H}^{i}_{\tca}
    ~+~ i \fracm12\, \d^{ab} \Bc_a^{\tca} \pa_\t \Bc_{b \tca} \\
    &~+~\Big[\, \fracm12 \, ({\rP}^{(+)})^{ab}  \,  ({\rP}^{(-)})^{cd} \, \cp'''_{\ca\tcb_1\tcb_2\tcb_3}(\Bvp) \, \Bj^{\ca}_{a} \Bc^{\tcb_1}_{b} \Bc^{\tcb_2}_{c} \Bc^{\tcb_3}_{d}  \\
    &~~~~~~ ~-~ i \,  ({\rP}^{(+)}\g^{\m})^{ab} \, \cp''_{\ca\tcb_1\tcb_2}(\Bvp) \, \bT_{\m}^{\tcb_1} \Bj^{\ca}_{a} \Bc^{\tcb_2}_{b}  \\
    &~~~~~~ ~+~ \fracm18 \, ({\rP}^{(+)})^{ab}  \,  ({\rP}^{(-)})^{cd}  \, \BF^{\ca} \, \cp''''_{\ca\tcb_1\tcb_2\tcb_3\tcb_4}(\Bvp) \, \Bc^{\tcb_1}_{a} \Bc^{\tcb_2}_{b} \Bc^{\tcb_3}_{c} \Bc^{\tcb_4}_{d}  \\
    &~~~~~~ ~-~ i \fracm14 \, (\g^{5}\g^{\m})^{ab} \, \BF^{\ca} \, \cp'''_{\ca\tcb_1\tcb_2\tcb_3}(\Bvp) \, \bT_{\m}^{\tcb_1} \, \Bc^{\tcb_2}_{a} \Bc^{\tcb_3}_{b}  \\
    &~~~~~~ ~+~ i \, ({\rP}^{(-)}\g^{0})^{ab} \, \BF^{\ca} \, \cp''_{\ca\tcb_1\tcb_2}(\Bvp) \, \Bc^{\tcb_1}_{a} \pa_{\t} \Bc^{\tcb_2}_{b}  \\
    &~~~~~~ ~-~ \fracm12 \, \BF^{\ca} \, \cp''_{\ca\tcb_1\tcb_2}(\Bvp) \, \bT_{\m}^{\tcb_1} \bT^{\m \, \tcb_2} ~+~ {\rm h.\, c.}\, \Big] \\
    &~+~ \fracm12 \, (\rP^{(+)}){}^{ab} \,  (\rP^{(-)}){}^{cd} ~ {\Bar\cp}''^{\ca}{}_{\tcb_1\tcb_2}({\bm \varphi}) \, \cp''_{\ca\tcb_3\tcb_4}({\bm \varphi}) ~ {\bm\chi}_{a}^{\tcb_1} {\bm\chi}_{b}^{\tcb_2} {\bm\chi}_{c}^{\tcb_3} {\bm\chi}_{d}^{\tcb_4}  ~~~.
\end{split}
\end{equation}

\subsection{CS + VS + nVS-B}

From Equation (\ref{equ:offshell-nVS-B}) as well as the chiral supermultiplet Lagrangian (\ref{equ:chiral-L}) and vector supermultiplet Lagrangian (\ref{equ:VS-L}), we follow the above specified dimension reduction technique and obtain the 1D, $\mathcal{N}=4$ off-shell Lagrangian as below.
\begin{equation}
\begin{split}
    \cl^{\rm (off-shell)}_{{\rm {CS+ VS + nVS-B}}} ~=&~ 
    ~ \fracm{1}{2} \pa_{\t} \BF^{\ca} \pa_{\t} \bBF_{\ca} 
    ~+~ \fracm{1}{2} \bX^{\ca} \bbX_{\ca}
    ~+~ i \, \fracm{1}{2} \, \d^{ab} \Bj^{\ca}_{a} \pa_{\t} \Bj_{b\ca}\\
    &~-~ \fracm{1}{2} \bF^{\vcb}_{0i} \bF_{\vcb}^{0i}  ~+~ i\,  \fracm{1}{2} \, \d^{ab} {\bm \l}^{\vcb}_a \partial_\t {\bm \l}_b{}_{\vcb}~+~ \fracm{1}{2} {\bm {\rm d}}^{\vcb}{\bm {\rm d}}_{\vcb}   \\
    &~+~ \Big\{\,i \fracm12 \, \bX^{\ca} \, \cf_{\ca} (\cj_{11}) 
    ~+~ i \cf'_{\ca\vcb_1\vcb_2}(\cj_{11})\,(\rP^{(+)}\g^{0i})^{ab}\,\bm{\psi}_a^{\ca}\, \bF_{0i}^{\vcb_1}\,\Bl_b^{\vcb_2} \\
    &~-~ \cf'_{\ca\vcb_1\vcb_2}(\cj_{11})\,(\rP^{(+)})^{ab}\,\bm{\psi}_a^{\ca}\, {\bm {\rm d}}^{\vcb_1}\,\Bl_b^{\vcb_2} \\
    &~+~ \bm{\Phi}^{\ca}\,\cf''_{\ca\vcb_1\vcb_2\vcb_3\vcb_4}(\cj_{11})\,\Big[\, -~ \fracm12 (\rP^{(+)}\g^{0i}\g^{0j})^{ab}\,\bF_{0i}^{\vcb_1}\,\bF_{0j}^{\vcb_3} \\
    &~+~ i \,(\rP^{(+)}\g^{0i})^{ab}\,{\bm {\rm d}}^{\vcb_1}\,\bF_{0i}^{\vcb_3}
    ~-~ \fracm12 \, (\rP^{(+)})^{ab}\,{\bm {\rm d}}^{\vcb_1}\,{\bm {\rm d}}^{\vcb_3}\,
    \Big]\,\Bl_a^{\vcb_2}\,\Bl_b^{\vcb_4} \\
    &~+~ \bm{\Phi}^{\ca}\,\cf'_{\ca\vcb_1\vcb_2}(\cj_{11})\,\Big[\, i \,(\rP^{(-)}\g^{0})^{ab}\,(\pa_{\t}\Bl_a^{\vcb_1})\,\Bl_b^{\vcb_2} ~-~ \fracm12 \,{\bm {\rm d}}^{\vcb_1}\,{\bm {\rm d}}^{\vcb_2}\\
    &~+~ \fracm12 \,\bF_{0i}^{\vcb_1}\,\bF^{0i}{}^{\vcb_2}\,\Big]
  ~+~ {\rm {h.\, c.}} ~ \Big\} ~~~.
\end{split}
\end{equation}

From Equation (\ref{equ:onshell-nVS-B}), the projected 1D, $\mathcal{N}=4$ on-shell Lagrangian is 
\begin{equation}
\begin{split}
    \cl^{(\rm on-shell)}_{{\rm {CS+ VS + nVS-B}}}  ~=&~ 
    ~  \fracm{1}{2} \, \pa_{\t} \BF^{\ca} \pa_{\t} \bBF_{\ca}
    ~+~ i  \, \fracm{1}{2} \, \d^{ab} \Bj^{\ca}_a \pa_\t  \Bj_{b \, \ca} \\
    &~-~ \fracm{1}{2} \bF^{\vcb}_{0i} \bF_{\vcb}^{0i}  ~+~ i\,  \fracm{1}{2} \, \d^{ab} {\bm \l}^{\vcb}_a \partial_\t {\bm \l}_b{}_{\vcb}  \\
    &~+~ \Big[\,  i  \,\cf'_{\ca\vcb_1\vcb_2}(\cj_{11})\,(\rP^{(+)}\g^{0i})^{ab}\,\bm{\psi}_a^{\ca}\, \bF_{0i}^{\vcb_1}\,\Bl_b^{\vcb_2} \\
    &~~~~~~ ~-~ \fracm12 \bm{\Phi}^{\ca}\,\cf''_{\ca\vcb_1\vcb_2\vcb_3\vcb_4}(\cj_{11})\, (\rP^{(+)}\g^{0i}\g^{0j})^{ab}\,\bF_{0i}^{\vcb_1}\,\bF_{0j}^{\vcb_3} \Bl_a^{\vcb_2}\,\Bl_b^{\vcb_4}\\
    &~~~~~~ ~+~ i \,\bm{\Phi}^{\ca}\,\cf'_{\ca\vcb_1\vcb_2}(\cj_{11})\,(\rP^{(-)}\g^{0})^{ab}\,(\pa_{\t}\Bl_a^{\vcb_1})\,\Bl_b^{\vcb_2} \\
    &~~~~~~ ~+~ \fracm12 \,\bm{\Phi}^{\ca}\,\cf'_{\ca\vcb_1\vcb_2}(\cj_{11})\,  \bF_{0i}^{\vcb_1}\,\bF^{0i}{}^{\vcb_2}
    ~+~ {\rm h.\,c.}     \,\Big] \\
    &~-~  \fracm12 \, \sum_{i,j=1}^{P}\,  (-1)^{i} \,  \k^{(i)\ca}{}_{\vcb_1\cdots\vcb_{2i}}\, \k^{(j)*}_{\ca\vcc_1\cdots\vcc_{2j}}\, \\
    &~~~~~~ ~\times~\prod_{k=1}^{i} \prod_{l=1}^{j} (\rP^{(+)})^{ a_{k}b_{k}} (\rP^{(-)})^{c_{l}d_{l}} \Bl_{a_{k}}^{\vcb_{2k-1}} \Bl_{b_{k}}^{\vcb_{2k}} \Bl_{c_{l}}^{\vcc_{2l-1}}\Bl_{d_{l}}^{\vcc_{2l}}\\
    &~-~ \fracm12\,\frac{\Omega_{\vca}\Omega_{\vcb}}{\d_{\vca\vcb}-\cy_{\vca\vcb}}  ~~~,
\end{split}
\end{equation}
where 
\begin{equation}
    \begin{split}
        \cy_{\vcb\vcd} ~=&~ 
        -~ \bm{\Phi}^{\ca}\,\cf''_{\ca\vcb\vcb_2\vcd\vcb_4}(\cj_{11})\,\cj_{11}^{\vcb_2\vcb_4} 
        ~+~ \bm{\Phi}^{\ca}\,\cf'_{\ca\vcb\vcd}(\cj_{11}) \\
        &~+~ \Bar{\bm{\Phi}}^{\ca}\,\Bar{\cf}''_{\ca\vcb\vcb_2\vcd\vcb_4}(\Bar{\cj}_{11})\,\Bar{\cj}_{11}^{\vcb_2\vcb_4} 
        ~+~ \Bar{\bm{\Phi}}\,\Bar{\cf}'_{\ca\vcb\vcd}(\Bar{\cj}_{11}) ~~~,
    \end{split}
\end{equation}
and
\begin{equation}
    \begin{split}
        \Omega_{\vcb}~=&~ \Big[\,\cf'_{\ca\vcb\vcc}(\cj_{11})\,(\rP^{(+)})^{ab}-\Bar{\cf}'_{\ca\vcb\vcc}(\Bar{\cj}_{11})\,(\rP^{(-)})^{ab}\,\Big]\, \Bj_a^{\ca}\,\Bl_b^{\vcc} \\
        &~-~ i  \,\Big[\,\bm{\Phi}^{\ca}\,\cf''_{\ca\vcb\vcc_2\vcc_3\vcc_4}(\cj_{11})\,(\rP^{(+)}\g^{0i})^{ab}+\Bar{\bm{\Phi}}^{\ca}\,\Bar{\cf}''_{\ca\vcb\vcc_2\vcc_3\vcc_4}(\Bar{\cj}_{11})\,(\rP^{(-)}\g^{0i})^{ab}\,\Big]\,\bF_{0i}^{\vcc_3}\Bl_a^{\vcc_2}\Bl_b^{\vcc_4}
    \end{split}
\end{equation}

\subsection{CS + TS + nTS-B}

From Equation (\ref{equ:offshell-nTS-B}) as well as the chiral supermultiplet Lagrangian (\ref{equ:chiral-L}) and the tensor supermultiplet Lagrangian (\ref{equ:TS-L}), dimension reduction gives the 1D, $\mathcal{N}=4$ off-shell Lagrangian,
\begin{equation}
\begin{split}
    \cl^{\rm (off-shell)}_{{\rm {CS+ TS + nTS-B}}} ~=&~ 
    ~ \fracm{1}{2} \pa_{\t} \BF^{\ca} \pa_{\t} \bBF_{\ca} 
    ~+~ \fracm{1}{2} \bX^{\ca} \bbX_{\ca}
    ~+~ i \, \fracm{1}{2} \, \d^{ab} \Bj^{\ca}_{a} \pa_{\t} \Bj_{b\ca}\\
    &~+~ \fracm12 \pa_\t \bm{\varphi}^{\tca} \pa_\t \bm{\varphi}_{\tca} ~+~ \fracm12 \bm{H}_{i}^{\tca} \bm{H}^{i}_{\tca}
    ~+~ i \fracm12 \, \d^{ab} \bm{\chi}^{\tca}_a \pa_\t \bm{\chi}_{b\tca} \\
    &~+~\Big[\,i \fracm12 \, \bX^{\ca} \, \cf_{\ca} (\cj_{22})
    ~-~ i \,({\rP}^{(+)}\g^{\m})^{ab} \, \cf'_{\ca\tcb_1\tcb_2} (\cj_{22}) \,\bm{\psi}^{\ca}_a \, \bT_{\m}^{\tcb_1} \,  \Bc^{\tcb_2}_b \\
    &~-~ \fracm12 \, (\rP^{(-)}){}^{ab} \, \cf''_{\ca\tcb_1\tcb_2\tcb_3\tcb_4} (\cj_{22}) \, \BF^{\ca} \, \bT_{\m}^{\tcb_1} \, \bT^{\m \, \tcb_3} \, \Bc^{\tcb_2}_a\, \Bc^{\tcb_4}_b \\ 
    &~+~ i \, ({\rP}^{(-)}\g^{0})^{ab} \, \cf'_{\ca\tcb_1\tcb_2} (\cj_{22}) \, \BF^{\ca}\,  \Bc^{\tcb_1}_a\, (\pa_{\t} \Bc^{\tcb_2}_b) \\
    &~-~ \fracm12 \, \cf'_{\ca\tcb_1\tcb_2} (\cj_{22}) \, \BF^{\ca}\, \bT_{\m}^{\tcb_1} \, \bT^{\m \, \tcb_2}
     ~+~ {\rm {h.\, c.}}~ \Big] ~~~.
\end{split}
\end{equation}

From Equation (\ref{equ:onshell-nTS-B}), the projected 1D, $\mathcal{N}=4$ on-shell Lagrangian is 
\begin{equation}
\begin{split}
    \cl^{(\rm on-shell)}_{{\rm {CS+ TS + nTS-B}}}  ~=&~ 
    -~ \fracm12 \,  \sum_{i,j=1}^{P}\, (-1)^{i} \, \k^{(i)\ca}{}_{\tcb_1\cdots\tcb_{2i}}\, \k^{(j)*}_{\ca\tcc_1\cdots\tcc_{2j}}\,  \\
    &~~~ ~\times~ \prod_{k=1}^{i} \prod_{l=1}^{j} (\rP^{(-)})^{a_{k}b_{k}} (\rP^{(+)})^{ c_{l}d_{l}} \Bc_{a_{k}}^{\tcb_{2k-1}} \Bc_{b_{k}}^{\tcb_{2k}} \Bc_{c_{l}}^{\tcc_{2l-1}} \Bc_{d_{l}}^{\tcc_{2l}}    \\
    &~+~  \fracm{1}{2} \, \pa_{\t} \BF^{\ca} \pa_{\t} \bBF_{\ca}
    ~+~ i  \, \fracm{1}{2} \, \d^{ab}\Bj^{\ca}_a \pa_\t  \Bj_{b \, \ca} \\
    &~+~ \fracm12 \pa_\t \bm{\varphi}^{\tca} \pa_\t \bm{\varphi}_{\tca} ~+~ \fracm12 \bm{H}_{i}^{\tca} \bm{H}^{i}_{\tca}
    ~+~ i \fracm12 \, \d^{ab} \Bc_a^{\tca} \pa_\t \Bc_{b \tca} \\
    &~+~ \Big[ ~-~ i \, ({\rP}^{(+)}\g^{\m})^{ab}  \, \cf'_{\ca\tcb_1\tcb_2}(\cj_{22}) \,\bm{\psi}^{\ca}_a \, \bT_{\m}^{\tcb_1} \,\Bc^{\tcb_2}_b \\
    &~~~~~~~ ~-~ \fracm12 \, (\rP^{(-)}){}^{ab} \, \cf''_{\ca\tcb_1\tcb_2\tcb_3\tcb_4}(\cj_{22}) \, \BF^{\ca} \, \bT^{\m}{}^{\tcb_1} \, \bT_{\m}^{\tcb_3} \,\Bc^{\tcb_2}_a \,\Bc^{\tcb_4}_b \\ 
    &~~~~~~~ ~+~i \, ({\rP}^{(-)}\g^{0})^{ab}  \, \cf'_{\ca\tcb_1\tcb_2}(\cj_{22}) \, \BF^{\ca} \, \Bc^{\tcb_1}_a \, (\pa_{\t}\Bc^{\tcb_2}_b) \\
    &~~~~~~~ ~-~ \fracm12  \, \cf'_{\ca\tcb_1\tcb_2}(\cj_{22}) \, \BF^{\ca} \, \bT^{\m}{}^{\tcb_1} \, \bT_{\m}^{\tcb_2}
    ~+~ {\rm {h.\, c.}}\,\Big] ~~~.
\end{split}
\end{equation}

\newpage

\section{Story \& Conclusion}
\label{sec:story}

If one reviews the 1D, $\mathcal{N}=1$ and $\mathcal{N}=2$ SYK models in literature, one realizes that both of the models, in superspace, are written solely in terms of fermionic superfields.
In the construction of all $\mathcal{N}=4$ models\footnote{It should be noted that these models can also be described as possessing $\mathcal{N}=(4,4)$ supersymmetry as they also descend from heterotic-type models in 2D. }, however, none of the models utilize fermionic superfields solely. This is also true for the $\cn = 4$ models in \cite{SUSYSYK2}. In fact, when we integrate over the whole superspace (3PT and nPT-A types), the Lagrangians are solely constructed via bosonic superfields. Only when we integrate over the chiral half of superspace (nPT-B types), can we accommodate fermionic superfields.

In \cite{SUSYSYK1}, the authors mentioned that the $\mathcal{N}=4$ SYK model in \cite{SUSYSYK2} contains dynamical bosons, and it would be interesting to discover $\mathcal{N}=4$ models without them. 
Our response is as follows.
For all known superfields in linear representations, if one requires dynamical fermions and SYK interaction terms, for systems with $\mathcal{N}>2$ supercharges, the current study of this work indicates an impossibility to eliminate dynamical bosons.  However,
this question is under continuing study and we have not arrived at a no-go theorem.

Let us further elaborate on this statement. 
In each of the supermultiplets (chiral, vector, tensor) we used to construct our SYK models, dynamical bosons appear in the free theory. 
If one review the simplest model with SYK interactions, for example (\ref{eqn:1Doff3PTA}), one see interaction terms like
\begin{equation}
    \cl^{\rm (off-shell)}_{\rm 3PT} ~=~ -~ \fracm12 \,\k{}_{{\ca}{\cb}{\cc}} \, ( \pa_\t \bBF^{\ca} ) ( \pa_\t \BF^{\cb} ) \BF^{\cc}  
    ~+~ \cdots
\end{equation}
Since $\rD^{2}\Bar{\rD}^{2} \sim \pa \pa$, and the two time derivatives would have to distribute among three or more bosonic superfields for SYK-type interactions, so at least one of them would not carry derivatives. These terms prohibit the possibility of applying the usual adinkra trick \cite{adnk11} of redefining $\pa_\t \BF \rightarrow b$.
Therefore, dynamical bosons must appear in all of the models discussed.

This is consistent with the findings in both the 1D, $\mathcal{N}=4$ models in \cite{SUSYSYK2} and the 2D, $\mathcal{N}=2$ models in \cite{SUSYSYK6}. In fact, dynamical bosons are required for computing the correlation functions in the 2D, $\mathcal{N}=2$ supersymmetric SYK models in \cite{SUSYSYK6}.


We also want to comment on how to assign randomness to these models. Let us first take a look at the quartic SYK terms in our on-shell Lagrangians,
\begin{align}
    \cl^{\rm (on-shell)}_{\rm CS+3PT} ~=&~ ~ \fracm12 \, \k^{\ce}{}_{\ca\cb} \, \k^{*}_{\ce\cc\cd} \, (\rP^{(+)})^{ab} \, (\rP^{(-)})^{cd} ~ \Bj_{a}^{\ca} \Bj_{b}^{\cb} \Bj_{c}^{\cc} \Bj_{d}^{\cd} ~+~ \cdots  ~~,
\label{eq:4pt1u}
\end{align}
\begin{align}
    \cl^{\rm (on-shell)}_{\rm CS + nCS-A} ~=&~ ~ 2 \, \k^{(2) \, \ce}{}_{\ca\cb} \, \k^{(2)*}_{\ce\cc\cd} \, (\rP^{(-)})^{ab} \, (\rP^{(+)})^{cd} \, \Bj_{a}^{\ca} \Bj_{b}^{\cb} \Bj_{c}^{\cc} \Bj_{d}^{\cd} ~+~ \cdots ~~,
\label{eq:4pt2u}
\end{align}
\begin{align}
    \cl^{\rm (on-shell)}_{\rm CS+ TS + nTS-A} ~=&~ ~ 2 \, \k^{(2) \, \ce}{}_{\tca\tcb} \, \k^{(2)*}_{\ce\tcc\tcd} \, (\rP^{(-)}){}^{ab} \,  (\rP^{(+)}){}^{cd} ~ {\bm\chi}_{a}^{\tca} {\bm\chi}_{b}^{\tcb} {\bm\chi}_{c}^{\tcc} {\bm\chi}_{d}^{\tcd} ~+~ \cdots {~\,~~~} ~~.
\label{eq:4pt3u}    
\end{align}
In the above examples, we have SYK terms of the form
\begin{equation}
    J_{\ca\cb\cc\cd} \, (\rP^{(\pm)})^{ab} \, (\rP^{(\mp)})^{cd} ~ \Bj_{a}^{\ca} \Bj_{b}^{\cb} \Bj_{c}^{\cc} \Bj_{d}^{\cd}  ~~~,
\label{eq:SpnRs}    
\end{equation}
where
\begin{equation}
    J_{\ca\cb\cc\cd} ~\sim~ \k^{\ce}{}_{\ca\cb} \, \k^{*}_{\ce\cc\cd} ~~~.
\end{equation}
Note that $J_{\ca\cb\cc\cd}$ resembles the form of $J_{ijkl}$ in Equation (\ref{eqn:Jform2}), and $\k_{\ca\cb\cc}$ is the analogue of $C_{ijk}$, which can be taken as independent Gaussian random complex numbers.
A Wishart matrix is constructed as $W = H H^{\dagger}$, where $H$ is a matrix with Gaussian random entries, and $W$ is Hermitian and positive semi-definite \cite{rmt2}. Thus $\k_{\ca\cb\cc}$ being Gaussian random implies $J_{\ca\cb\cc\cd}$ being Wishart-Laguerre random \cite{rmt1,rmt2,SUSYSYK9,SUSYSYK10}. This implies that our 1D, $\mathcal{N}=4$ models have the same distinctive feature as the 1D, $\mathcal{N}=1$ and $\mathcal{N}=2$ models constructed in \cite{SUSYSYK1}.
One subtle difference though, is that when we start constructing our models, we utilize bosonic superfields instead of fermionic superfields. Therefore, $\k_{\ca\cb\cc}$ is symmetric in the last two indices, i.e. $\k_{\ca\cb\cc} = \k_{\ca\cc\cb}$, unlike $C_{ijk}$ which is totally antisymmetric.
However, let us also point out that the 1D spinors in Equation (\ref{eq:SpnRs}) carry pairs of ``isospin'' indices
$\ca \, a $ since the spinor-type indices $a,\,b,\, \dots $ become isospin indices upon reduction to a one dimensional 
model. Thus, $\k_{\ca\cb\cc}\, (\rP^{(\pm)})^{b c} $ = $-\,(\rP^{(\pm)})^{c b} \,  \k_{\ca\cc\cb}$ which is appropriate
for Wishart-Laguerre random matrices.

Another point to note is the $(\rP^{(\pm)})^{ab} \, (\rP^{(\mp)})^{cd}$ factors.
In our work, we use four component notation. Two component notation translates to four component notation via $\rD^2 \Bar{\rD}^2 \sim \rD^{a} \rD^{(+)}_{a} \rD^{b} \rD^{(-)}_{a}$, so we have $\Bj\Bj\Bar{\Bj}\Bar{\Bj} \sim (\rP^{(+)})^{ab} \, (\rP^{(-)})^{cd} ~ \Bj_{a} \Bj_{b} \Bj_{c} \Bj_{d}$. Therefore we see the analogy between the $\mathcal{N}=2$ and the $\mathcal{N}=4$ cases.

Finally, it is obvious the terms describing interactions of the fermions in Equation (\ref{eq:4pt2u}) 
and 
Equation (\ref{eq:4pt3u})
are the same. This also the true for the fermionic interactions in
Equation (\ref{eq:4pt1u}). To see this one simply needs to make the redefinition of the 
coupling constant Equation (\ref{eq:4pt1u}) according to: $   \k^{\ce}{}_{\ca\cb}  ~\to ~ 2 \, \k^{(2)* \, \ce}{}_{\ca\cb} $
along with reordering of quadratic pairs of the fermions.  So all three models describe the same
pure four point fermion interaction.

Now let us turn to the 2$(q-1)$-pt SYK interactions. Similar to the quartic interactions, our $\k^{(i)}_{\ca\hcb_{1}\cdots\hcb_{2i}}$ are analogues of $C_{j k_{1} \cdots k_{2i}}$ in $\mathcal{N}=1$ and $\mathcal{N}=2$ SYK models in literature. We constructed these models from a polynomial of chiral currents, so in the on-shell Lagrangians there are cross terms from multiplications of two polynomials. To restrict them to SYK models, let us set all except one $\k$'s to zero, so we only get one diagonal term. 
\begin{align}
\begin{split}
    \cl^{(\rm on-shell)}_{\rm CS+ VS + nVS-B} ~=&~ -~ \fracm12 \, (-1)^{i} \, \k^{(i)\ca}{}_{\vcb_1\cdots\vcb_{2i}} \,  \k^{(i)*}_{\ca\vcc_1\cdots\vcc_{2i}} \, \prod_{k=1}^{i} \, (\rP^{(+)})^{a_{k}b_{k}} (\rP^{(-)})^{c_{k}d_{k}} \Bl_{a_{k}}^{\vcb_{2k-1}} \Bl_{b_{k}}^{\vcb_{2k}} \Bl_{c_{k}}^{\vcc_{2k-1}} \Bl_{d_{k}}^{\vcc_{2k}} \\
    &~+~ \cdots
\end{split} \\
\begin{split}
    \cl^{(\rm on-shell)}_{{\rm {CS+ TS + nTS-B}}}  ~=&~ -~ \fracm12 \,  (-1)^{i} \, \k^{(i)\ca}{}_{\tcb_1\cdots\tcb_{2i}}\, \k^{(i)*}_{\ca\tcc_1\cdots\tcc_{2i}}\, \prod_{k=1}^{i} \, (\rP^{(-)})^{a_{k}b_{k}} (\rP^{(+)})^{c_{k}d_{k}} \Bc_{a_{k}}^{\tcb_{2k-1}} \Bc_{b_{k}}^{\tcb_{2k}} \Bc_{c_{k}}^{\tcc_{2k-1}} \Bc_{d_{k}}^{\tcc_{2k}} \\
    &~+~ \cdots 
\end{split} 
\end{align}
Again, we can assign Guassian random distribution to $\k^{(i)}_{\ca\hcb_{1}\cdots\hcb_{2i}}$ variables, as what we do for $C_{j k_{1} \cdots k_{2i}}$. The difference is $\k$'s are totally symmetric in all the indices except the first one, while $C$'s are totally antisymmetric. But again, with the spinor indices on the projection matrices, we'll get back antisymmetry as what we dicussed for the quartic interactions.
The SYK coupling
\begin{equation}
    J_{\cb_{1}\cdots\cb_{2i}\cc_{1}\cdots\cc_{2i}} ~\sim~ \k^{(i) \, \ca}{}_{\cb_{1}\cdots\cb_{2i}} \, \k^{(i) *}_{\ca\cc_{1}\cdots\cc_{2i}} ~~~,
\end{equation}
thus exhibit Wishart-Laguerre randomness.



Apart from the fermionic interaction terms that involve the same type of fermions, there are also mixings of fermions from different supermultiplets in some models. Below we show these interactions from a pair of models where we exchange the vector supermultiplet with the tensor supermultiplet

\begin{equation}
    \cl^{\rm (on-shell)}_{\rm CS+TS+nTS-A}
    ~=~ 3 \, \k^{(3)}_{\ca\tcb_1\tcb_2\tcb_3} \,  ({\rP}^{(+)})^{ab}  \,  ({\rP}^{(-)})^{cd} \, \Bj^{\ca}_{a} \, \Bc^{\tcb_1}_{b} \, \Bc^{\tcb_2}_{c} \, \Bc^{\tcb_3}_{d} ~+~ {\rm h.\,c.} ~+~ \cdots
\label{eq:4ptCT}    
\end{equation}

\begin{equation}
\begin{split}
    \cl^{\rm (on-shell)}_{\rm CS+VS+nVS-B}
    ~=~ -~ \fracm12\,\Big[\, & \k^{(1)}_{\ca\vcb\vcc}(\rP^{(+)})^{ab} ~-~ \k^{(1)*}_{\ca\vcb\vcc}(\rP^{(-)})^{ab}\,\Big] \\
    & ~\times~ \Big[\,\k^{(1)}{}_{\cd}{}^{\vcb}{}_{\vce}(\rP^{(+)})^{cd} ~-~ \k^{(1)*}{}_{\cd}{}^{\vcb}{}_{\vce}(\rP^{(-)})^{cd}\,\Big]\, \Bj_a^{\ca}\,\Bl_b^{\vcc}\, \Bj_c^{\cd}\,\Bl_d^{\vce} ~+~ \cdots 
\label{eq:4ptCV}    
\end{split}
\end{equation}

\noindent
and here it is clear that the pure four point fermionic interactions in Equation (\ref{eq:4ptCT}) and
Equation (\ref{eq:4ptCV}) are very different.  One other pair of systems of equations that are candidates for such a possibility is provided by the results in Equations (\ref{eq:nCSA})
and (\ref{eq:nTSA})
which suggests a duality between a system of $m$ + 1 chiral supermultiplets and
$m$ tensor supermultiplets.  Construction of a ``master action" where the elimination of some auxiliary extra fields leading to
results in Equation (\ref{eq:nCSA}) in
one order while the
elimination of some auxiliary extra fields leading to
results in Equation
(\ref{eq:nTSA})
in a different order
needs to be explored.
This opens up a pathway to an intriguing question. 

Many years ago \cite{TwsTd1,TwsTd2}, it was pointed out that by
reducing the chiral supermultiplet and the vector supermultiplet from consideration in the 
4D, $\cal N$ = 1 domain to the 2D, $\cal N$ = 2 domain led to the discovery of the pair of distinct 
2D, $\cal N$ = 2 representations.  In other words, both ``chiral supermultiplets" and ``twisted chiral supermultiplets"
emerged as co-equal representations that when combined with the study of non-linear $\s$-models implied
a pairing of K\"ahler manifolds, one with coordinates described by chiral supermultiplets, and one with coordinates described by twisted chiral supermultiplets.  This was, perhaps, the earliest precursor of the concept of ``mirror symmetry.''  Clearly,
the similar emergence of pairs of SYK-like models shown in 
Equation (\ref{eq:4ptCV}) and Equation (\ref{eq:4ptCT})
raises the question of whether there can be manifestations of mirror symmetry in this domain?

Given that our approach has emphasized a starting point in 4D, $\cal N $ = 1 models, one can also contemplate only reduction to 2D, $\cal N$ = 2 models. This is the realm of superstring theory.  Thus, another possibility to explore is to
investigate are there interesting string theories that can be ``uplifted" SYK models to the two dimensional domain.

There's another point to note for future work. When one looks at the off-shell component Lagrangians, one sees many terms, and might think that the derivation of effective actions in bilocal fields would be very complicated. However, one should remember that the derivation of effective actions can be recasted in superspace \cite{SUSYSYK1}, and they are written in bilocal superfields \cite{SUSYSYK8}. Since in all of our ${\cal N}$ = 4 models we only have one superfield vertex, integrating over quenched disorder should give neat results.

\newpage



\vspace{.75in}

\noindent
{\bf {Acknowledgements}}\\[.1in] \indent

The research of S.\ J.\ G., Y.\ Hu, and S.-N.\ Mak is supported 
in part by the endowment of the Ford Foundation Professorship of Physics at Brown 
University and they gratefully acknowledge the support of the Brown Theoretical Physics 
Center. 
The authors thank Antal Jevicki, Jeff Murugan, Marcus Spradlin and Anastasia Volovich for discussions.

\vspace{.75in}

\noindent
{\bf {Added Note in Proof}}\\[.1in] \indent

Recall in Section \ref{subsec:AAD}, we reviewed the 1D, $\cn = 4$ supersymmetric SYK-type model in \cite{SUSYSYK2}. The off-shell component Lagrangian is given in Equation (\ref{eqn:AADcomp}),
\begin{equation}
\begin{split}
    \cl^{\rm (off-shell)} ~=&~ \pa_{\t} \Bar{\phi}_{\a} \pa_{\t} \phi_{\a} ~+~ \Bar{\psi}_{\a} \pa_{\t} \psi_{\a} ~-~ \Bar{F}_{\a} F_{\a} \\
    &~+~ \Big[~ \O_{\a\b\g} \, \big(~ \phi_{\a} \phi_{\b} F_{\g} ~+~ \psi_{\a} \e \psi_{\b} \phi_{\g} ~\big) ~+~ {\rm h.\,c.} ~\Big] ~~~.
\end{split}
\end{equation}
If one go on-shell, the equation of motion for $F$ is
\begin{equation}
    \Bar{F}_{\g} ~=~ \O_{\a\b\g} \, \phi_{\a} \phi_{\b} ~~~,
\end{equation}
and one would find the interaction terms to be
\begin{equation}
    \cl_{\rm int}^{\rm (on-shell)} ~=~ \O_{\a\b\g} \Bar{\O}_{\a\d\e} \, \phi_{\b} \phi_{\g} \Bar{\phi}_{\d} \Bar{\phi}_{\e} ~+~ \Big[~ \O_{\a\b\g} \psi_{\a} \e \psi_{\b} \phi_{\g} ~+~ {\rm h.\,c.} ~\Big] ~~~.
\end{equation}
It should be noted that this model is a supersymmetric bosonic SYK-type model. Therefore, our paper gives the first supersymmetrizations of fermionic SYK models with $\cn = 4$ supersymmetries.

\newpage
\appendix
\section{Explicit Calculations for ${\cal N} = 1$ Hamiltonian\label{appen:Hcalculation}}

Start from the supercharge \cite{SUSYSYK1}
\begin{equation}
    Q ~=~ i \sum_{1\leq i<j<k \leq N} C_{ijk} \psi^{i} \psi^{j} \psi^{k} ~~~,
\end{equation}
and the Hamiltonian is given by
\begin{equation}
\begin{split}
     {\cal H} ~=&~ Q^{2} ~=~ \fracm12\, \{ Q,\,Q\}\\
     ~=&~ -\fracm12\,\sum_{1\leq i<j<k \leq N}\sum_{1\leq l<m<n \leq N} C_{ijk}C_{lmn}\,\{\psi^{i} \psi^{j} \psi^{k},\,\psi^{l} \psi^{m} \psi^{n}\} ~~~.
\end{split}
\end{equation}
Then we can explicitly discuss three different cases ${\cal H} = H_1 + H_2 + H_3$. 
\begin{enumerate}
    \item $i = l, j = m, k = n$, 
    \begin{equation}
        \begin{split}
            H_1 ~=&~ -\fracm12\,\sum_{1\leq i<j<k \leq N} C^2_{ijk}\,2\,\psi^{i} \psi^{j} \psi^{k} \psi^{i} \psi^{j} \psi^{k}\\
            ~=&~ \sum_{1\leq i<j<k \leq N} C^2_{ijk}\,(\psi^{i})^2 (\psi^{j})^2 (\psi^{k})^2 \\
            ~=&~ \fracm18\, \sum_{1\leq i<j<k \leq N} C^2_{ijk} ~~~.
        \end{split}
    \end{equation}
    \item one of $\{ i,j,k\}$ = one of $\{ l,m,n\}$,
    \begin{equation}
        \begin{split}
            H_2 ~=&~ -\fracm12\,\sum_{a} \sum_{1\leq j<k<m<n \leq N}\,\fracm{4!}{2!2!}\, C_{ajk}C_{amn}\,2\,\psi^{a} \psi^{j} \psi^{k} \psi^{a} \psi^{m} \psi^{n}\\
            ~=&~ -\sum_{a} \sum_{1\leq j<k<m<n \leq N}\,\fracm{4!}{2!2!}\, C_{ajk}C_{amn}\,(\psi^{a})^2 \psi^{j} \psi^{k} \psi^{m} \psi^{n}\\
            ~=&~ -\fracm{4!}{8}\,\sum_{a} \sum_{1\leq j<k<m<n \leq N}\, C_{ajk}C_{amn}\, \psi^{j} \psi^{k} \psi^{m} \psi^{n}\\
            ~=&~ -\fracm{1}{8}\,\sum_{a} \sum_{1\leq j<k<m<n \leq N}\, C_{a[jk}C_{mn]a}\, \psi^{j} \psi^{k} \psi^{m} \psi^{n} ~~~.
        \end{split}
    \end{equation}
    \item one of $\{ i,j,k\}$ = one of $\{ l,m,n\}$, $H_3 = 0$. 
\end{enumerate}

Therefore, we have 
\begin{equation}
    {\cal H} ~=~ \fracm18\, \sum_{1\leq i<j<k \leq N} C^2_{ijk}~-~\fracm{1}{8}\,\sum_{a} \sum_{1\leq j<k<m<n \leq N}\, C_{a[jk}C_{mn]a}\, \psi^{j} \psi^{k} \psi^{m} \psi^{n} ~~~.
\end{equation}

For the $E_0$ term, in order to further convince ourselves that the $\fracm18$ coefficient is correct, we can also look at an example and set $N=3$. When $N=3$, there is only one term in the supercharge,
\begin{equation}
    Q ~=~ i\,C_{123}\,\psi^{1} \psi^{2} \psi^{3} ~~~,
\end{equation}
and the Hamiltonian is 
\begin{equation}
    {\cal H} ~=~ i\,C_{123}\,\psi^{1} \psi^{2} \psi^{3} i\,C_{123}\,\psi^{1} \psi^{2} \psi^{3} ~=~ \fracm18\,C^2_{123} ~~~.
\end{equation}

\newpage
\section{On-Shell Lagrangian Derivation Examples}
In this Appendix, we will show three on-shell Lagrangian calculation examples: CS+3PT model, CS+nCS-A model, and CS+VS+nVS-B model.  

We use the standard approach: 1.) write the part of off-shell Lagrangian that includes all terms involving auxiliary fields; 2.) do the variation and get the equations of motion (EoMs); 3.) solve the EoMs and substitute the solution to the original Lagrangian. 

\subsection{CS + 3PT-A\label{appen:CS+3PT-A}}


In this section, we will present the step-by-step derivations towards the on-shell Lagrangian for the CS+3PT model. 
There are only two fields $\bX$ and $\bbX$ in this model are auxiliary. 
The part of the Lagrangian $\cl_{{\rm {CS}}} + \cl_{3{\rm {PT}}}$ containing the auxiliary fields $\bX$ and $\bbX$ reads
\begin{equation}
\begin{split}
    \mathcal{L}_{\bX} ~=&~ \fracm12 \bX^{\ca} {\Bar\bX}_{\ca} \\
    &~-~ \fracm12 \, \k_{\ca\cb\cc} \, \Big\{~ {\Bar\bX}^{\ca} \bX^{\cb} {\bm \Phi}^{\cc} ~+~ i \, (\rP^{(+)})^{ab} {\Bar\bX}^{\ca} \Bj_{a}^{\cb} \Bj_{b}^{\cc}  ~\Big\}  \\
    &~-~ \fracm12 \, \k^{*}_{\ca\cb\cc} \, \Big\{~ \bX^{\ca} {\Bar\bX}^{\cb} {\Bar {\bm \Phi}}^{\cc} ~+~ i \, (\rP^{(-)})^{ab} \bX^{\ca} \Bj_{a}^{\cb} \Bj_{b}^{\cc}  ~\Big\}  ~~~.
\end{split}
\end{equation}
The equations of motion are
\begin{align}
    \bX^{\ca} ~\big[~ \d_{\ca\cd} ~-~ \k_{\cd\ca\cc} {\bm\Phi}^{\cc} ~-~ \k^{*}_{\ca\cd\cc} {\Bar{\bm\Phi}}^{\cc} ~\big] ~=&~ i \k_{\cd\cb\cc} \, (\rP^{(+)})^{ab} \Bj_{a}^{\cb} \Bj_{b}^{\cc}  ~~~, \\
    {\Bar\bX}^{\ca} ~\big[~ \d_{\ca\cd} ~-~ \k^{*}_{\cd\ca\cc} {\Bar{\bm\Phi}}^{\cc} ~-~ \k_{\ca\cd\cc} {\bm\Phi}^{\cc} ~\big] ~=&~ i \k^{*}_{\cd\cb\cc} \, (\rP^{(-)})^{ab} \Bj_{a}^{\cb} \Bj_{b}^{\cc} ~~~,
\end{align}
where they are conjugate of each other. Solving them, we have
\begin{align}
    \bX^{\ca} ~=&~ \frac{ i \k_{\cd\cb\cc} \, (\rP^{(+)})^{ab} \Bj_{a}^{\cb} \Bj_{b}^{\cc} }{ \d_{\ca\cd} ~-~ \k_{\cd\ca\cg} {\bm\Phi}^{\cg} ~-~ \k^{*}_{\ca\cd\cg} {\Bar{\bm\Phi}}^{\cg} }  ~~~,  \\
    {\Bar\bX}^{\ca} ~=&~ \frac{ i \k^{*}_{\cd\cb\cc} \, (\rP^{(-)})^{ab} \Bj_{a}^{\cb} \Bj_{b}^{\cc} }{ \d_{\ca\cd} ~-~ \k^{*}_{\cd\ca\cg} {\Bar{\bm\Phi}}^{\cg} ~-~ \k_{\ca\cd\cg} {\bm\Phi}^{\cg} } ~~~.
\end{align}
By substituting the equations of motion and going on-shell, the auxiliary Lagrangian becomes
\begin{equation}
\begin{split}
{\cal L}_{\bX} ~=&~ - i \, \frac12 \, \k_{\ca\cb\cc} \, (\rP^{(+)})^{ab} {\Bar\bX}^{\ca} \Bj_{a}^{\cb} \Bj_{b}^{\cc} \\
~=&~ \frac12 \, \frac{\k_{\ca\cb\cc} \, \k^{*}_{\cd\ce\cf} \, (\rP^{(+)})^{ab} \, (\rP^{(-)})^{cd} }{\d_{\cd\ca} ~-~
\k^{*}_{\cd\ca\cg} {\Bar{\bm\Phi}}^{\cg} ~-~ \k_{\ca\cd\cg} {\bm\Phi}^{\cg} } ~ \Bj_{a}^{\cb} \Bj_{b}^{\cc}
\Bj_{c}^{\ce} \Bj_{d}^{\cf} \\
~=&~ \frac12 \, {\Big[} ~  \k_{\ca\cb\cc} \, \k^{*\ca}{}_{\ce\cf} \,+\, \k_{\ca\cb\cc} \, \k^{*}_{\cd\ce\cf} \,
\k^{*\ca\cd}{}_{\cg} \,  {\Bar{\bm\Phi}}^{\cg} \, + \,  \k_{\ca\cb\cc} \, \k^{*}_{\cd\ce\cf} \, \k^{\cd\ca}{}_{\cg}
\,  {\bm\Phi}^{\cg}   \\
&{~~~~} ~+~ \co (\k^{n+2} {\bm\Phi}^n) \,      \Big]    \, (\rP^{(+)})^{ab} \, (\rP^{(-)})^{cd}
 ~  \Bj_{a}^{\cb} \Bj_{b}^{\cc} \Bj_{c}^{\ce} \Bj_{d}^{\cf}  ~~~,
\end{split}
\end{equation}
if we expand the non-linear term. The first term is the SYK term. Note that if we define
\begin{equation}
    \cy_{\ca\cb} ~=~ \k^{*}_{\ca\cb\cc} {\Bar{\bm\Phi}}^{\cc} ~+~ \k_{\cb\ca\cc} {\bm \Phi}^{\cc} ~~~,
\end{equation}
we can write 
\begin{equation}
\begin{split}
    {\cal L}_{\bX} ~=&~ 
    \frac12 \, \frac{\k_{\ca\cb\cc} \, \k^{*}_{\cd\ce\cf} \, (\rP^{(+)})^{ab} \, (\rP^{(-)})^{cd} }{\d_{\cd\ca} ~-~ \cy_{\cd\ca} } ~ \Bj_{a}^{\cb} \Bj_{b}^{\cc} \Bj_{c}^{\ce} \Bj_{d}^{\cf} ~~~.\\
\end{split}
\end{equation}

The final on-shell Lagrangian is 
\begin{equation}
\begin{split}
    \cl^{\rm (on-shell)}_{\rm CS+3PT} ~=&~ 
    -~ \fracm{1}{2} \pa_{\m} \BF^{\ca} \pa^{\m} \bBF_{\ca} 
    ~+~ i \, \fracm{1}{2} (\g^{\m})^{ab} \Bj^{\ca}_{a} \pa_{\m} \Bj_{b\ca} \\
    &~+~\fracm12 \,\k{}_{{\ca}{\cb}{\cc}} \Big[\,
    ( \pa^\m \Bar{\bm\Phi}^{\ca} ) ( \pa_\m {\bm\Phi}^{\cb} ) {\bm\Phi}^{\cc}  
    ~+~ i \, 2 \, (\rP^{(-)}\g^{\m})^{ab} \, (\pa_{\mu}{\bm {\psi}}{}^{\ca}_a) {\bm {\psi}}{}^{\cb}_b {\bm\Phi}^{\cc}   \,\Big] \\
    &~+~\fracm12 \,\k^{*}_{{\ca}{\cb}{\cc}} \Big[\,
    ( \pa^\m {\bm\Phi}^{\ca} ) ( \pa_\m \Bar{\bm\Phi}^{\cb} ) \Bar{\bm\Phi}^{\cc}  
    ~+~ i \, 2 \, (\rP^{(+)}\g^{\m})^{ab} \, (\pa_{\mu}{\bm {\psi}}{}^{\ca}_a) {\bm {\psi}}{}^{\cb}_b \Bar{\bm\Phi}^{\cc}   \,\Big] \\
    &~+~\frac12 \, \frac{\k_{\ca\cb\cc} \, \k^{*}_{\cd\ce\cf} \, (\rP^{(+)})^{ab} \, (\rP^{(-)})^{cd} }{\d_{\cd\ca} ~-~ \cy_{\cd\ca} } ~ \Bj_{a}^{\cb} \Bj_{b}^{\cc} \Bj_{c}^{\ce} \Bj_{d}^{\cf}  ~~~.
    \end{split}
\end{equation}

\subsection{CS + nCS-A\label{appen:CS+nCS-A}}

In this section, we will present the step-by-step derivations towards the on-shell Lagrangian for the CS+nCS-A model. 
There are only two fields $\bX$ and $\bbX$ in this model are auxiliary. 

The part of the Lagrangian $\cl_{\rm CS} + \cl_{n{\rm CS-A}}$ containing the auxiliary field $\bX$ and $\bbX$ is
\begin{equation}
\begin{split}
    \cl_{\bX} ~=&~ \fracm12 \bX^{\ca} \bbX_{\ca} 
    ~-~ \fracm12 \, \cp'_{\ca\cb} (\bBF) \, \bX^{\ca} \bbX^{\cb}
    ~-~ \fracm12 \, \Bar{\cp}'_{\ca\cb} (\BF) \, \bbX^{\ca} \bX^{\cb} \\
    &~-~ i \, \fracm12 \, \cp''_{\ca\cb_1\cb_2} (\bBF) \, (\rP^{(-)})^{ab} \, \bX^{\ca} \, \Bj_{a}^{\cb_1} \Bj_{b}^{\cb_2}
    ~-~ i \, \fracm12 \, \Bar{\cp}''_{\ca\cb_1\cb_2} (\BF) \, (\rP^{(+)})^{ab} \, \bbX^{\ca} \, \Bj_{a}^{\cb_1} \Bj_{b}^{\cb_2} ~~~.
\end{split}
\end{equation}
The equations of motion are
\begin{align}
    \bX^{\ca} ~=~ i ~ \frac{ \Bar{\cp}''_{\cd\cb_{1}\cb_{2}} (\BF) }{ \d_{\cd\ca} ~-~ \Bar{\cp}'_{\cd\ca} (\BF) ~-~ \cp'_{\ca\cd} (\bBF) } ~ (\rP^{(+)})^{ab} \, \Bj_{a}^{\cb_{1}} \Bj_{b}^{\cb_{2}} ~~~,
\end{align}
and its conjugate. The Lagrangian $\cl_{\bX}$ then becomes
\begin{equation}
    \cl_{\bX} ~=~ \frac12 ~ \frac{ \Bar{\cp}''_{\ca\cb_{1}\cb_{2}} (\BF) \, \cp''_{\cd\cc_{1}\cc_{2}} (\bBF) }{ \d_{\ca\cd} ~-~ \Bar{\cp}'_{\ca\cd} (\BF) ~-~ \cp'_{\cd\ca} (\bBF) } ~ (\rP^{(+)})^{ab} \, (\rP^{(-)})^{cd} \, \Bj_{a}^{\cb_{1}} \Bj_{b}^{\cb_{2}} \Bj_{c}^{\cc_{1}} \Bj_{d}^{\cc_{2}} ~~~.
\end{equation}

Hence, the final on-shell Lagrangian is 
\begin{equation}
\begin{split}
    \cl^{\rm (on-shell)}_{{\rm CS} + n{\rm {CS-A}}} 
    ~=&  
    ~-~  \fracm{1}{2} \, \pa^{\m} \BF^{\ca} \pa_{\m} \bBF_{\ca}
    ~+~ i  \, \fracm{1}{2} \,(\g^\m)^{ab} \Bj^{\ca}_a \pa_\m  \Bj_{b \, \ca} \\
    &~+~\Big[ ~-~ \fracm12 \cp'_{\ca\cb}({\Bar{\bm \Phi}})\,\BF^{\ca}\,\Box\,\Bar{\BF}^{\cb}
    ~-~ \fracm12 \cp''_{\ca\cb_1\cb_2}({\Bar{\bm \Phi}})\,\BF^{\ca}\,(\pa_{\m}\Bar{\BF}^{\cb_1})\,(\pa^{\m}\Bar{\BF}^{\cb_2}) \\
    &~~~~~~~ ~-~ i\,\cp'_{\ca\cb}({\Bar{\bm \Phi}})\,(\rP^{(+)}\g^{\m})^{ab}\,{\bm \psi}_a^{\ca}\,\pa_{\mu}{\bm \psi}_b^{\cb} \\
    &~~~~~~~ ~-~ i\,\cp''_{\ca\cb_1\cb_2}({\Bar{\bm \Phi}})\,(\rP^{(+)}\g^{\m})^{ab}\,{\bm \psi}_a^{\ca}\,(\pa_{\mu}{\bm \bBF}^{\cb_1})\,{\bm \psi}_b^{\cb_2} 
    ~+~ {\rm h.\, c.} \, \Big] \\
    &~+~ \frac12 ~ \frac{ \Bar{\cp}''_{\ca\cb_{1}\cb_{2}} (\BF) \, \cp''_{\cd\cc_{1}\cc_{2}} (\bBF) }{ \d_{\ca\cd} ~-~ \Bar{\cp}'_{\ca\cd} (\BF) ~-~ \cp'_{\cd\ca} (\bBF) } ~ (\rP^{(+)})^{ab} \, (\rP^{(-)})^{cd} \, \Bj_{a}^{\cb_{1}} \Bj_{b}^{\cb_{2}} \Bj_{c}^{\cc_{1}} \Bj_{d}^{\cc_{2}} ~~~.
\end{split}
\end{equation}

\subsection{CS + VS + nVS-B\label{appen:CS+VS+nVS-B}}

In this section, we will present the step-by-step derivations towards the on-shell Lagrangian for the CS + VS + nVS-B model. 
There are three fields $\bX$, $\bbX$, and ${\bm {\rm d}}$ in this model are auxiliary.


From $\cl_{{\rm {CS}}} + \cl_{{\rm {VS}}}+ \cl_{n{\rm {VS-B}}}$, the Lagrangian that contains the auxiliary fields $\bX$ and $\bbX$ is
\begin{equation}
   \mathcal{L}_{\bX} ~=~ \fracm12 \bX^{\ca} {\Bar\bX}_{\ca} ~+~ i \fracm12 \, \bX^{\ca} \, \cf_{\ca} (\cj_{11}) ~-~ i \fracm12 \, \Bar{\bX}^{\ca} \, \Bar{\cf}_{\ca} (\Bar{\cj}_{11})    ~~~,
\end{equation}
and the equations of motion are
\begin{equation}
    \begin{split}
       \bX_{\ca}  ~=&~ i \, \Bar{\cf}_{\ca} (\Bar{\cj}_{11}) ~~~, \\
       \overline{\bX}_{\ca} ~=&~ - i \, \cf_{\ca} (\cj_{11}) ~~~.
    \end{split}
\end{equation}
Therefore, the $\bX$-part of the on-shell Lagrangian is
\begin{equation}
\begin{split}
    \cl_{\bX} ~=&~ -~ \fracm12 \, \cf^{\ca} (\cj_{11}) \, \Bar{\cf}_{\ca} (\Bar{\cj}_{11}) \\
    ~=&~ -~ \fracm12 \, \sum_{i,j=1}^{P}\,  (-1)^{i} \,  \k^{(i)\ca}{}_{\vcb_1\cdots\vcb_{2i}}\, \k^{(j)*}_{\ca\vcc_1\cdots\vcc_{2j}}\, 
    \prod_{k=1}^{i} \prod_{l=1}^{j} (\rP^{(+)})^{ a_{k}b_{k}} (\rP^{(-)})^{c_{l}d_{l}} \Bl_{a_{k}}^{\vcb_{2k-1}} \Bl_{b_{k}}^{\vcb_{2k}} \Bl_{c_{l}}^{\vcc_{2l-1}} \Bl_{d_{l}}^{\vcc_{2l}}  ~~~,
\end{split}
\end{equation}

The Lagrangian that contains the auxiliary field ${\bm {\rm d}}$ is
\begin{equation}
\begin{split}
    \mathcal{L}_{{\bm {\rm d}}} ~=&~  \fracm{1}{2} \, {\bm {\rm d}}^{\vcb}{\bm {\rm d}}_{\vcb} \\
    &~-~ \cf'_{\ca\vcb_1\vcb_2}(\cj_{11})\,(\rP^{(+)})^{ab}\,\bm{\psi}_a^{\ca}\, {\bm {\rm d}}^{\vcb_1}\,\Bl_b^{\vcb_2} 
    ~+~ \Bar{\cf}'_{\ca\vcb_1\vcb_2}(\Bar{\cj}_{11})\,(\rP^{(-)})^{ab}\,\bm{\psi}_a^{\ca}\, {\bm {\rm d}}^{\vcb_1}\,\Bl_b^{\vcb_2} \\
    &~+~ \bm{\Phi}^{\ca}\,\cf''_{\ca\vcb_1\vcb_2\vcb_3\vcb_4}(\cj_{11})\,\Big[\,   i \fracm12 \,(\rP^{(+)}\g^{\a\b})^{ab}\,{\bm {\rm d}}^{\vcb_1}\,\bF_{\a\b}^{\vcb_3}
    ~-~ \fracm12 \, (\rP^{(+)})^{ab}\,{\bm {\rm d}}^{\vcb_1}\,{\bm {\rm d}}^{\vcb_3}\,
    \Big]\,\Bl_a^{\vcb_2}\,\Bl_b^{\vcb_4}\\
    &~-~\Bar{\bm{\Phi}}^{\ca}\,\Bar{\cf}''_{\ca\vcb_1\vcb_2\vcb_3\vcb_4}(\Bar{\cj}_{11})\,\Big[\,  - i\fracm12 \,(\rP^{(-)}\g^{\a\b})^{ab}\,{\bm {\rm d}}^{\vcb_1}\,\bF_{\a\b}^{\vcb_3}
    ~-~ \fracm12 \, (\rP^{(-)})^{ab}\,{\bm {\rm d}}^{\vcb_1}\,{\bm {\rm d}}^{\vcb_3}\,
    \Big]\,\Bl_a^{\vcb_2}\,\Bl_b^{\vcb_4}\\
    &~-~ \fracm12\,\bm{\Phi}^{\ca}\,\cf'_{\ca\vcb_1\vcb_2}(\cj_{11})\,{\bm {\rm d}}^{\vcb_1}\,{\bm {\rm d}}^{\vcb_2} ~-~ \fracm12\,\Bar{\bm{\Phi}}^{\ca}\,\Bar{\cf}'_{\ca\vcb_1\vcb_2}(\Bar{\cj}_{11})\,{\bm {\rm d}}^{\vcb_1}\,{\bm {\rm d}}^{\vcb_2} ~~~.
    \end{split}
\end{equation}
The equation of motion for $\brd$ is
\begin{equation}
    \begin{split}
        \brd^{\vcd}~=&~ \frac{\Omega_{\vcb}}{\d_{\vcb\vcd}-\cy_{\vcb\vcd}} ~~~, 
    \end{split}
\end{equation}
where 
\begin{equation}
    \begin{split}
        \cy_{\vcb\vcd} ~=&~ 
        -~ \bm{\Phi}^{\ca}\,\cf''_{\ca\vcb\vcb_2\vcd\vcb_4}(\cj_{11})\,\cj_{11}^{\vcb_2\vcb_4} 
        ~+~ \bm{\Phi}^{\ca}\,\cf'_{\ca\vcb\vcd}(\cj_{11}) \\
        &~+~ \Bar{\bm{\Phi}}^{\ca}\,\Bar{\cf}''_{\ca\vcb\vcb_2\vcd\vcb_4}(\Bar{\cj}_{11})\,\Bar{\cj}_{11}^{\vcb_2\vcb_4} 
        ~+~ \Bar{\bm{\Phi}}\,\Bar{\cf}'_{\ca\vcb\vcd}(\Bar{\cj}_{11}) ~~~,
    \end{split}
\end{equation}
satisfying 
\begin{equation}
    \cy_{\vcb\vcd} ~=~ \cy_{\vcd\vcb} ~~~,~~~ (\cy_{\vcb\vcd})^* ~=~ \cy_{\vcb\vcd} ~~~,
\end{equation}
and
\begin{equation}
    \begin{split}
        \Omega_{\vcb}~=&~ \Big[\,\cf'_{\ca\vcb\vcc}(\cj_{11})\,(\rP^{(+)})^{ab}-\Bar{\cf}'_{\ca\vcb\vcc}(\Bar{\cj}_{11})\,(\rP^{(-)})^{ab}\,\Big]\, \Bj_a^{\ca}\,\Bl_b^{\vcc} \\
        &~-~ i \fracm12 \,\Big[\,\bm{\Phi}^{\ca}\,\cf''_{\ca\vcb\vcc_2\vcc_3\vcc_4}(\cj_{11})\,(\rP^{(+)}\g^{\m\n})^{ab}+\Bar{\bm{\Phi}}^{\ca}\,\Bar{\cf}''_{\ca\vcb\vcc_2\vcc_3\vcc_4}(\Bar{\cj}_{11})\,(\rP^{(-)}\g^{\m\n})^{ab}\,\Big]\,\bF_{\m\n}^{\vcc_3}\Bl_a^{\vcc_2}\Bl_b^{\vcc_4} ~~~.
    \end{split}
\end{equation}
Therefore, the ${\bm {\rm d}}$-part of the on-shell Lagrangian is
\begin{equation}
\begin{split}
    \cl_{{\bm {\rm d}}} ~=&~ -~ \fracm{1}{2} \, \brd^{\vcb} \O_{\vcb}  \\
    ~=&~ -\fracm12\,\frac{\Omega_{\vca}\Omega_{\vcb}}{\d_{\vca\vcb}-\cy_{\vca\vcb}} ~~~.
\end{split}
\end{equation}

The final on-shell Lagrangian is given by
\begin{equation}
\begin{split}
    \cl^{(\rm on-shell)}_{{\rm {CS+ VS + nVS-B}}}  ~=&
    ~-~  \fracm{1}{2} \, \pa^{\m} \BF^{\ca} \pa_{\m} \bBF_{\ca}
    ~+~ i  \, \fracm{1}{2} \,(\g^\m)^{ab} \Bj^{\ca}_a \pa_\m  \Bj_{b \, \ca} \\
    &~-~ \fracm{1}{4} \bF^{\vcb}_{\mu\nu} \bF_{\vcb}^{\mu\nu}  ~+~ i\,  \fracm{1}{2} \, (\gamma^\mu)^{a b} {\bm \l}^{\vcb}_a \partial_\mu {\bm \l}_b{}_{\vcb}  \\
    &~+~ \Big[\,  i \fracm12 \,\cf'_{\ca\vcb_1\vcb_2}(\cj_{11})\,(\rP^{(+)}\g^{\m\n})^{ab}\,\bm{\psi}_a^{\ca}\, \bF_{\m\n}^{\vcb_1}\,\Bl_b^{\vcb_2} \\
    &~~~~~~ ~-~ \fracm18 \bm{\Phi}^{\ca}\,\cf''_{\ca\vcb_1\vcb_2\vcb_3\vcb_4}(\cj_{11})\, (\rP^{(+)}\g^{\m\n}\g^{\a\b})^{ab}\,\bF_{\m\n}^{\vcb_1}\,\bF_{\a\b}^{\vcb_3} \Bl_a^{\vcb_2}\,\Bl_b^{\vcb_4}\\
    &~~~~~~ ~+~ i \,\bm{\Phi}^{\ca}\,\cf'_{\ca\vcb_1\vcb_2}(\cj_{11})\,(\rP^{(-)}\g^{\m})^{ab}\,(\pa_{\mu}\Bl_a^{\vcb_1})\,\Bl_b^{\vcb_2} \\
    &~~~~~~ ~+~ \fracm14 \,\bm{\Phi}^{\ca}\,\cf'_{\ca\vcb_1\vcb_2}(\cj_{11})\, \Big(\, \bF_{\m\n}^{\vcb_1}\,\bF^{\m\n}{}^{\vcb_2} 
    ~-~ i \fracm12 \, \e^{\m\n\a\b}\,\bF_{\m\n}^{\vcb_1}\,\bF_{\a\b}^{\vcb_2}   \,\Big) \\
    &~~~~~~ 
    ~+~ {\rm h.\,c.}     \,\Big] \\
    &~-~  \fracm12 \, \sum_{i,j=1}^{P}\,  (-1)^{i} \,  \k^{(i)\ca}{}_{\vcb_1\cdots\vcb_{2i}}\, \k^{(j)*}_{\ca\vcc_1\cdots\vcc_{2j}}\, \\
    &~~~~~~ ~\times~\prod_{k=1}^{i} \prod_{l=1}^{j} (\rP^{(+)})^{ a_{k}b_{k}} (\rP^{(-)})^{c_{l}d_{l}} \Bl_{a_{k}}^{\vcb_{2k-1}} \Bl_{b_{k}}^{\vcb_{2k}} \Bl_{c_{l}}^{\vcc_{2l-1}}\Bl_{d_{l}}^{\vcc_{2l}}\\
    &~-~ \fracm12\,\frac{\Omega_{\vca}\Omega_{\vcb}}{\d_{\vca\vcb}-\cy_{\vca\vcb}}  ~~~.
\end{split}
\end{equation}

\newpage
\section{An Example of Prohibited Action: nVS-A}
In Chapter \ref{sec:n-point}, we constructed the nCS-A and nTS-A models. We did not construct a similar action with the vector supermultiplet. 
In this appendix, we will show how the nVS-A fails to be a proper Lagrangian. 

The nVS-A is constructed via the integration over the whole superspace which involves a $q$-point superfield interaction among one chiral superfield and a polynomial in ${\bm {\rm d}}$ in the vector supermultiplet. The explicit form is given by
\begin{equation}
    \cl_{n{\rm {VS-A}}} ~=~ \fracm18 \, \rD^{a} \rD^{(+)}_{a} \, \rD^{b} \rD^{(-)}_{b} ~ \Big[~  \BF^{\ca} \cp_{\ca} ({\bm {\rm d}}) ~\Big] 
    ~+~ {\rm {h.\, c.}}  ~~~.
\end{equation}
The polynomial takes the form
\begin{equation}
    \cp_{\ca}({\bm {\rm d}}) ~=~ \sum_{i=1}^{P} \k^{(i)}_{\ca\vcb_1\cdots\vcb_i} \prod_{k=1}^{i} {\bm {\rm d}}^{\vcb_k}  ~~~,
\end{equation}
where $\k^{(i)}_{\ca\vcb_1\cdots\vcb_{i}}$'s are arbitrary coefficients, and the degree of the polynomial is $P$. 
Obviously, $\vcb_1$ to $\vcb_i$ indices for any $1\leq i \leq (P-1)$ on the coefficient $\k^{(i)}_{\ca\vcb_{1}\cdots\vcb_{i}}$ are symmetric.
We then have
\begin{align}
    \cp''_{\ca\vcb_1\vcb_2}({\bm {\rm d}}) ~=&~ 2\, \k^{(2)}_{\ca\vcb_1\vcb_2}  ~+~ \sum_{j=3}^{P} j(j-1) \, \k^{(j)}_{\ca\vcb_1\cdots\vcb_{j}} \prod_{k=3}^{j} {\bm {\rm d}}^{\vcb_k}  ~~~,  \\
    \cp'''_{\ca\vcb_1\vcb_2\vcb_3}({\bm {\rm d}}) ~=&~ 6\, \k^{(3)}_{\ca\vcb_1\vcb_2\vcb_3}  ~+~ \sum_{j=4}^{P} j (j-1) (j-2) \k^{(j)}_{\ca\vcb_1\cdots\vcb_{j}} \prod_{k=4}^{j} {\bm {\rm d}}^{\vcb_k}  ~~~, \\
    \cp''''_{\ca\vcb_1\vcb_2\vcb_3\vcb_4}({\bm {\rm d}}) ~=&~ 24\,\k^{(4)}_{\ca\vcb_1\vcb_2\vcb_3\vcb_4}  ~+~ \sum_{j=5}^{P} j (j-1) (j-2) (j-3) \k^{(j)}_{\ca\vcb_1\cdots\vcb_{j}} \prod_{k=5}^{j} {\bm {\rm d}}^{\vcb_k}  ~~~.
\end{align}

Then we can proceed and obtain the component Lagrangian as 
\begin{equation}
    \cl_{n{\rm {VS-A}}} 
    ~=~ -~ i \fracm12 \, ({\rP}^{(+)}\g^{\m}\g^{\n})^{ab}  \bX^{\ca} \, \cp''_{\ca\vcb_1\vcb_2}({\bm {\rm d}}) \, (\pa_{\m}\,\Bl^{\vcb_1}_{a})\,(\pa_{\n}\,\Bl^{\vcb_2}_{b}) ~+~ \cdots ~~~,
\end{equation}
where the first term is already problematic since it leads to dynamics for the fermions with derivatives acting on them which cannot be get rid of through integration by parts.

\newpage
\section{Numerical Values for Gamma Matrices\label{appen:convention}}
In this appendix, we give the numerical values of the matrices that appear in 1D actions. 

\begin{align}
    (\g^0)^{ab} ~=&~ 
    \begin{pmatrix}
    1 & 0 & 0 & 0\\
    0 & 1 & 0 & 0 \\
    0 & 0 & 1 & 0\\
    0 & 0 & 0 & 1
    \end{pmatrix}~~,~~
    C^{ab} ~=~ \begin{pmatrix}
    0&-1&0&0\\
1&0&0&0\\
0&0&0&1\\
0&0&-1&0\\
    \end{pmatrix}~~,\\
    (\rP^{(+)})^{ab} ~=&~ 
    \begin{pmatrix}
    0&$-\fracm12$&$\fracm{i}{2}$&0\\
$\fracm12$&0&0&$\fracm{i}{2}$\\
$-\fracm{i}{2}$&0&0&$\fracm12$\\
0&$-\fracm{i}{2}$&$-\fracm12$&0
\end{pmatrix}~~,~~
    (\rP^{(-)})^{ab} ~=~ 
    \begin{pmatrix}
    0&$-\fracm12$&$-\fracm{i}{2}$&0\\
$\fracm12$&0&0&$-\fracm{i}{2}$\\
$\fracm{i}{2}$&0&0&$\fracm12$\\
0&$\fracm{i}{2}$&$-\fracm12$&0
\end{pmatrix}~~,~~\\
    (\rP^{(+)}\g^0)^{ab} ~=&~ 
    \begin{pmatrix}
    $\fracm12$&0&0&$-\fracm{i}{2}$\\
0&$\fracm12$&$\fracm{i}{2}$&0\\
0&$-\fracm{i}{2}$&$\fracm12$&0\\
$\fracm{i}{2}$&0&0&$\fracm12$
\end{pmatrix}~~,~~
    (\rP^{(-)}\g^0)^{ab} ~=~ 
    \begin{pmatrix}
    $\fracm12$&0&0&$\fracm{i}{2}$\\
0&$\fracm12$&$-\fracm{i}{2}$&0\\
0&$\fracm{i}{2}$&$\fracm12$&0\\
$-\fracm{i}{2}$&0&0&$\fracm12$
\end{pmatrix}~~,~~
\end{align}

$(\rP^{(\pm)}\g^i)^{ab}$: 
\begin{align}
    (\rP^{(+)}\g^1)^{ab} ~=&~ 
\begin{pmatrix}
    $-\fracm12$&0&0&$\fracm{i}{2}$\\
0&$\fracm12$&$\fracm{i}{2}$&0\\
0&$-\fracm{i}{2}$&$\fracm12$&0\\
$-\fracm{i}{2}$&0&0&$-\fracm12$
\end{pmatrix}~~,~~
    (\rP^{(-)}\g^1)^{ab} ~=~ \begin{pmatrix}
    $-\fracm12$&0&0&$-\fracm{i}{2}$\\
0&$\fracm12$&$-\fracm{i}{2}$&0\\
0&$\fracm{i}{2}$&$\fracm12$&0\\
$\fracm{i}{2}$&0&0&$-\fracm12$
\end{pmatrix}~~,~~\\
    (\rP^{(+)}\g^2)^{ab} ~=&~ 
    \begin{pmatrix}
    0&$\fracm{i}{2}$&$-\fracm12$&0\\
$-\fracm{i}{2}$&0&0&$-\fracm12$\\
$-\fracm12$&0&0&$\fracm{i}{2}$\\
0&$-\fracm12$&$-\fracm{i}{2}$&0
\end{pmatrix}~~,~~
    (\rP^{(-)}\g^2)^{ab} ~=~ \begin{pmatrix}
    0&$-\fracm{i}{2}$&$-\fracm12$&0\\
$\fracm{i}{2}$&0&0&$-\fracm12$\\
$-\fracm12$&0&0&$-\fracm{i}{2}$\\
0&$-\fracm12$&$\fracm{i}{2}$&0
\end{pmatrix}~~,~~\\
    (\rP^{(+)}\g^3)^{ab} ~=&~ 
    \begin{pmatrix}
    0&$\fracm12$&$\fracm{i}{2}$&0\\
$\fracm12$&0&0&$-\fracm{i}{2}$\\
$-\fracm{i}{2}$&0&0&$-\fracm12$\\
0&$\fracm{i}{2}$&$-\fracm12$&0
\end{pmatrix}~~,~~
    (\rP^{(-)}\g^3)^{ab} ~=~ 
    \begin{pmatrix}
    0&$\fracm12$&$-\fracm{i}{2}$&0\\
$\fracm12$&0&0&$\fracm{i}{2}$\\
$\fracm{i}{2}$&0&0&$-\fracm12$\\
0&$-\fracm{i}{2}$&$-\fracm12$&0
\end{pmatrix}~~.
\end{align}

$(\rP^{(\pm)}\g^{0i})^{ab}$: 
\begin{align}
    (\rP^{(+)}\g^{01})^{ab} ~=&~ \begin{pmatrix}
    0&$\fracm12$&$-\fracm{i}{2}$&0\\
$\fracm12$&0&0&$\fracm{i}{2}$\\
$-\fracm{i}{2}$&0&0&$\fracm12$\\
0&$\fracm{i}{2}$&$\fracm12$&0
\end{pmatrix}~~,~~
    (\rP^{(-)}\g^{01})^{ab} ~=~ \begin{pmatrix}
    0&$\fracm12$&$\fracm{i}{2}$&0\\
$\fracm12$&0&0&$-\fracm{i}{2}$\\
$\fracm{i}{2}$&0&0&$\fracm12$\\
0&$-\fracm{i}{2}$&$\fracm12$&0
\end{pmatrix}~~,~~\\
    (\rP^{(+)}\g^{02})^{ab} ~=&~ \begin{pmatrix}
    $\fracm{i}{2}$&0&0&$-\fracm12$\\
0&$\fracm{i}{2}$&$\fracm12$&0\\
0&$\fracm12$&$-\fracm{i}{2}$&0\\
$-\fracm12$&0&0&$-\fracm{i}{2}$
\end{pmatrix}~~,~~
    (\rP^{(-)}\g^{02})^{ab} ~=~ \begin{pmatrix}
    $-\fracm{i}{2}$&0&0&$-\fracm12$\\
0&$-\fracm{i}{2}$&$\fracm12$&0\\
0&$\fracm12$&$\fracm{i}{2}$&0\\
$-\fracm12$&0&0&$\fracm{i}{2}$
\end{pmatrix}~~,~~\\
    (\rP^{(+)}\g^{03})^{ab} ~=&~ \begin{pmatrix}
    $\fracm12$&0&0&$\fracm{i}{2}$\\
0&$-\fracm12$&$\fracm{i}{2}$&0\\
0&$\fracm{i}{2}$&$\fracm12$&0\\
$\fracm{i}{2}$&0&0&$-\fracm12$
\end{pmatrix}~~,~~
    (\rP^{(-)}\g^{03})^{ab} ~=~ \begin{pmatrix}
    $\fracm12$&0&0&$-\fracm{i}{2}$\\
0&$-\fracm12$&$-\fracm{i}{2}$&0\\
0&$-\fracm{i}{2}$&$\fracm12$&0\\
$-\fracm{i}{2}$&0&0&$-\fracm12$
\end{pmatrix}~~.
\end{align}

$(\rP^{(\pm)}\g^{0i}\g^{0j})^{ab}$: 
\begin{align}
    (\rP^{(+)}\g^{01}\g^{01})^{ab} ~=&~ \begin{pmatrix}
    0&$-\fracm12$&$\fracm{i}{2}$&0\\
$\fracm12$&0&0&$\fracm{i}{2}$\\
$-\fracm{i}{2}$&0&0&$\fracm12$\\
0&$-\fracm{i}{2}$&$-\fracm12$&0
\end{pmatrix}~~,~~
    (\rP^{(-)}\g^{01}\g^{01})^{ab} ~=~ \begin{pmatrix}
    0&$-\fracm12$&$-\fracm{i}{2}$&0\\
$\fracm12$&0&0&$-\fracm{i}{2}$\\
$\fracm{i}{2}$&0&0&$\fracm12$\\
0&$\fracm{i}{2}$&$-\fracm12$&0
\end{pmatrix}~~,~~\\
    (\rP^{(+)}\g^{02}\g^{01})^{ab} ~=&~ \begin{pmatrix}
    $\fracm{i}{2}$&0&0&$-\fracm12$\\
0&$-\fracm{i}{2}$&$-\fracm12$&0\\
0&$-\fracm12$&$\fracm{i}{2}$&0\\
$-\fracm12$&0&0&$-\fracm{i}{2}$
\end{pmatrix}~~,~~
    (\rP^{(-)}\g^{02}\g^{01})^{ab} ~=~ \begin{pmatrix}
    $-\fracm{i}{2}$&0&0&$-\fracm12$\\
0&$\fracm{i}{2}$&$-\fracm12$&0\\
0&$-\fracm12$&$-\fracm{i}{2}$&0\\
$-\fracm12$&0&0&$\fracm{i}{2}$
\end{pmatrix}~~,~~\\
    (\rP^{(+)}\g^{03}\g^{01})^{ab} ~=&~ \begin{pmatrix}
    $\fracm12$&0&0&$\fracm{i}{2}$\\
0&$\fracm12$&$-\fracm{i}{2}$&0\\
0&$-\fracm{i}{2}$&$-\fracm12$&0\\
$\fracm{i}{2}$&0&0&$-\fracm12$
\end{pmatrix}~~,~~
    (\rP^{(-)}\g^{03}\g^{01})^{ab} ~=~ \begin{pmatrix}
    $\fracm12$&0&0&$-\fracm{i}{2}$\\
0&$\fracm12$&$\fracm{i}{2}$&0\\
0&$\fracm{i}{2}$&$-\fracm12$&0\\
$-\fracm{i}{2}$&0&0&$-\fracm12$
\end{pmatrix}~~,~~\\
    (\rP^{(+)}\g^{01}\g^{02})^{ab} ~=&~ \begin{pmatrix}
    $-\fracm{i}{2}$&0&0&$\fracm12$\\
0&$\fracm{i}{2}$&$\fracm12$&0\\
0&$\fracm12$&$-\fracm{i}{2}$&0\\
$\fracm12$&0&0&$\fracm{i}{2}$
\end{pmatrix}~~,~~
    (\rP^{(-)}\g^{01}\g^{02})^{ab} ~=~ \begin{pmatrix}
    $\fracm{i}{2}$&0&0&$\fracm12$\\
0&$-\fracm{i}{2}$&$\fracm12$&0\\
0&$\fracm12$&$\fracm{i}{2}$&0\\
$\fracm12$&0&0&$-\fracm{i}{2}$
\end{pmatrix}~~,~~\\
    (\rP^{(+)}\g^{02}\g^{02})^{ab} ~=&~ \begin{pmatrix}
    0&$-\fracm12$&$\fracm{i}{2}$&0\\
$\fracm12$&0&0&$\fracm{i}{2}$\\
$-\fracm{i}{2}$&0&0&$\fracm12$\\
0&$-\fracm{i}{2}$&$-\fracm12$&0
\end{pmatrix}~~,~~
    (\rP^{(-)}\g^{02}\g^{02})^{ab} ~=~ \begin{pmatrix}
    0&$-\fracm12$&$-\fracm{i}{2}$&0\\
$\fracm12$&0&0&$-\fracm{i}{2}$\\
$\fracm{i}{2}$&0&0&$\fracm12$\\
0&$\fracm{i}{2}$&$-\fracm12$&0
\end{pmatrix}~~,~~\\
    (\rP^{(+)}\g^{03}\g^{02})^{ab} ~=&~ \begin{pmatrix}
    0&$\fracm{i}{2}$&$\fracm12$&0\\
$\fracm{i}{2}$&0&0&$-\fracm12$\\
$\fracm12$&0&0&$\fracm{i}{2}$\\
0&$-\fracm12$&$\fracm{i}{2}$&0
\end{pmatrix}~~,~~
    (\rP^{(-)}\g^{03}\g^{02})^{ab} ~=~ \begin{pmatrix}
    0&$-\fracm{i}{2}$&$\fracm12$&0\\
$-\fracm{i}{2}$&0&0&$-\fracm12$\\
$\fracm12$&0&0&$-\fracm{i}{2}$\\
0&$-\fracm12$&$-\fracm{i}{2}$&0
\end{pmatrix}~~,~~\\
    (\rP^{(+)}\g^{01}\g^{03})^{ab} ~=&~ \begin{pmatrix}
    $-\fracm12$&0&0&$-\fracm{i}{2}$\\
0&$-\fracm12$&$\fracm{i}{2}$&0\\
0&$\fracm{i}{2}$&$\fracm12$&0\\
$-\fracm{i}{2}$&0&0&$\fracm12$
\end{pmatrix}~~,~~
    (\rP^{(-)}\g^{01}\g^{03})^{ab} ~=~ \begin{pmatrix}
    $-\fracm12$&0&0&$\fracm{i}{2}$\\
0&$-\fracm12$&$-\fracm{i}{2}$&0\\
0&$-\fracm{i}{2}$&$\fracm12$&0\\
$\fracm{i}{2}$&0&0&$\fracm12$
\end{pmatrix}~~,~~\\
    (\rP^{(+)}\g^{02}\g^{03})^{ab} ~=&~ \begin{pmatrix}
    0&$-\fracm{i}{2}$&$-\fracm12$&0\\
$-\fracm{i}{2}$&0&0&$\fracm12$\\
$-\fracm12$&0&0&$-\fracm{i}{2}$\\
0&$\fracm12$&$-\fracm{i}{2}$&0
\end{pmatrix}~~,~~
    (\rP^{(-)}\g^{02}\g^{03})^{ab} ~=~ \begin{pmatrix}
    0&$\fracm{i}{2}$&$-\fracm12$&0\\
$\fracm{i}{2}$&0&0&$\fracm12$\\
$-\fracm12$&0&0&$\fracm{i}{2}$\\
0&$\fracm12$&$\fracm{i}{2}$&0
\end{pmatrix}~~,~~\\
    (\rP^{(+)}\g^{03}\g^{03})^{ab} ~=&~ \begin{pmatrix}
    0&$-\fracm12$&$\fracm{i}{2}$&0\\
$\fracm12$&0&0&$\fracm{i}{2}$\\
$-\fracm{i}{2}$&0&0&$\fracm12$\\
0&$-\fracm{i}{2}$&$-\fracm12$&0
\end{pmatrix}~~,~~
    (\rP^{(-)}\g^{03}\g^{03})^{ab} ~=~ \begin{pmatrix}
    0&$-\fracm12$&$-\fracm{i}{2}$&0\\
$\fracm12$&0&0&$-\fracm{i}{2}$\\
$\fracm{i}{2}$&0&0&$\fracm12$\\
0&$\fracm{i}{2}$&$-\fracm12$&0
\end{pmatrix}~~.
\end{align}

$(\g^5\g^{\mu})^{ab}$:
\begin{align}
    (\g^5\g^{0})^{ab} ~=&~ \begin{pmatrix}
    0&0&0&-i\\
0&0&i&0\\
0&-i&0&0\\
i&0&0&0
\end{pmatrix}~~,~~
    (\g^5\g^{1})^{ab} ~=~ \begin{pmatrix}
    0&0&0&i\\
0&0&i&0\\
0&-i&0&0\\
-i&0&0&0
\end{pmatrix}~~,~~\\
    (\g^5\g^{2})^{ab} ~=&~ \begin{pmatrix}
    0&i&0&0\\
-i&0&0&0\\
0&0&0&i\\
0&0&-i&0
\end{pmatrix}~~,~~
    (\g^5\g^{3})^{ab} ~=~ \begin{pmatrix}
    0&0&i&0\\
0&0&0&-i\\
-i&0&0&0\\
0&i&0&0
\end{pmatrix}~~.
\end{align}

\newpage
$$~~$$

\end{document}